%% file: main.tex
\documentclass{jfm}
\usepackage{iftex}
\ifPDFTeX
  \pdfoutput=1
\fi

\usepackage{graphicx}
\ifPDFTeX\usepackage{epstopdf}\fi
\usepackage{newtxtext}
\usepackage{newtxmath}
\usepackage{natbib}
\usepackage{booktabs}
\usepackage[dvipsnames,table]{xcolor}
\usepackage{bm}
\usepackage{amsmath,amssymb}
\usepackage{hyperref}
\hypersetup{colorlinks=true, urlcolor=blue, citecolor=black, linkcolor=black}

\graphicspath{{Figures/}}

\newcommand{\Nu}{\text{Nu}}

\renewcommand{\Pr}{\text{Pr}}
\renewcommand{\Re}{\text{Re}}

\newcommand{\dprime}{\prime \prime}
\newcommand{\uupp}{\widetilde{u^{\dprime }u^{\dprime }}}

\newcommand{\ttpp}{\widetilde{\theta^{\dprime }\theta^{\dprime }}}

\title{Pressure-strain redistribution as the mechanism for dissimilar heat transfer under spanwise wall oscillation waveforms}

\author{Lionel Agostini\aff{1} \corresp{\email{lionel.agostini@cnrs.fr}}
  \and C\'edric Flageul\aff{1}}

\affiliation{%
\aff{1}Institut Pprime, CNRS UPR 3346, Universit\'e de Poitiers, ISAE-ENSMA, Poitiers, France}

\begin{document}
\maketitle

%% ==================================================================
\input{abstract}


%% ==================================================================
\input{introduction}
\input{methodology}
\input{results}

\input{conclusion}

%% ==================================================================
\input{appendix}

\clearpage

%% ==================================================================
\backsection[Acknowledgements]{This project was provided with computing HPC and storage resources by GENCI at
TGCC thanks to the grants AD012A14284, A0172A07624 and A0152A07624 on the supercomputer Joliot
Curie's SKL partition.}

\backsection[Funding]{This work was supported by the French National Research Agency (ANR) under ANR-23-CE46-0004.}

%% ==================================================================

\input{main.bbl}
\end{document}

%% file: abstract.tex
\begin{abstract}
Spanwise wall oscillation can break the Reynolds analogy, enhancing convective heat transfer more than the accompanying drag, for suitable actuation parameters and waveform shapes; this preferential enhancement is termed dissimilar heat transfer. The companion study of \citet{guerin_PBO_2026} optimised both the actuation parameters and the waveform shape, establishing that an optimised quasi-plateau waveform attains an analogy factor $\overline{A}_n\approx 1.09$ at $\Pr=1$, and hypothesised that this enhancement is structural in origin, attributable to the absence of a pressure-strain redistribution channel in the temperature variance equation; the underlying mechanism has not hitherto been verified. The present investigation addresses this gap through phase-resolved analysis of the stochastic variance transport budgets of the streamwise velocity and temperature variances, derived from direct numerical simulation of turbulent channel flow at $\Re_\tau = 200$ and $\Pr=1$. The time-mean budgets reveal a conspicuous sign reversal of the pressure-strain term, from a modest drain in the unactuated buffer layer to a substantial near-wall source under actuation, a modification with no analogue in the scalar equation, though one that is locally dissipated and confers no net advantage in the mean. The phase-resolved budgets trace the dissimilarity to the diagonal pressure-strain redistribution, which acts through two temporally sequenced expressions. The first, at the Stokes-strain reversal, is a pronounced drain of streamwise variance by $\Pi_{uu}$ with no counterpart in the temperature variance, curtailing the velocity production whilst the scalar production continues undiminished. The second, during the quasi-steady plateau phases, is the maintenance by $\Pi_{vv}$ of the wall-normal fluctuation common to both fluxes, upon which a small mean-gradient asymmetry preferentially maintains the scalar flux. The off-diagonal pressure-strain correlations, examined through closing flux-magnitude budgets, are found to favour neither flux, so that the breaking of the analogy is governed by the diagonal redistribution alone. As the dissimilarity-producing action is concentrated within the reversal and plateau phases rather than at the Stokes-layer penetration maxima, the controlling parameter is proposed to be the duty cycle of the self-sustaining process \citep{agostini_duty_cycle_2026}, i.e.\ the fraction of the oscillation period spent in quasi-steady lingering, rather than the penetration depth. This supports the argument of \citet{guerin_PBO_2026} and offers a resolution of the paradox whereby increased penetration depth is not accompanied by increased dissimilarity.
\end{abstract}

%% file: introduction.tex
\section{Introduction}
\label{sec:intro}

The Reynolds analogy constitutes a fundamental relationship in wall-bounded turbulent flows, establishing that the turbulent transport of momentum and thermal energy proceed by similar mechanisms when the Prandtl number $\Pr = \nu/\alpha$ is unity, i.e. when the molecular diffusivities of momentum and heat are identical. Under this condition, the turbulent Prandtl number $\Pr_t = \nu_t/\alpha_t$ takes values close to unity throughout the buffer and log layers, as established by the direct numerical simulations (DNS) of \citet{kasagi_direct_1992}, \citet{kawamura_dns_1998} and \citet{kawamura_dns_1999} for a wide range of Prandtl numbers, and further confirmed at substantially higher Reynolds numbers by \citet{alcantara_dns_2021} and \citet{pirozzoli_dns_2022}; the implication is that, at $\Pr = 1$, the two transport mechanisms increase or decrease together by approximately the same amount under any modification of the flow. The condition $\Pr = 1$ accordingly constitutes the most demanding baseline for the generation of dissimilarity, as any departure from the Reynolds analogy must arise purely from turbulent redistribution rather than from molecular-level anisotropy between the scalar and momentum diffusivities. In numerous engineering applications, however, the objective is to enhance convective heat transfer whilst minimising the attendant drag penalty, or, alternatively, to reduce drag without sacrificing thermal performance. The ability to selectively break the Reynolds analogy, hereafter termed dissimilar heat transfer (DHT), is therefore of substantial practical importance, with direct implications for the design of energy-efficient thermal management systems, including concentrated solar power receivers \citep{ho_review_2014}.

The degree of dissimilarity is quantified through the analogy factor $\overline{A}_n = (\Nu/\Nu_0)/(C_f/C_{f,0})$, wherein $\Nu$ denotes the Nusselt number, $C_f$ the skin-friction coefficient, and the subscript $0$ refers to the unactuated reference flow at the same bulk Reynolds number; values exceeding unity therefore indicate preferential thermal enhancement relative to drag. The viability of breaking the Reynolds analogy at $\Pr = 1$, even in the canonical configuration in which the mean transport equations for the streamwise momentum and the scalar are rendered formally identical through a uniform volumetric heat source matched to the mean pressure gradient, was first established by \citet{hasegawa_dissimilar_2011} through suboptimal control of wall blowing and suction in a fully developed turbulent channel. The latter authors demonstrated, via the Fr\'{e}chet differential of the velocity and temperature responses to the actuation, that the divergence-free constraint imposed upon the velocity field, by introducing a pressure-mediated coupling between the three velocity components from which the passive scalar is structurally exempt, suffices in itself to engender a dissimilar response of the two fields, the resultant travelling-wave-like optimum attaining analogy factors as large as $\overline{A}_n \approx 1.5$. The broader taxonomy of routes through which the analogy may be broken was subsequently delineated by \citet{kasagi_control_2012}, who classified the sources of dissimilarity into two groups, namely those acting upon the averaged transport equations, through dissimilar source terms, a Prandtl number departing from unity, or dissimilar thermal boundary conditions, and those acting upon the fluctuating quantities, through the divergence-free constraint, the Prandtl-number dependence of the scalar fluctuations, or dissimilar boundary conditions imposed upon the velocity and scalar fluctuations. In the canonical configuration considered by the latter authors, in which the mean equations are rendered identical and the Prandtl number is fixed at unity with no-slip walls throughout, every averaged-equation route is closed by construction, and the divergence-free constraint upon the velocity field is thereby left as the sole admissible route to dissimilarity. This structural argument, namely that the pressure term present in the momentum equations and absent from the scalar equation furnishes the fundamental route to dissimilarity at $\Pr = 1$, is the cornerstone of the present investigation. The wall blowing and suction considered above is, however, only one of several actuation strategies through which this structural route has been exploited to achieve DHT, further examples including the optimised transpiration of \citet{yamamoto_optimal_2013}, the streamwise travelling wave-like wall deformation of \citet{uchino_dissimilarity_2017} and the surface modifications, such as riblets, examined by \citet{rouhi_riblet_2022}. In the more specific context of spanwise wall-based actuation, it is worth distinguishing between two fundamentally different routes through which DHT may be achieved, as this distinction bears directly upon the mechanism investigated in the present work.

The first route operates in the drag-reduction regime and exploits a molecular-diffusivity asymmetry between the momentum and scalar fields at $\Pr > 1$. Through DNS at $\Re_\tau = 590$ and $\Pr \in [0.71, 20]$, \citet{Rouhi_Hultmark_Smits_2025} demonstrated that spanwise wall oscillations operating in the drag-reduction regime can reduce heat transfer more effectively than drag at Prandtl numbers greater than unity; at $\Pr = 7.5$, the heat-transfer reduction was found to exceed the drag reduction by a substantial margin. The mechanism responsible is the differential thinning of the conductive sublayer relative to the viscous sublayer at elevated Prandtl number, which displaces the energetically dominant thermal scales closer to the wall and renders them more susceptible to Stokes-layer modulation than their momentum counterparts. At $\Pr = 1$, however, the conductive and viscous sublayers are coincident by definition, and this sublayer-thinning mechanism is entirely absent. This corollary has been confirmed directly by \citet{nabae_dissimilar_2026}, who demonstrated through DNS of turbulent Couette flow at $\Pr = 1$ that spanwise wall-oscillation travelling waves operating in the drag-reduction regime produce $\overline{A}_n \approx 1$ across the entire parameter space examined: the Reynolds analogy is thereby preserved.

The second route, and the one investigated in the present work, requires a structural asymmetry between the governing transport equations themselves and operates in the drag-increase regime at $\Pr = 1$. Purely temporal, spatially-uniform spanwise wall oscillations (SWO) constitute a well-characterised technique whose parameter space has been mapped extensively in the drag-reduction literature \citep{choi_mechanism_2001,quadrio_critical_2004,quadrio_streamwise-travelling_2009,leschziner_review_2020,ricco_review_2021}. The relevance of this body of work to the present scalar-transport problem, although these studies addressed neither heat transfer nor the analogy factor, resides in the dynamical agent through which the actuation operates: the near-wall streaks. The latter have been shown to be reorganised, attenuated and intermittently regenerated by the cyclic Stokes-layer straining \citep{touber_near-wall_2012,yakeno_modification_2014,agostini_turbulence_2015}. The streaks and the quasi-streamwise vortices of the self-sustaining process constitute the primary mechanism of wall-normal turbulent mixing in the buffer layer \citep{hamilton_regeneration_1995,waleffe_self-sustaining_1997,schoppa_coherent_2002,jimenez_near-wall_2013}, and the same near-wall structures correlate the velocity and scalar fields closely in the unactuated flow \citep{abe_correlation_2009}; their modulation by SWO is therefore expected, in principle, to act upon both the momentum and the scalar transfer simultaneously. That this expectation is borne out in the heat-transfer problem at $\Pr = 1$ has been established by \citet{guerin_preferential_2024}, whose DNS at $\Re_\tau = 180$ demonstrated that SWO operating in the drag-reduction regime attenuates both drag and heat transfer in concert, the Reynolds analogy being thereby preserved; the streak-suppression and weakening of ejection-sweep events impose the same response upon the two transport mechanisms. The latter observation is consistent with the earlier large-eddy simulations of \citet{fang_heat_2010}, who reported, for sinusoidal SWO at $T^+ \approx 104$ and amplitudes up to $W^+ \approx 19$, that the quadrant contributions to the Reynolds shear stress $\overline{u^{\prime}v^{\prime}}$ and the wall-normal turbulent heat flux $\overline{t^{\prime}v^{\prime}}$ vary in concert under the actuation, and that the correlation between the phase-filtered skin-friction coefficient and Stanton number remains as high as $0.90$ across the entire range of amplitudes examined; the Reynolds analogy is thereby retained throughout the drag-reduction regime. It is only when the oscillation period is extended into the drag-increase regime that the two responses decouple, as established by \citet{guerin_preferential_2024}: at $T^+=500$ and $W^+=30$, a configuration which increases drag by $7.7\%$, the heat transfer is found to increase by $14.9\%$, resulting in $\overline{A}_n = 1.064$. This result constitutes the first demonstration that a purely temporal spanwise wall motion can break the Reynolds analogy at $\Pr = 1$; as molecular-diffusivity differences are excluded under this condition and the source term difference (constant and variable, spatially, for the momentum and for the scalar, respectively) was shown to have no substantial dissimilar effect, the analogy breaking must necessarily originate from a structural asymmetry between the momentum and scalar transport equations themselves, of the kind first identified by \citet{hasegawa_dissimilar_2011} in the context of wall transpiration.

The result of \citet{guerin_preferential_2024} established the feasibility of DHT through SWO at a single, non-optimised operating point; however, two questions were left unanswered: whether the dissimilarity can be maximised through an appropriate choice of the waveform parameters, and through which physical mechanism the actuation breaks the Reynolds analogy. The first question has been addressed in the companion study of \citet{guerin_PBO_2026}, in which the first systematic optimisation of SWO waveforms for DHT was established through a Policy-Based Optimisation framework coupled with Large-Eddy Simulation (PBO-LES). That investigation identified two principal results pertinent to the present work. First, sinusoidal optimisation at maximum amplitude ($T^+=325$, $W^+=40$) achieves $\overline{A}_n = 1.091$, which constitutes the sinusoidal optimum against which non-sinusoidal waveforms are hereafter compared. Second, waveform optimisation at $25\%$ reduced amplitude ($T^+=350$, $W^+=30$) identifies a quasi-plateau configuration, namely a near-square-wave profile characterised by extended phases of nearly constant wall velocity separated by rapid reversals, which achieves a dissimilarity ($\overline{A}_n = 1.087$) almost identical to the sinusoidal baseline whilst delivering it at appreciably lower energetic cost: the control input is reduced by $7.4\%$ and the absolute heat-transfer enhancement, defined as the dimensional change in the Nusselt number relative to the unactuated reference, $\Delta\Nu = \Nu - \Nu_0$, is increased by $37.9\%$ over the sinusoidal optimum. The implication is that waveform topology constitutes a more effective design variable for DHT than amplitude increase. The mechanism through which this quasi-plateau waveform achieves preferential thermal enhancement was further hypothesised by \citet{guerin_PBO_2026} to involve the pressure-strain redistribution $\Pi_{uu}$, which arises in the velocity variance budget through the divergence-free constraint ($\nabla \cdot \bm{u} = 0$) and has no analogue in the passive scalar equation; this hypothesis was, however, advanced on qualitative grounds, and was explicitly identified by the authors themselves as requiring quantitative verification through full computation of the phase-resolved variance transport budgets. Furnishing precisely that verification is the principal aim of the present investigation.

A complementary motivation for the budget-level analysis is provided by a further observation reported in the companion study, namely the decoupling of DHT performance from the wall-normal penetration depth of the Stokes layer. In the SWO drag-reduction literature, the Stokes-layer protrusion height $\ell_{0.01}^+$, defined as the wall-normal distance at which the phase-averaged spanwise velocity attains $1\%$ of its wall-imposed amplitude, has long been advanced as a primary indicator of control effectiveness, on the premise that a deeper penetration of the spanwise momentum into the buffer layer produces a more substantial disruption of the streak-regeneration cycle \citep{quadrio_critical_2004,touber_near-wall_2012,rouhi_les_2022}. As the streak-regeneration cycle is, by the argument advanced above, also the agent of scalar mixing in the buffer layer, the natural extrapolation of this drag-reduction intuition to the heat-transfer problem would predict that DHT enhancement should likewise increase with $\ell_{0.01}^+$; the optimisation results of \citet{guerin_PBO_2026}, however, contradict this prediction. The sinusoidal optimum at $W^+=40$ achieves the highest dissimilarity ($\overline{A}_n=1.091$) despite possessing the shallowest protrusion ($\ell_{0.01}^+ = 54.18$), whilst the sinusoidal case at $W^+=30$ exhibits the deepest protrusion ($\ell_{0.01}^+ = 71.80$) and concurrently the lowest dissimilarity ($\overline{A}_n=1.068$); the quasi-plateau optimum at $W^+=30$ falls in between on both metrics. The implication is that the spatial measure of Stokes-layer penetration neither ranks the configurations correctly with respect to DHT performance nor selects the optimal period, and that the controlling mechanism for the analogy breaking resides not in the vertical reach of the spanwise momentum but, rather, in a feature of its temporal organisation. The qualitative resolution proposed in the companion study, namely the duration of the lingering plateau phases, over which the Stokes strain evolves only slowly, i.e.\ the duty-cycle ratio of the self-sustaining process \citep{agostini_duty_cycle_2026}, is the working hypothesis to be tested at the budget level in the present work. To the best of the present authors' knowledge, no prior investigation has compared the velocity and temperature variance budgets directly under SWO actuation, nor has the phase-resolved evolution of the pressure-strain terms been examined under non-sinusoidal waveforms; the connection between waveform topology, pressure-strain redistribution and DHT performance therefore remains unestablished.

The present study addresses these open issues through phase-resolved analysis of the stochastic variance transport budgets for the optimal quasi-plateau waveform of \citet{guerin_PBO_2026}. DNS is performed at $\Re_\tau = 200$ and $\Pr = 1$, and the phase-averaged budgets for both the streamwise velocity variance $\uupp$ and the temperature variance $\ttpp$ are computed over $25$ complete actuation cycles. Three specific objectives are pursued: first, to identify, through direct examination of the budgets, the term or combination of terms whose differential modulation by the actuation is responsible for the production asymmetry between the velocity and temperature variances, the absence of a pressure-strain channel in the scalar equation being the candidate to be tested; second, to characterise the temporal organisation of this differential modulation across the actuation cycle, with a view to establishing how the lingering plateau phases act upon the budgets; and, third, to articulate the chain of causation connecting the structural asymmetry of the transport equations to the macroscopic dissimilarity indicator $\overline{A}_n$, so that the mechanism may be transferred to other waveform topologies and operating conditions in subsequent work. The resulting causal chain is assembled as an annotated phase timeline in figure~\ref{fig:mechanism_schematic}, which the reader is invited to consult as a road map throughout the analysis.

The remainder of this paper is organised as follows. Section~\ref{sec:methodology} presents the governing equations, the variance transport framework and the numerical configuration. Section~\ref{sec:results} presents the results, beginning with a time-averaged characterisation of the statistical modifications induced by the actuation, before proceeding through the phase-resolved analysis from the coherent fields and stochastic fluctuations to the variance budgets and pressure-strain terms. Section~\ref{sec:conclusion} summarises the principal results and discusses their implications.

%% file: methodology.tex
\section{Problem formulation and numerical methodology}
\label{sec:methodology}

\subsection{Governing equations and configuration}
\label{sec:methodology:governing}

The flow configuration consists of a fully developed turbulent channel flow of half-height $\delta$, driven by a body force enforcing a constant streamwise mass-flow rate and subjected to spanwise wall oscillations, with a passive scalar transported at $\Pr = 1$. The streamwise, wall-normal and spanwise directions are denoted by $x$, $y$ and $z$, with corresponding velocity components $u$, $v$ and $w$; the temperature field is denoted by $\theta$. Periodic boundary conditions are applied in the streamwise and spanwise directions, statistical homogeneity in these directions being thereby ensured, whilst the wall-normal direction is bounded by two parallel walls at $y=\pm\delta$. The non-dimensional incompressible Navier--Stokes equations and the advection--diffusion equation for the temperature, expressed in wall units defined by the unactuated reference friction velocity $u_{\tau,0}$, kinematic viscosity $\nu$ and reference friction temperature $\theta_{\tau,0}$, are
\begin{equation}
	\frac{\partial u_i^+}{\partial x_i^+} = 0, \qquad
	\frac{\partial u_i^+}{\partial t^+} + u_j^+\frac{\partial u_i^+}{\partial x_j^+} = -\frac{\partial p^+}{\partial x_i^+} + \frac{\partial^2 u_i^+}{\partial x_j^{+}\partial x_j^{+}} + f_i^+, \qquad
	\frac{\partial \theta^+}{\partial t^+} + u_j^+\frac{\partial \theta^+}{\partial x_j^+} = \frac{1}{\Pr}\frac{\partial^2 \theta^+}{\partial x_j^{+}\partial x_j^{+}} + q^+,
	\label{eq:gov_eq}
\end{equation}
where $f_i = \left( f_x, 0, 0\right)$ is the spatially constant body-force term that enforces a constant streamwise mass-flow rate and $q$ the volumetric heat source that enforces a constant bulk temperature. The friction Reynolds number $\Re_\tau = u_{\tau,0}\delta / \nu = 200$ is identical to that of the companion study \citep{guerin_PBO_2026}. As the actuation modifies the wall shear stress at fixed mass-flow rate, the friction velocity of the actuated flow differs from that of the unactuated reference; all wall-unit quantities reported in the present work, in both the actuated and unactuated cases, are normalised by the single unactuated reference set $(u_{\tau,0},\,\theta_{\tau,0})$, so that the two flows are rendered directly comparable upon a common scale.

The thermal boundary condition adopted at both walls is of mixed type (MBC) and warrants a brief recapitulation, as the structural distinction between the velocity and temperature equations exploited throughout the present work is itself rooted in the form of the wall conditions. The MBC fixes the wall temperature $\theta_w$ to a constant value, taken here as $\theta_w=0$, whilst the volumetric heat source $q$ is adjusted so that the bulk temperature $\theta_b = \delta^{-1}\!\int_0^{\delta}\!\overline{\theta}\,\mathrm{d}y$ remains constant in time. This choice provides a stationary thermal statistical state without imposing a uniform heat flux at the wall: the instantaneous wall heat flux is permitted to fluctuate with the turbulent and coherent activity, whilst its time-average is set by global energy balance against $q^+$. The MBC therefore mirrors, on the thermal side, the constant mass-flow-rate condition that governs the velocity field through $f_i^+$, the integrated wall-normal gradient of each mean profile being thereby conserved by the boundary specification, an integral constraint that is invoked at several points in the discussion that follows. The complete derivation of the MBC formulation, including the specific form of $q^+$ and its sensitivity at $\Pr=1$, is provided in \citet{guerin_preferential_2024,guerin_PBO_2026} and in the references therein.

\begin{figure}
	\centering
	\includegraphics[width=0.7\textwidth]{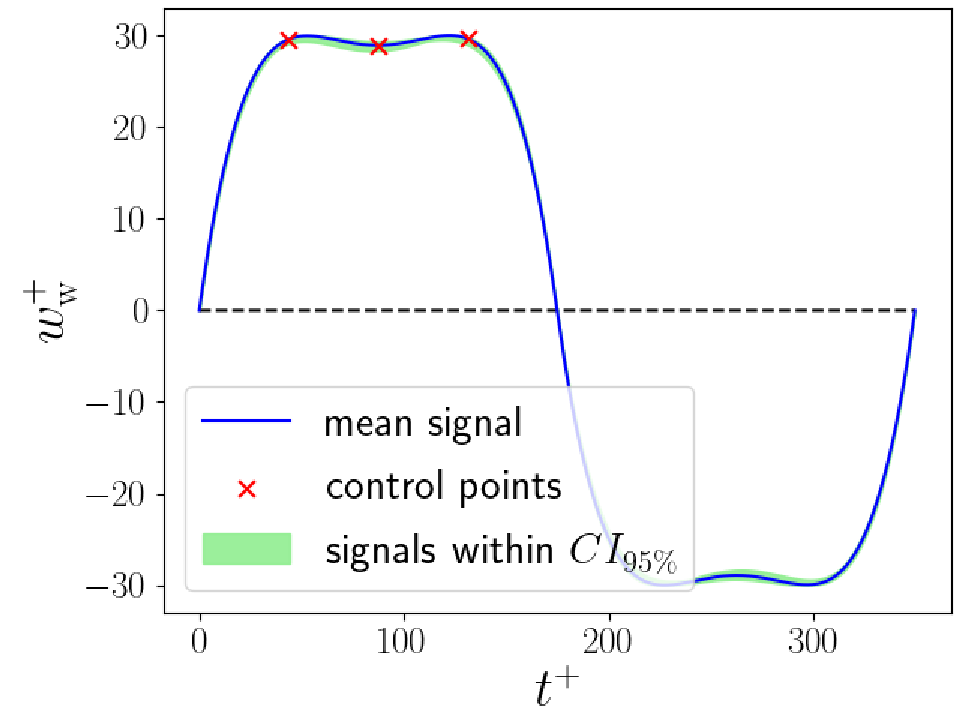}
	\caption{Quasi-plateau waveform $w^+_w(t^*)$ employed in the present work, reproduced from \citet{guerin_PBO_2026}. The wall spanwise velocity is held nearly constant at $\pm W^+_\text{max} = \pm 30$ over extended plateau intervals and transitions rapidly between the two extrema during short reversal intervals; the cycle phase $t^* = t/T \in [0,1]$ is normalised by the actuation period $T^+ = 350$.}
	\label{fig:waveform}
\end{figure}

The spanwise wall oscillation is imposed through the boundary condition $w_w(t) = W^+_\text{max}\, g(t/T)$, where $g$ denotes the waveform function, $W^+_\text{max}$ the maximum amplitude, and $T$ the oscillation period. The waveform $g$ employed throughout the present work is the quasi-plateau topology identified by \citet{guerin_PBO_2026} as the optimum of the four-dimensional waveform optimisation, parameterised through Lagrangian polynomial interpolation. The complete analytical expression and the polynomial coefficients are reported in \citet{guerin_PBO_2026}; for the purposes of the present analysis, it is sufficient to recall that this waveform comprises two extended plateau intervals of nearly constant spanwise wall velocity at $\pm W^+_\text{max}$ separated by short reversal intervals over which the velocity transitions rapidly between the two plateau values, as illustrated in figure~\ref{fig:waveform}. The plateau intervals are accordingly designated as the lingering phases of the cycle, in which the Stokes strain rate $\partial^2\widetilde{w}/\partial y\,\partial t$ is small over the buffer-layer region, whilst the reversal intervals concentrate the Stokes-strain activity into narrow temporal bands. This mixed derivative $\partial^2\widetilde{w}/\partial y\,\partial t$ is the rate of change of the Stokes strain and constitutes the gate of the duty-cycle modulation of the self-sustaining process by spanwise oscillation \citep{agostini_duty_cycle_2026}: where its magnitude is large, at the reversals, the self-sustaining process is interrupted; where it is small, through the lingering plateaux, the process recommences. It is the temporal counterpart of the acceleration parameter $a^+ = W^+/T^+$ shown there to govern the drag-reduction efficiency, so that the lingering fraction introduced below is the duty-cycle ratio expressed in the present waveform. The phase-resolved structure of $\widetilde{w}^+(y^+,t^*)$ generated by this waveform, including the spatio-temporal organisation of the lingering and reversal intervals, is presented in \S\,\ref{sec:results:coherent} (figure~\ref{fig:stokes_wall}\textit{a}) and is exploited throughout the analysis as the coherent forcing acting upon the variance budgets. The optimal quasi-plateau waveform is characterised by $T^+ = 350$ and $W^+ = 30$ and achieves $\overline{A}_n = 1.087$, the Nusselt number being enhanced by $+22.3\%$ and the skin-friction coefficient by $+12.4\%$ relative to the unactuated reference.

The governing equations are integrated using the high-order finite-difference solver \textit{Xcompact3d} \citep{bartholomew_xcompact3d_2020}, in which the spatial derivatives are evaluated through sixth-order compact schemes on a Cartesian mesh and the temporal advancement is effected through a third-order Runge--Kutta scheme; the divergence-free constraint is enforced through a fractional-step method, the associated Poisson equation for the pressure being solved spectrally by means of three-dimensional fast Fourier transforms. The complete numerical configuration, including the validation of the solver for the present class of actuated thermal channel flows, is documented in \citet{guerin_preferential_2024,guerin_PBO_2026}.

Direct numerical simulations are performed within a computational domain of streamwise, wall-normal and spanwise extents $(L_x, L_y, L_z) = (24\delta,\, 2\delta,\, 6\delta)$, discretised on a grid comprising $400 \times 221 \times 200$ points in the streamwise, wall-normal and spanwise directions, respectively. The resulting spatial resolution is $\Delta x^+ \approx 10.7$ and $\Delta z^+ \approx 5.3$ in the homogeneous directions, whilst the wall-normal spacing is stretched from $\Delta y^+_\text{min} \approx 0.43$ at the wall to $\Delta y^+_\text{max} \approx 6.2$ at the channel centreline, this resolution being sufficient to resolve the near-wall turbulent and thermal scales at the present Reynolds and Prandtl numbers. Statistical convergence is ensured through collection over $N = 25$ complete oscillation cycles following an initial transient of five cycles, the latter having been verified to be sufficient for the phase-averaged wall quantities to attain a statistically reproducible cycle, and corresponding to a total integration time of $8750$ viscous time units for the phase-averaged quantities.

\subsection{Triple decomposition and phase averaging}
\label{sec:methodology:triple}

In turbulent flows subjected to deterministic periodic actuation, a triple decomposition is employed whereby the instantaneous velocity field $u_i(\bm{x}, t)$ is separated into three distinct components:
\begin{equation}
	u_i(\bm{x}, t) = \underbrace{\overline{u}_i(\bm{x})}_{\text{Time average}} + \underbrace{\hat{u}_i(\bm{x}, t)}_{\text{Periodic fluctuation}} + \underbrace{u_i^{\dprime}(\bm{x}, t)}_{\text{Stochastic fluctuation}}
	\label{eq:triple_decomp}
\end{equation}
where $\overline{u}_i$ denotes the time average over the entire observation period. In the present channel flow configuration, statistical homogeneity in the streamwise and spanwise directions permits all averaging operations, whether temporal, phase-resolved, or statistical, to be performed simultaneously over time and the homogeneous directions; thus $\overline{u}_i$ depends only on the wall-normal coordinate $y$, and all fluctuating quantities are likewise averaged over $x$, $z$, and $t$ (or phase) with no spatial dependence in the homogeneous directions retained. The same decomposition is applied to the temperature field $\theta$. The phase-averaging operator $\langle \cdot \rangle_{\phi}$ is introduced to isolate the coherent motion locked to the actuation cycle. For an actuation period $T$, the phase average of a quantity $f$ at a specific phase $\phi$ ($0 \le \phi < T$) is defined as
\begin{equation}
	\widetilde{f}(\bm{x}, \phi) \equiv \langle f(\bm{x}, t) \rangle_{\phi} = \frac{1}{N} \sum_{n=0}^{N-1} f(\bm{x}, \phi + nT)
	\label{eq:phase_avg}
\end{equation}
The phase-averaged field thus comprises both the time-mean and the coherent periodic component, $\widetilde{u}_i = \overline{u}_i + \hat{u}_i$. Throughout the analysis that follows, the cycle phase is reported in normalised form through $t^* \equiv \phi/T \in [0,1)$, the actuation cycle being thereby mapped onto the unit interval irrespective of the period $T$. The stochastic turbulent fluctuation is defined as
\begin{equation}
	u_i^{\dprime}(\bm{x}, t) = u_i(\bm{x}, t) - \widetilde{u}_i(\bm{x}, t)
	\label{eq:stochastic_fluct}
\end{equation}
By construction, $\langle u_i^{\dprime} \rangle_\phi \equiv 0$ at every phase; the phase average of the products $\widetilde{u_i^{\dprime} u_j^{\dprime}}$, however, is non-zero and varies throughout the actuation cycle, reflecting the modulation of turbulence intensity by the Stokes strain.

A simplification specific to the present forcing should be noted before proceeding to the variance budgets. For a spatially homogeneous spanwise actuation, the coherent wall-normal velocity $\hat{v}$ vanishes identically throughout the domain: the phase-averaged continuity equation, applied to a flow statistically homogeneous in the streamwise and spanwise directions, reduces to $\partial\hat{v}/\partial y = 0$, and the no-penetration condition at the wall then leads to $\hat{v} \equiv 0$ everywhere. As a consequence, the coherent cross-product $\hat{u}\hat{v}$ is identically zero, and the stochastic shear stress $\widetilde{u^{\dprime}v^{\dprime}}$ equals the total phase-averaged Reynolds shear stress $\widetilde{u'v'}$ for the present configuration. The same conclusion applies to the wall-normal turbulent heat flux $\widetilde{v^{\dprime}\theta^{\dprime}}$, since $\hat{v}\equiv 0$ removes the coherent contribution $\hat{v}\hat{\theta}$ irrespective of the behaviour of $\hat{\theta}$. It is to be emphasised that this equivalence holds for the wall-normal cross-products alone, and not for the diagonal variances: as the coherent streamwise velocity $\hat{u}$ does not vanish, $\widetilde{u'u'} = \hat{u}\hat{u} + \widetilde{u^{\dprime}u^{\dprime}}$ retains a coherent contribution, and the distinction between $u_i'$ and $u_i^{\dprime}$ is, for the streamwise variance, without consequence only insofar as the cross-gradient production term is concerned.

\subsection{Variance transport equations}
\label{sec:methodology:budgets}

The analysis that follows is centred upon the variance transport equations, which describe how the streamwise velocity variance $\uupp$ and the temperature variance $\ttpp$ change in space and time under the action of the turbulent motion. These equations decompose the rate of change of each variance into contributions from four distinct physical processes: production, which generates variance from the mean gradients; pressure-strain redistribution, which reallocates energy amongst velocity components through the action of the pressure field; dissipation, which removes variance through molecular viscosity and diffusivity; and diffusion, which transports variance from one region of the flow to another. The fundamental insight motivating the present investigation is that the velocity variance budget contains a pressure-strain term that has no counterpart in the temperature variance budget, a structural difference that must underlie any dissimilarity between the two fields at $\Pr = 1$.

The budgets are presented throughout in their phase-resolved form, which retains the time-derivative on the left-hand side and describes how each contribution evolves through the actuation cycle. The time-mean budget, which describes the balance between source and sink terms in the statistically stationary state and governs the structure of the actuated flow, is then recovered as a particular case: averaging the phase-resolved equation over a complete actuation cycle, the left-hand-side time-derivative vanishes identically, the phase-averaged variance returning to the same value at the end of each period in the statistically periodic state, and every right-hand-side term reduces to its long-time mean. The time-mean operator $\overline{(\cdot)}$ employed here denotes simultaneous averaging over the streamwise direction, the spanwise direction and the entire observation period, in accordance with the statistical homogeneity established in \S\,\ref{sec:methodology:triple}, and is thereby mathematically equivalent to averaging the phase-resolved field over a complete actuation cycle. The budgets are formulated in terms of the stochastic fluctuation $u_i^{\dprime}$, since this isolates the genuinely turbulent contribution from the coherent component induced by the actuation; the equivalence $\widetilde{u^{\dprime}v^{\dprime}} = \widetilde{u'v'}$ established in \S\,\ref{sec:methodology:triple} ensures that the production terms, which depend on the wall-normal cross-products, are unaffected by this choice.

The phase-resolved budgets for the streamwise velocity variance $\widetilde{u^{\dprime}u^{\dprime}}^+$ and the temperature variance $\widetilde{\theta^{\dprime}\theta^{\dprime}}^+$ are
\begin{equation}
	\frac{\partial \widetilde{u^{\dprime}u^{\dprime}}^+}{\partial t^+} =
	\underbrace{-2 \widetilde{u^{\dprime}v^{\dprime}}^+ \frac{\partial \widetilde{u}^+}{\partial y^+}}_{P_{uu}^+}
	+ \underbrace{2\, \widetilde{p^{\dprime +} \frac{\partial u^{\dprime +}}{\partial x^+}}}_{\Pi_{uu}^+}
	- \underbrace{2\, \widetilde{\frac{\partial u^{\dprime +}}{\partial x_k^+} \frac{\partial u^{\dprime +}}{\partial x_k^+}}}_{\epsilon_{uu}^+}
	+ \underbrace{\frac{\partial}{\partial y^+} \left( \frac{\partial \widetilde{u^{\dprime}u^{\dprime}}^+}{\partial y^+} - \widetilde{u^{\dprime}u^{\dprime}v^{\dprime}}^+ \right)}_{D_{uu}^+}
	\label{eq:budget_uu}
\end{equation}
\begin{equation}
	\frac{\partial \widetilde{\theta^{\dprime}\theta^{\dprime}}^+}{\partial t^+} =
	\underbrace{-2\, \widetilde{v^{\dprime}\theta^{\dprime}}^+ \frac{\partial \widetilde{\theta}^+}{\partial y^+}}_{P_{\theta\theta}^+}
	- \underbrace{2\, \widetilde{\frac{\partial \theta^{\dprime +}}{\partial x_k^+} \frac{\partial \theta^{\dprime +}}{\partial x_k^+}}}_{\epsilon_{\theta\theta}^+}
	+ \underbrace{\frac{\partial}{\partial y^+} \left( \frac{\partial \widetilde{\theta^{\dprime}\theta^{\dprime}}^+}{\partial y^+} - \widetilde{v^{\dprime}\theta^{\dprime}\theta^{\dprime}}^+ \right)}_{D_{\theta\theta}^+}
	\label{eq:budget_thth}
\end{equation}
Each term is identified by the bracketed label appearing beneath it: $P$ denotes the production, defined as the product of the turbulent flux and the mean gradient of the corresponding scalar or velocity component; $\Pi$ denotes the pressure-strain redistribution, present in the velocity budget alone; $\epsilon$ denotes the dissipation, defined in pseudo-dissipation (homogeneous) form $\epsilon_{uu}^+ = 2\,\overline{(\partial u^{\dprime+}/\partial x_k^+)(\partial u^{\dprime+}/\partial x_k^+)}$ with the residual viscous-transport contribution absorbed into the diffusion term; and $D$ denotes the diffusion, comprising the molecular and turbulent transport of variance, written above as the wall-normal Laplacian of the variance augmented by the triple-correlation transport. In the scalar budget, the molecular diffusion appears with coefficient unity, the diffusivity $1/\Pr$ of the governing equation~\eqref{eq:gov_eq} reducing to unity at $\Pr = 1$. The budgets close to within the residual reported in \S\,\ref{sec:results:time_avg} under these conventions. The corresponding time-mean budgets, which underpin the time-averaged analysis of \S\,\ref{sec:results:time_avg}, follow by setting the left-hand-side time-derivative to zero and replacing the phase average $\widetilde{(\cdot)}$ by the time-mean $\overline{(\cdot)}$ throughout, in accordance with the cyclic stationarity noted above.

The fundamental structural asymmetry between the velocity and scalar budgets is now apparent from a direct comparison of equations~\eqref{eq:budget_uu} and~\eqref{eq:budget_thth}: the streamwise velocity variance budget contains a pressure-strain term $\Pi_{uu}$, which arises through the divergence-free constraint ($\nabla \cdot \bm{u} = 0$) imposed upon the velocity field and which redistributes energy amongst the three velocity components, whereas no analogous redistribution mechanism appears in the temperature variance equation, the passive scalar evolving under pure advection--diffusion dynamics. A more detailed analysis of the tensor-rank origin of this asymmetry, in the framework of the rapid/slow decomposition of the fluctuating pressure, is provided in appendix~\ref{app:rapid_slow}.

The pressure-strain correlation tensor is defined as
\begin{equation}
	\Pi_{ij} = \left\langle p^{\dprime} \left( \frac{\partial u_i^{\dprime}}{\partial x_j} + \frac{\partial u_j^{\dprime}}{\partial x_i} \right) \right\rangle_\phi
	\label{eq:pi_ij}
\end{equation}
with diagonal components
\begin{equation}
	\Pi_{uu} = 2\left\langle p^{\dprime} \frac{\partial u^{\dprime}}{\partial x} \right\rangle_\phi, \qquad
	\Pi_{vv} = 2\left\langle p^{\dprime} \frac{\partial v^{\dprime}}{\partial y} \right\rangle_\phi, \qquad
	\Pi_{ww} = 2\left\langle p^{\dprime} \frac{\partial w^{\dprime}}{\partial z} \right\rangle_\phi,
	\label{eq:pi_diag}
\end{equation}
which satisfy the trace-free condition $\Pi_{uu} + \Pi_{vv} + \Pi_{ww} = 0$ as a direct consequence of the incompressibility constraint $\partial u_i^{\dprime}/\partial x_i = 0$, the sum of the diagonal pressure-strain components reducing to $2\,\langle p^{\dprime}\,\partial u_i^{\dprime}/\partial x_i\rangle_\phi$. It is to be noted that equation~\eqref{eq:pi_diag} defines the pressure-strain, that is the redistributive, part of the pressure correlation alone; the pressure-transport contribution, which takes the form of a divergence, vanishes identically for $\Pi_{uu}$ ($\Pi_{ww}$) under streamwise (spanwise) homogeneity, whilst for the $vv$ component budget it persists and is treated as a separate transport term. As only the $uu$ and $\theta\theta$ budgets are formulated explicitly in the present work, the components $\Pi_{vv}$ and $\Pi_{ww}$ entering the discussion solely through the trace-free constraint, the redistributive definition adopted here is sufficient for the analysis that follows. The off-diagonal terms relevant to the flux budgets are
\begin{equation}
	\Pi_{uv} = \left\langle p^{\dprime} \left( \frac{\partial u^{\dprime}}{\partial y} + \frac{\partial v^{\dprime}}{\partial x} \right) \right\rangle_\phi, \qquad
	\Pi_{v\theta} = \left\langle p^{\dprime} \frac{\partial \theta^{\dprime}}{\partial y} \right\rangle_\phi,
	\label{eq:pi_offdiag}
\end{equation}
where $\Pi_{uv}$ appears in the transport equation for $\widetilde{u^{\dprime}v^{\dprime}}$ and $\Pi_{v\theta}$ in that for $\widetilde{v^{\dprime}\theta^{\dprime}}$. In the present flow configuration, both turbulent fluxes are negative; the sign-reversed quantities $-\Pi_{uv}$ and $-\Pi_{v\theta}$ are therefore the ones acting as source terms for the respective flux magnitudes $-\widetilde{u^{\dprime}v^{\dprime}}$ and $-\widetilde{v^{\dprime}\theta^{\dprime}}$ when positive, and as sink terms when negative. As these flux magnitudes appear directly in the production terms $\widetilde{P}_{uu} = -2\widetilde{u^{\dprime}v^{\dprime}}\,\partial\widetilde{u}/\partial y$ and $\widetilde{P}_{\theta\theta} = -2\widetilde{v^{\dprime}\theta^{\dprime}}\,\partial\widetilde{\theta}/\partial y$, any differential action of $-\Pi_{uv}$ and $-\Pi_{v\theta}$ upon the two fluxes would propagate directly to the production asymmetry; whether the off-diagonal pressure in fact favours either flux is examined in \S\,\ref{sec:results:off_diagonal}, the supporting flux-magnitude budgets, together with the full phase-resolved transport equations for $\widetilde{u^{\dprime}v^{\dprime}}^+$ and $\widetilde{v^{\dprime}\theta^{\dprime}}^+$, being provided in appendix~\ref{app:offdiag}. The structural asymmetry on which the present analysis principally rests is rather the absence from the scalar variance budget of the diagonal pressure-strain redistribution that is present in the velocity budget.

%% file: results.tex
\section{Results and discussion}
\label{sec:results}

That a spanwise oscillation should enhance heat transfer more than it penalises friction, the favourable breaking of the Reynolds analogy quantified by $\overline{A}_n \approx 1.09$ \citep{guerin_PBO_2026}, is an anomaly to be explained at the level of the governing equations. The momentum and the passive scalar obey transport equations that differ in but one structural respect, namely the presence of the pressure in the former and its absence from the latter; \citet{guerin_PBO_2026} conjectured that this single difference is the origin of the dissimilarity, a conjecture advanced without direct verification. The present section accordingly undertakes to trace the dissimilarity to its dynamical source, following the variance budgets through successive levels of resolution and withholding the identification of the mechanism until the evidence requires it. A time-averaged characterisation of the statistical fingerprint of the actuation (\S\,\ref{sec:results:time_avg}) furnishes the first clues; the coherent response of the mean fields (\S\,\ref{sec:results:coherent}) and the phase modulation of the stochastic fluctuations (\S\,\ref{sec:results:stochastic}) narrow the search to the production terms, the phase-resolved behaviour of which (\S\,\ref{sec:results:production}) exposes a sharply localised asymmetry whose origin is then pursued. The trail leads to the diagonal pressure-strain redistribution and the rate at which the imposed Stokes strain is reversed (\S\,\ref{sec:results:pressure_strain}); the more obvious suspect, a pressure acting directly and preferentially upon the turbulent fluxes, is confronted and eliminated (\S\,\ref{sec:results:off_diagonal}, the supporting flux-magnitude budgets being deferred to appendix~\ref{app:offdiag}). The complete causal chain is assembled in the synthesis of \S\,\ref{sec:results:synthesis}.

\subsection{Time-averaged characterisation}
\label{sec:results:time_avg}

\begin{figure}
	\centering
	\begin{minipage}{0.32\textwidth}
		\centering
		\includegraphics[width=\textwidth]{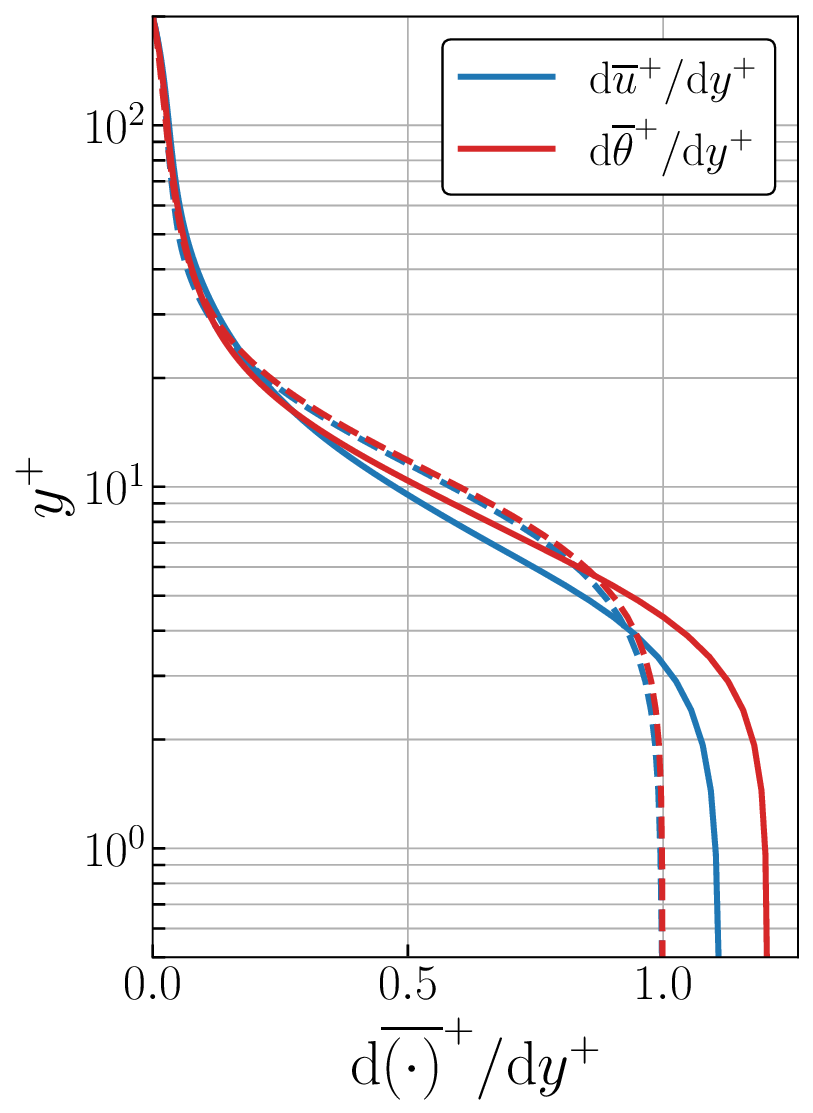}
		\vspace{0.2cm}
		\centerline{(\textit{a})}
	\end{minipage}
	\hfill
	\begin{minipage}{0.32\textwidth}
		\centering
		\includegraphics[width=\textwidth]{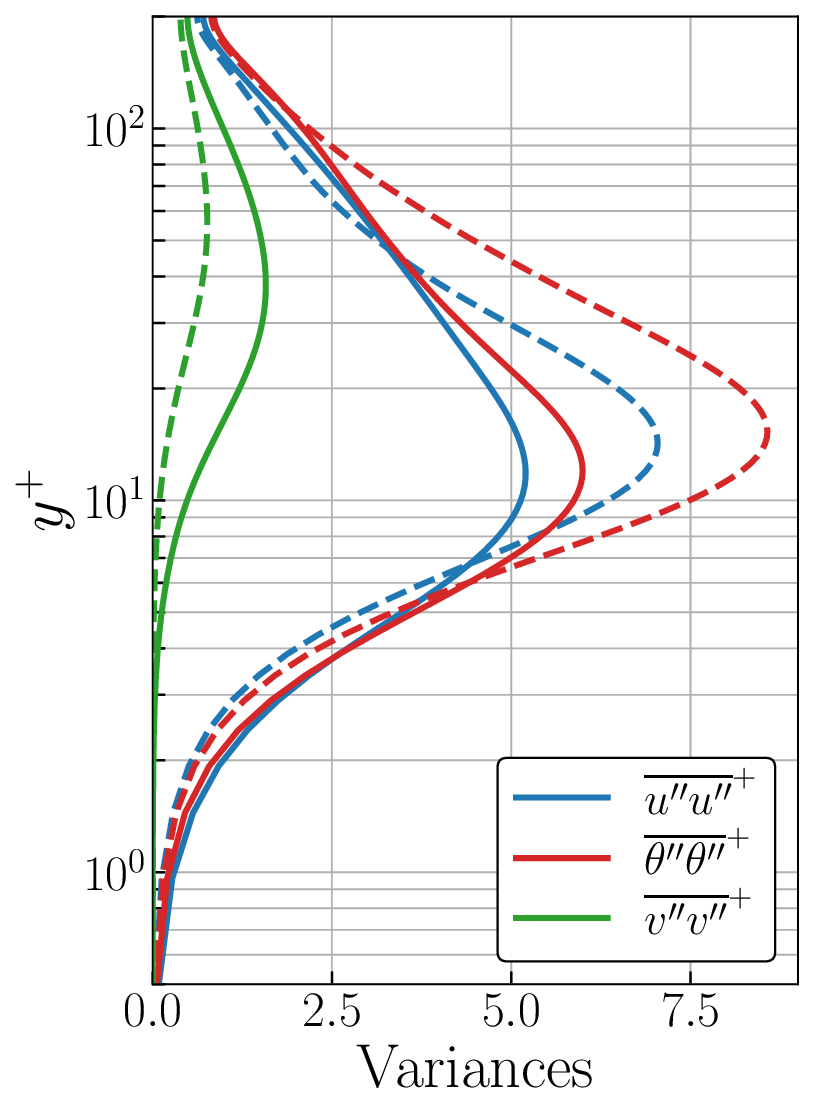}
		\vspace{0.2cm}
		\centerline{(\textit{b})}
	\end{minipage}
	\hfill
	\begin{minipage}{0.32\textwidth}
		\centering
		\includegraphics[width=\textwidth]{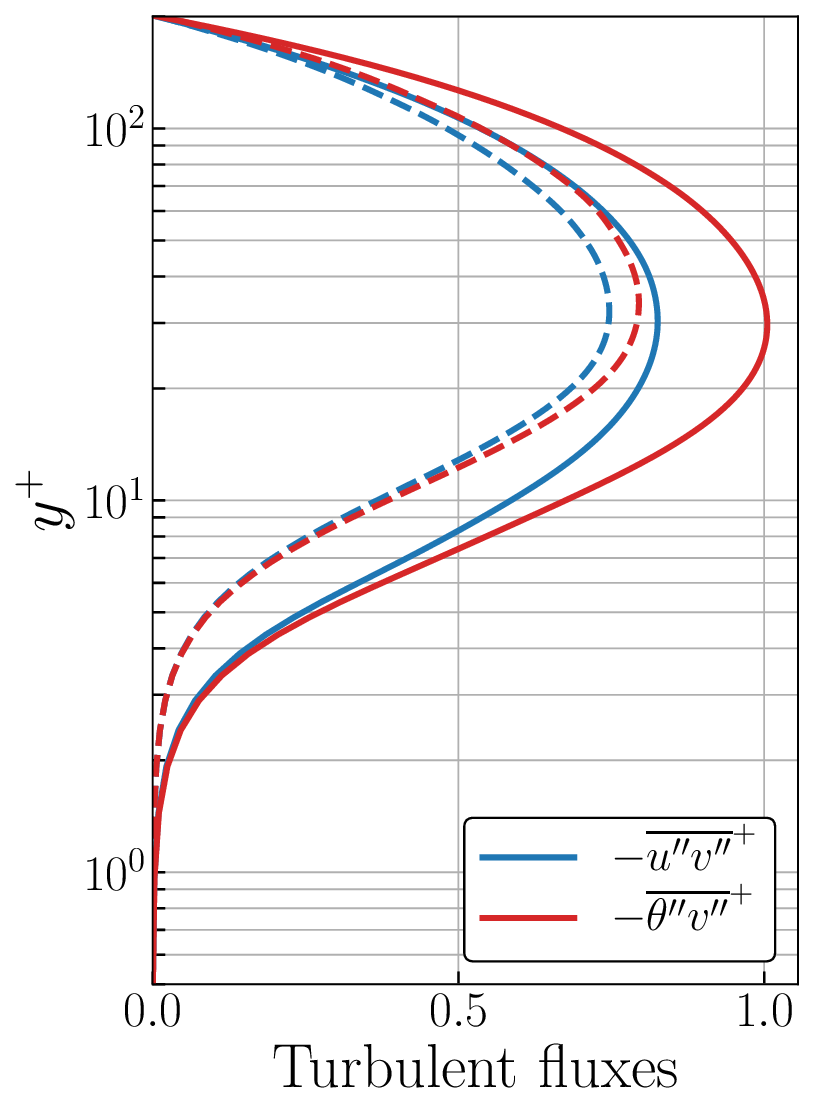}
		\vspace{0.2cm}
		\centerline{(\textit{c})}
	\end{minipage}
	\caption{Time-averaged wall-normal profiles comparing the unactuated baseline (dashed) and the actuated ($T^+=350$, $W^+=30$; solid) cases. Colour denotes the variable: momentum/velocity quantities in blue, scalar/thermal quantities in red, and the wall-normal velocity variance in green. (\textit{a}) Mean gradients $\mathrm{d}\overline{u}^+/\mathrm{d}y^+$ (blue) and $\mathrm{d}\overline{\theta}^+/\mathrm{d}y^+$ (red). (\textit{b}) Streamwise velocity variance $\overline{u^{\dprime}u^{\dprime}}^+$ (blue), thermal variance $\overline{\theta^{\dprime}\theta^{\dprime}}^+$ (red), and wall-normal velocity variance $\overline{v^{\dprime}v^{\dprime}}^+$ (green). (\textit{c}) Turbulent momentum flux $-\overline{u^{\dprime}v^{\dprime}}^+$ (blue) and turbulent scalar flux $-\overline{\theta^{\dprime}v^{\dprime}}^+$ (red).}
	\label{fig:time_avg_stats}
\end{figure}

The time-averaged modifications to the mean gradients, stochastic variances and turbulent fluxes are presented in figure~\ref{fig:time_avg_stats} for the unactuated baseline (dashed lines) and the actuated ($T^+=350$, $W^+=30$; solid lines) cases, each variable being identified by colour. The mean-gradient profiles (figure~\ref{fig:time_avg_stats}\textit{a}) reveal, as a primary effect of the actuation, an increase in the wall gradients of both the streamwise velocity and the temperature, consistent with an intensified wall-normal exchange driven by the Stokes strain. It is recalled that, under the mixed boundary condition formulation employed here, the non-dimensional temperature $\overline{\theta}^+$ vanishes at the wall and is positive in the interior, so that $\mathrm{d}\overline{\theta}^+/\mathrm{d}y^+|_w > 0$; this sign convention mirrors that of the streamwise velocity and renders the two mean-gradient profiles directly comparable throughout the discussion that follows. Away from the wall, this enhancement is accompanied by a reduction of both $\mathrm{d}\overline{u}^+/\mathrm{d}y^+$ and $\mathrm{d}\overline{\theta}^+/\mathrm{d}y^+$ in the buffer-layer region. This compensating reduction is required by integral constraints imposed by the boundary specification: the bulk velocity is held fixed by the constant-mass-flow-rate constraint enforced through the body force, in the case of the streamwise momentum, whilst the mixed-type thermal boundary condition fixes the bulk temperature through the uniform volumetric heat source, in the case of the scalar field. As the integral of each mean profile across the half-height is thereby conserved, an enhancement of the wall gradient must be offset by a compensating reduction elsewhere in the profile; the present data show that this compensation is concentrated in the buffer layer for both fields. Of particular significance is the differential response of the two wall gradients: whereas $\mathrm{d}\overline{u}^+/\mathrm{d}y^+|_w$ is enhanced relative to the baseline, $\mathrm{d}\overline{\theta}^+/\mathrm{d}y^+|_w$ undergoes a markedly stronger increase, reflecting directly the time-averaged dissimilarity ($\overline{A}_n \approx 1.09$) whereby the gain in Nusselt number exceeds that in skin friction. This decoupling of the two wall gradients constitutes the principal macroscopic signature of dissimilar heat transfer in the time-mean. The question of whether this decoupling is sustained throughout the actuation cycle, or arises preferentially during particular phases of the actuation, is deferred to the phase-resolved analysis of \S\,\ref{sec:results:coherent}.

The variance profiles (figure~\ref{fig:time_avg_stats}\textit{b}) are examined next. The streamwise velocity variance and the thermal variance both decrease, by approximately $26\%$ and $30\%$ respectively, their peaks migrating slightly inwards. The wall-normal variance $\overline{v^{\dprime}v^{\dprime}}^+$, by contrast, is strongly amplified, its peak approximately doubling whilst likewise shifting towards the wall. Herein lies an apparent paradox: the suppression of $\overline{u^{\dprime}u^{\dprime}}^+$ is the documented signature of drag-reducing SWO at moderate periods \citep{ricco_review_2021,quadrio_critical_2004}; the present configuration, however, lies in the drag-increase regime, the wall gradient of $\overline{u}^+$ having just been shown to be enhanced; the variance profile thus carries a drag-reduction signature whilst the wall gradients indicate that drag has been increased. The resolution is provided by the wall-normal turbulent fluxes, that is the fluxes $-\overline{u^{\dprime}v^{\dprime}}^+$ and $-\overline{\theta^{\dprime}v^{\dprime}}^+$ that transport momentum and heat across the region of steepest mean gradient. The amplification of $\overline{v^{\dprime}v^{\dprime}}^+$ noted above provides the kinematic basis for an enhancement of these fluxes, and the flux profiles in figure~\ref{fig:time_avg_stats}\textit{c} confirm directly that both $-\overline{u^{\dprime}v^{\dprime}}^+$ and $-\overline{\theta^{\dprime}v^{\dprime}}^+$ are amplified under the actuation: the scalar flux rises by approximately $30\%$ at its peak whilst the momentum flux increases by approximately $10\%$, with the scalar flux enhancement exceeding that of the momentum flux throughout the buffer-layer region. Moreover, both flux peaks migrate inwards in unison with those of the variances, an inward shift that concentrates the wall-normal mixing closer to the wall and is the kinematic counterpart of the enhancement of $\mathrm{d}\overline{u}^+/\mathrm{d}y^+|_w$ and $\mathrm{d}\overline{\theta}^+/\mathrm{d}y^+|_w$: the actuation thins the turbulent transport layer and intensifies the wall-normal exchange within it, the streamwise variance being depleted only as a kinematic consequence of the redistribution of fluctuation energy out of the streamwise component and into the wall-normal one. The turbulent momentum flux $-\overline{u^{\dprime}v^{\dprime}}^+$ directly determines the skin friction through the total stress balance, whilst the turbulent heat flux $-\overline{\theta^{\dprime}v^{\dprime}}^+$ plays the analogous role for the Nusselt number; both represent the wall-normal transport realised by the buffer-layer turbulence through sweep events ($v^{\dprime}<0$) that advect high-momentum, high-temperature fluid towards the wall and ejection events ($v^{\dprime}>0$) that lift low-momentum, cooler fluid outward. The spanwise actuation alters $C_f$ and $\Nu$ through the modulation of these two wall-normal exchange mechanisms, and the differential response of the two fluxes to this modulation underlies the observed dissimilarity.

The proportionally stronger enhancement of $-\overline{\theta^{\dprime}v^{\dprime}}^+$ relative to $-\overline{u^{\dprime}v^{\dprime}}^+$ accordingly constitutes the direct macroscopic precursor to the dissimilar heat transfer performance. The differential suppression of the two streamwise variances provides a complementary indication of the asymmetric response of the scalar and momentum fields to the actuation, with the scalar variance reduced slightly more than the velocity variance; the physical origin of both the flux asymmetry and the variance asymmetry is examined through the time-mean variance budgets in the remainder of the present subsection and through the phase-resolved analysis in the subsequent subsections.

\begin{figure}
	\centering
	\begin{minipage}{0.48\textwidth}
			\centering
			\includegraphics[width=\textwidth]{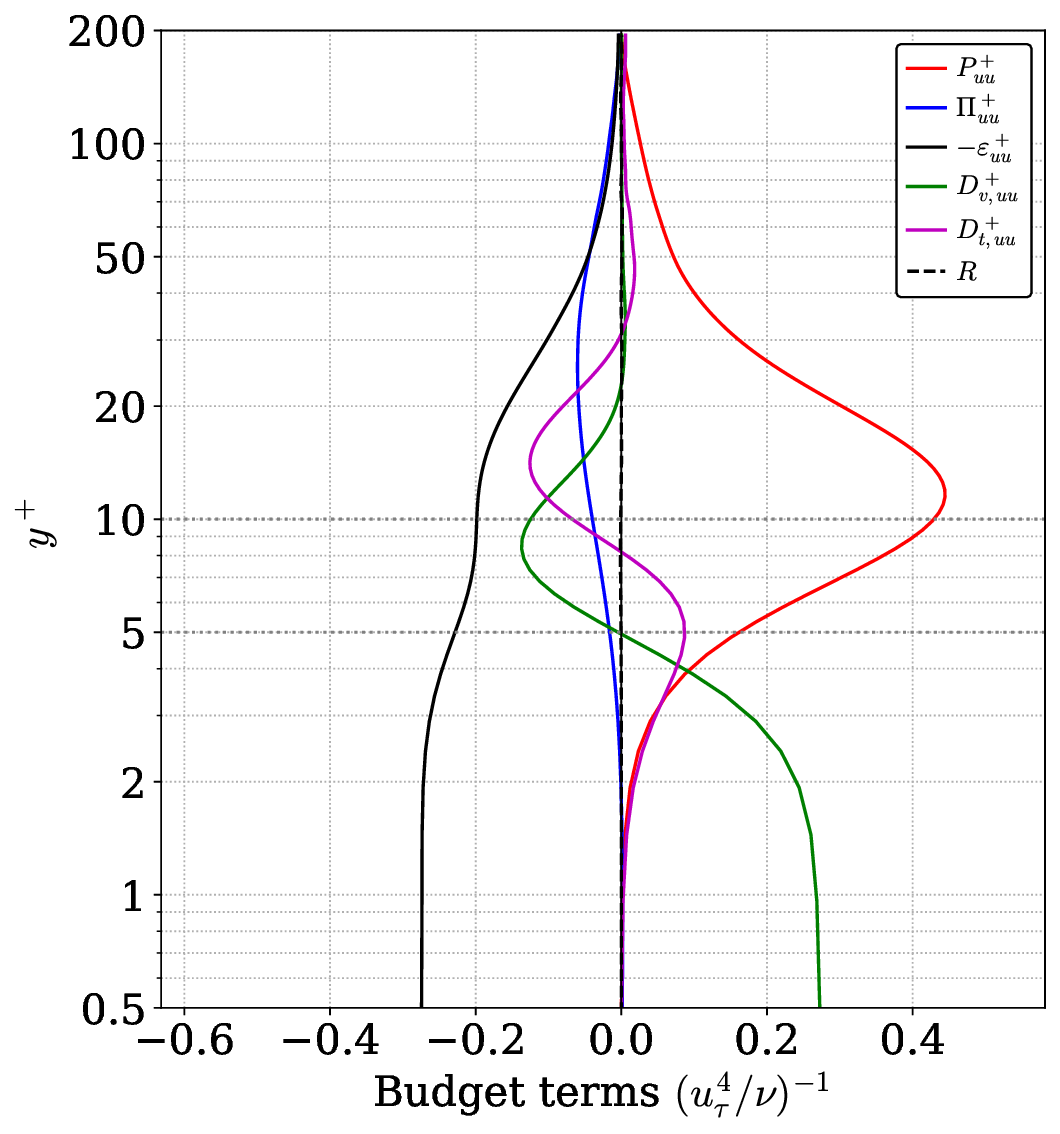}
			\vspace{0.2cm}
			\centerline{(\textit{a})}
		\end{minipage}
		\hfill
		\begin{minipage}{0.48\textwidth}
			\centering
			\includegraphics[width=\textwidth]{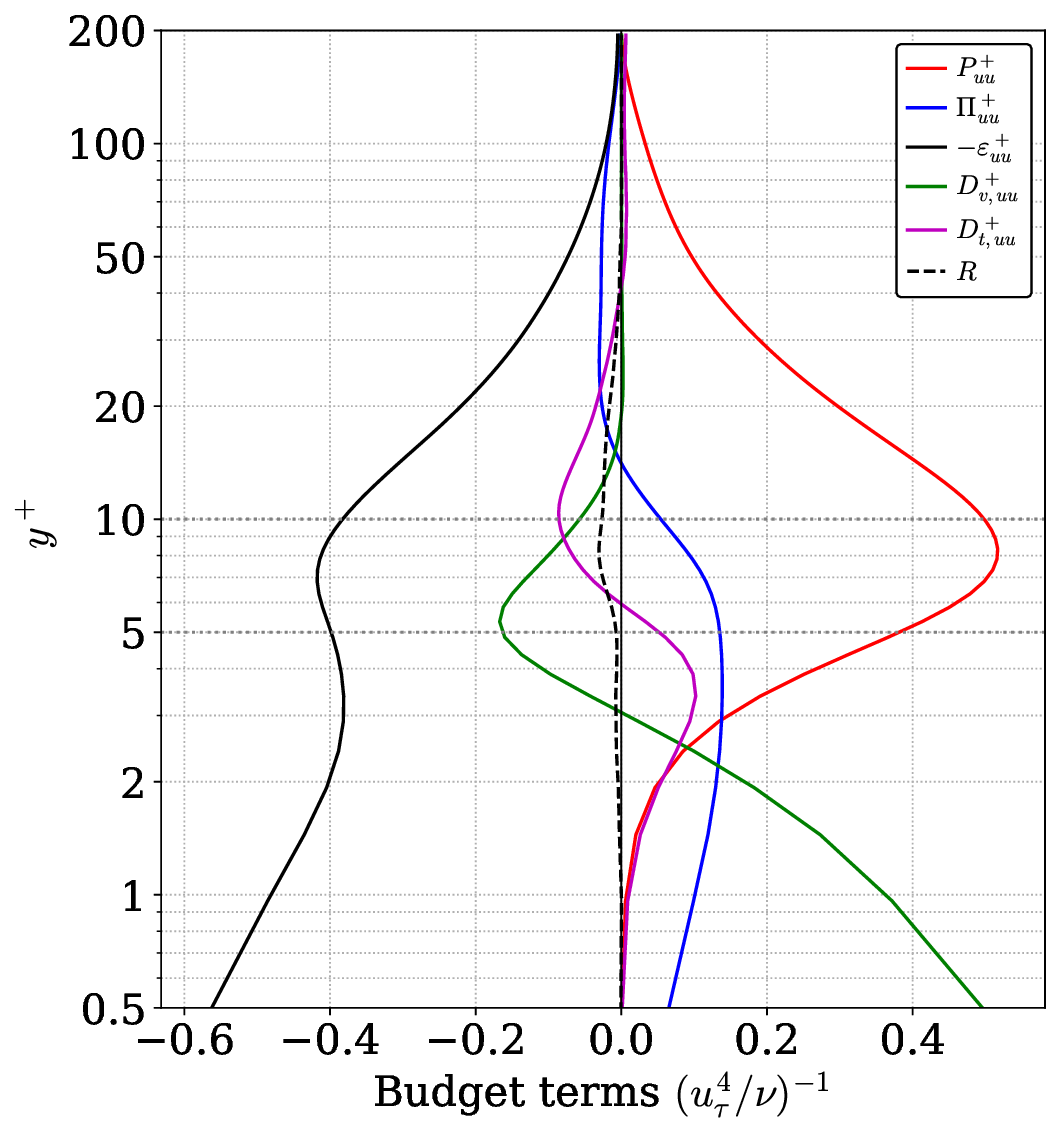}
			\vspace{0.2cm}
			\centerline{(\textit{b})}
		\end{minipage}
		\\[0.4cm]
		\begin{minipage}{0.48\textwidth}
			\centering
			\includegraphics[width=\textwidth]{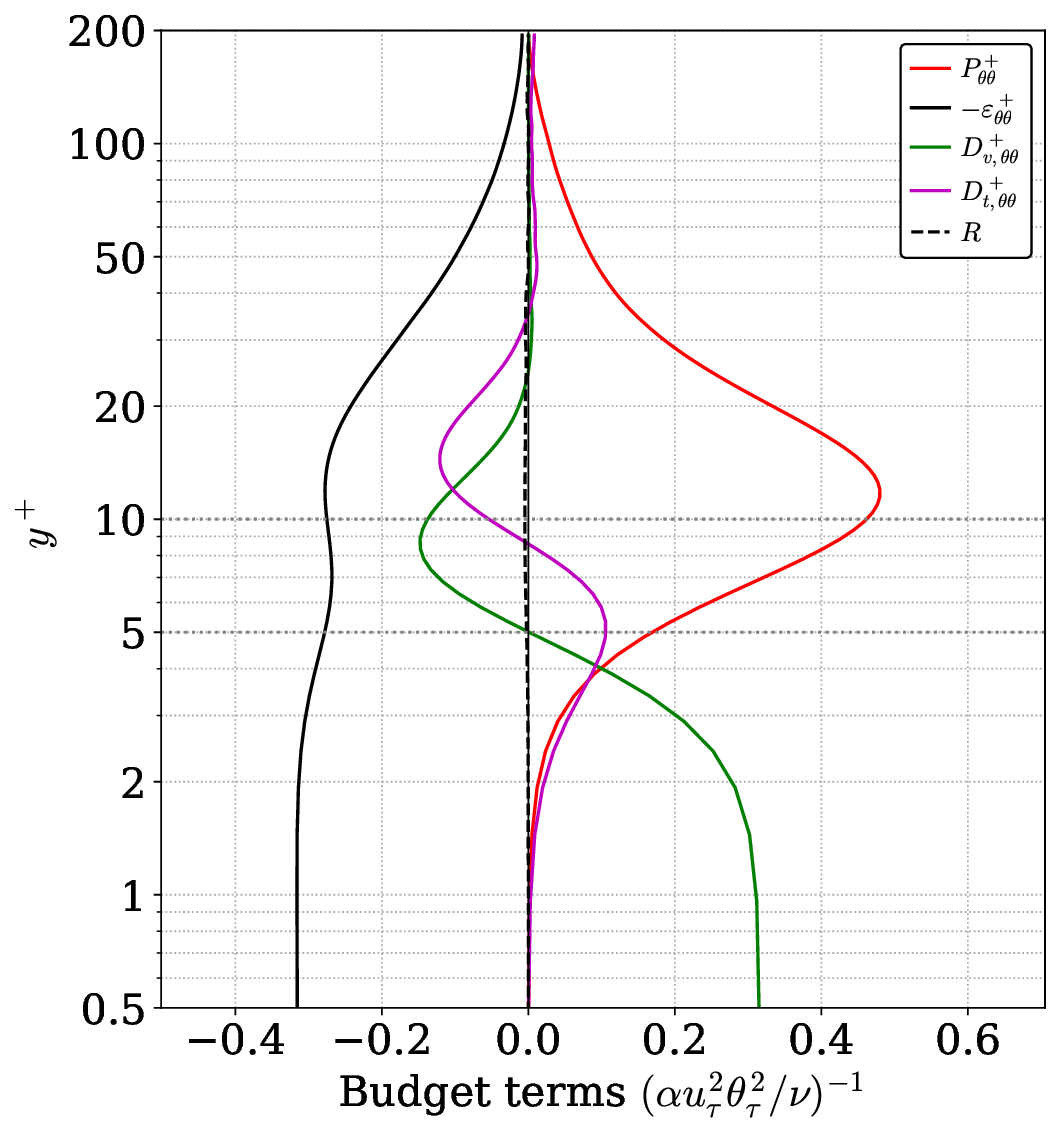}
			\vspace{0.2cm}
			\centerline{(\textit{c})}
		\end{minipage}
		\hfill
		\begin{minipage}{0.48\textwidth}
			\centering
			\includegraphics[width=\textwidth]{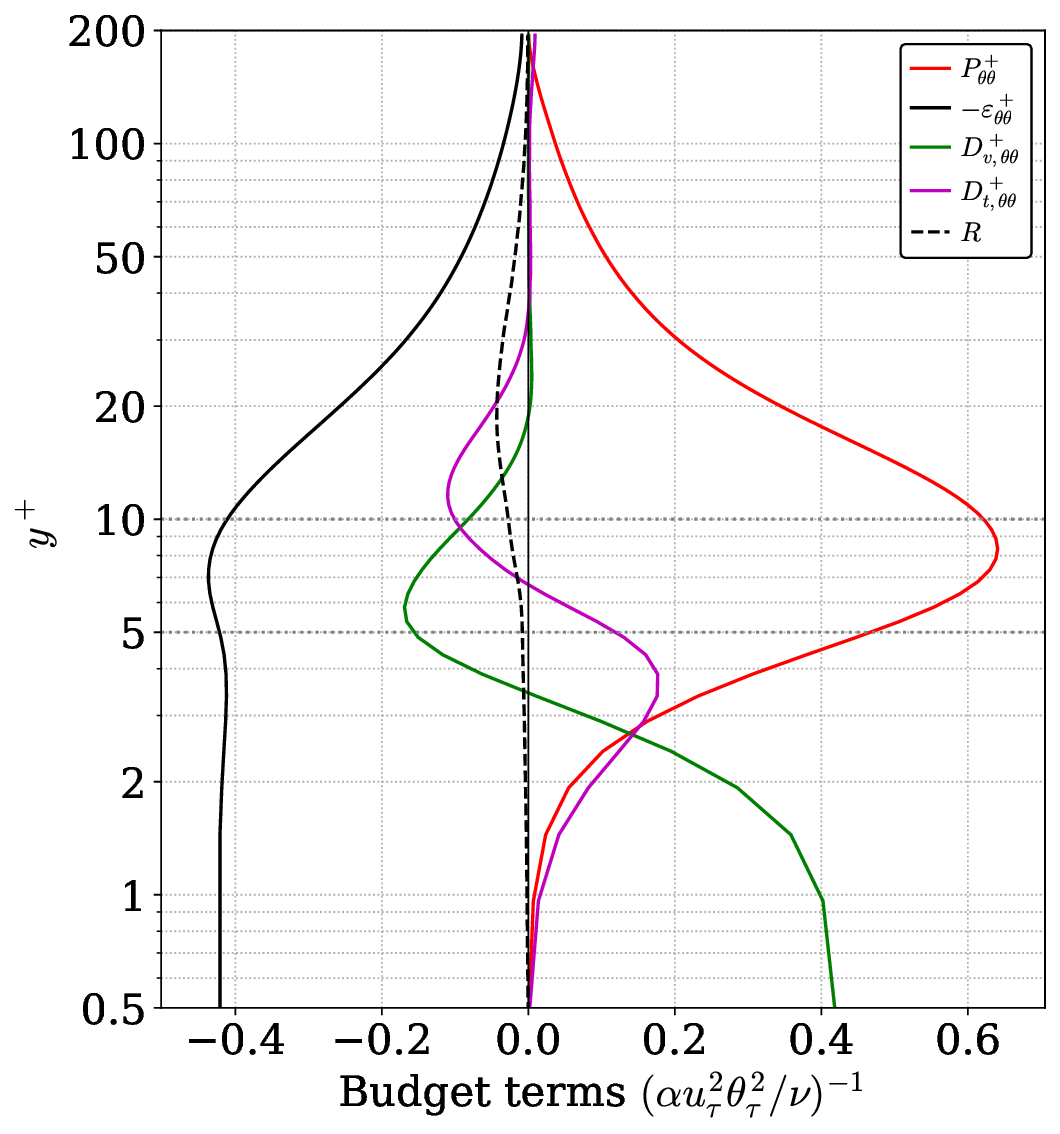}
			\vspace{0.2cm}
			\centerline{(\textit{d})}
	\end{minipage}
	\caption{Time-mean variance budgets for the unactuated baseline (left column) and the actuated case ($T^+=350$, $W^+=30$; right column). (\textit{a}, \textit{b}) Budget of $\overline{u^{\dprime}u^{\dprime}}^+$, normalised by $u_\tau^4/\nu$; terms shown are production $P_{uu}^+$, pressure-strain redistribution $\Pi_{uu}^+$, dissipation $-\epsilon_{uu}^+$, viscous diffusion $D_{v,uu}^+$, and turbulent diffusion $D_{t,uu}^+$. (\textit{c}, \textit{d}) Budget of $\overline{\theta^{\dprime}\theta^{\dprime}}^+$, normalised by $u_\tau^2\theta_\tau^2/\nu$; terms shown are production $P_{\theta\theta}^+$, dissipation $-\epsilon_{\theta\theta}^+$, viscous diffusion $D_{v,\theta\theta}^+$, and turbulent diffusion $D_{t,\theta\theta}^+$. The dashed line in each panel denotes the budget residual.}
	\label{fig:tm_budget}
\end{figure}

To identify the mechanism underlying the differential response of the two fields, the time-mean variance budgets of $\overline{u^{\dprime}u^{\dprime}}^+$ and $\overline{\theta^{\dprime}\theta^{\dprime}}^+$ are presented in figure~\ref{fig:tm_budget}, for both the unactuated baseline (panels \textit{a}, \textit{c}) and the actuated case (panels \textit{b}, \textit{d}); the budget residual, shown dashed in each panel, remains small relative to the dominant production and dissipation terms throughout the buffer layer. The two budgets are compared by means of the effective source available to each variance, defined as the algebraic sum of the production and redistribution terms on the right-hand side of the time-mean form of equations~\eqref{eq:budget_uu} and~\eqref{eq:budget_thth}: for the velocity, this comprises the production $P_{uu}^+$ together with the pressure-strain redistribution $\Pi_{uu}^+$, which augments the production where positive and depletes it where negative, and for the scalar it comprises the production $P_{\theta\theta}^+$ alone, as no pressure-strain term enters the scalar budget. This pairing renders the velocity and scalar variances comparable on the same basis, the additional source channel available to the velocity field being thereby accounted for explicitly. The unactuated budgets are examined first, as they provide the canonical reference against which the actuation-induced modifications are subsequently characterised. In the buffer layer, the unactuated velocity budget (panel \textit{a}) is dominated by production $P_{uu}^+$, which injects energy from the mean shear and attains its maximum of approximately $0.44$ at $y^+ \approx 12$, and by the competing sinks of dissipation $-\epsilon_{uu}^+$ and diffusion $D_{uu}^+ = D_{v,uu}^+ + D_{t,uu}^+$; at the production peak, the two sinks contribute in approximately equal measure, jointly balancing the production net of the pressure-strain drain. The pressure-strain redistribution $\Pi_{uu}^+$ is structurally present throughout the buffer layer, where it is negative and thus extracts energy from the streamwise component for redistribution to the wall-normal and spanwise components; however, its magnitude at the production peak is modest relative to dissipation and diffusion combined, amounting to approximately $10\%$ of $P_{uu}^+$ at this $\Re_\tau$. The role of $\Pi_{uu}^+$ in the unactuated state is therefore not that of a leading-order sink, but rather that of a structurally necessary inter-component redistribution channel that progressively depletes the streamwise variance relative to the scalar variance over wall-normal distance. The unactuated scalar budget (panel \textit{c}) is dominated by the production $P_{\theta\theta}^+$, which peaks at approximately $0.48$ near $y^+ \approx 12$, and the dissipation $-\epsilon_{\theta\theta}^+$, with the diffusion terms $D_{v,\theta\theta}^+$ and $D_{t,\theta\theta}^+$ playing a secondary role above the viscous sublayer; no pressure-strain term appears, as the passive scalar equation contains no pressure contribution and the scalar variance cannot be redistributed amongst components. The structural asymmetry between the two budgets is therefore that $\Pi_{uu}^+$ acts as an inter-component redistribution channel in the velocity budget with no analogue in the scalar budget, and that, in the unactuated state, this channel operates as a persistent drain upon the streamwise component. As a consequence, the effective source for the streamwise variance at the production peak, $P_{uu}^+ + \Pi_{uu}^+ \approx 0.40$, falls below the scalar production $P_{\theta\theta}^+ \approx 0.48$, characteristic of the unactuated buffer layer at $\Pr=1$ \citep[cf.][]{kasagi_direct_1992, hasegawa_dissimilar_2011}: the negative $\Pi_{uu}^+$ continuously redistributes streamwise energy to the cross-stream components whilst no equivalent drain operates upon the scalar field. The present analysis extends this static observation to the actuated state, the question being whether the same structural channel, once activated by the periodic Stokes-layer forcing, gives rise to a more substantial dissimilarity in the time-mean.

The actuated budgets (panels \textit{b}, \textit{d}) preserve the qualitative structure of the scalar equation, which remains devoid of a pressure-strain contribution; the velocity budget, by contrast, is materially altered, and the pressure-strain redistribution is considered first. Throughout the buffer-layer region $y^+ \lesssim 15$ the term $\Pi_{uu}^+$ (panel \textit{b}) has become positive, reaching $\Pi_{uu}^+ \approx +0.091$ at the production maximum ($y^+ \approx 8.3$). Having served as a modest sink of magnitude $\Pi_{uu,0}^+ \approx -0.045$ in the unactuated buffer layer, it now operates as a substantial near-wall source: a reversal of sign together with an approximately twofold increase in magnitude, the principal time-mean signature of the structural asymmetry being activated by the Stokes-layer forcing. At the production peak the source amounts to approximately $16\%$ of $P_{uu}^+$, substantial though not leading-order. The trace-free condition $\Pi_{uu} + \Pi_{vv} + \Pi_{ww} = 0$, which holds pointwise as a strict algebraic identity, requires that the energy supplying it be drawn from the wall-normal and spanwise components through $\Pi_{vv}$ and $\Pi_{ww}$; by which of the two, and how the redistribution is organised across the cycle, cannot be read from the time mean and is addressed in \S\,\ref{sec:results:pressure_strain}.

The streamwise and scalar productions are likewise modified. Both production peaks migrate inwards, to $y^+ \approx 8.3$, following the inward shift of the turbulent fluxes documented above, the production at each height being the product of the local flux and the local mean gradient. At this common peak the scalar production exceeds the velocity production, the ratio $P_{uu}^+/P_{\theta\theta}^+ \approx 0.80$; taken alone, this would suggest that the scalar field enjoys an intrinsic advantage. The pertinent comparison, however, is not between the two productions but between the two effective sources defined above: for the scalar, the production alone; for the velocity, the production augmented by its positive pressure-strain source, $P_{uu}^+ + \Pi_{uu}^+$. Including this channel raises the effective-source ratio to $(P_{uu}^+ + \Pi_{uu}^+)/P_{\theta\theta}^+ \approx 0.94$, against $0.83$ in the unactuated state: the actuated $\Pi_{uu}^+$ source narrows the gap, both relative to the velocity production alone and relative to the unactuated reference, without closing it. This added source channel, however, sharpens a question rather than settling one. With its production now augmented by a source the velocity budget did not previously possess, the streamwise variance might be expected to rise; instead $\overline{u^{\dprime}u^{\dprime}}^+$ falls by some $26\%$, as documented above. The resolution is that the additional energy is removed locally by an enhanced dissipation, $\epsilon_{uu}^+$ increasing approximately $2.1$-fold under the actuation, so that the budget re-closes at a reduced variance. In the time mean the activated $\Pi_{uu}^+$ thus effects a budgetary compensation, not a preferential enhancement of either field.

The sign reversal of $\Pi_{uu}^+$ is accordingly the most conspicuous time-mean signature of the structural asymmetry being activated by the actuation: a substantial pressure-strain source appears in the velocity budget with no equivalent on the scalar side. Its bearing on the analogy breaking, however, is the opposite of what its prominence suggests. The source supplies the velocity field a channel the scalar lacks; were it converted into variance, it would tend to equalise the two budgets and thus to \emph{oppose} the dissimilarity, not to create it. That it is instead dissipated locally means it does neither. In the time mean, then, the most visible modification of the budgets is silent on the origin of the dissimilarity, which the wall gradients and the fluxes have nonetheless shown to be real ($\overline{A}_n \approx 1.09$). The resolution must lie in the phase organisation that the time average conceals, the mean $\Pi_{uu}^+$ being, as will emerge, the residue of a quantity that alternates in sign over the cycle and whose dissimilarity-producing action is concentrated in particular phases. The conspicuous reversal is thus a clue without an immediate culprit. The phase-resolved investigation is accordingly undertaken in the subsections that follow, beginning with the coherent fields (\S\,\ref{sec:results:coherent}) and proceeding to the production asymmetry (\S\,\ref{sec:results:production}), whose dynamical origin is then pursued.

\subsection{Coherent response of the mean fields}
\label{sec:results:coherent}

\begin{figure}
	\centering
	\begin{minipage}{0.48\textwidth}
		\centering
		\includegraphics[width=\textwidth]{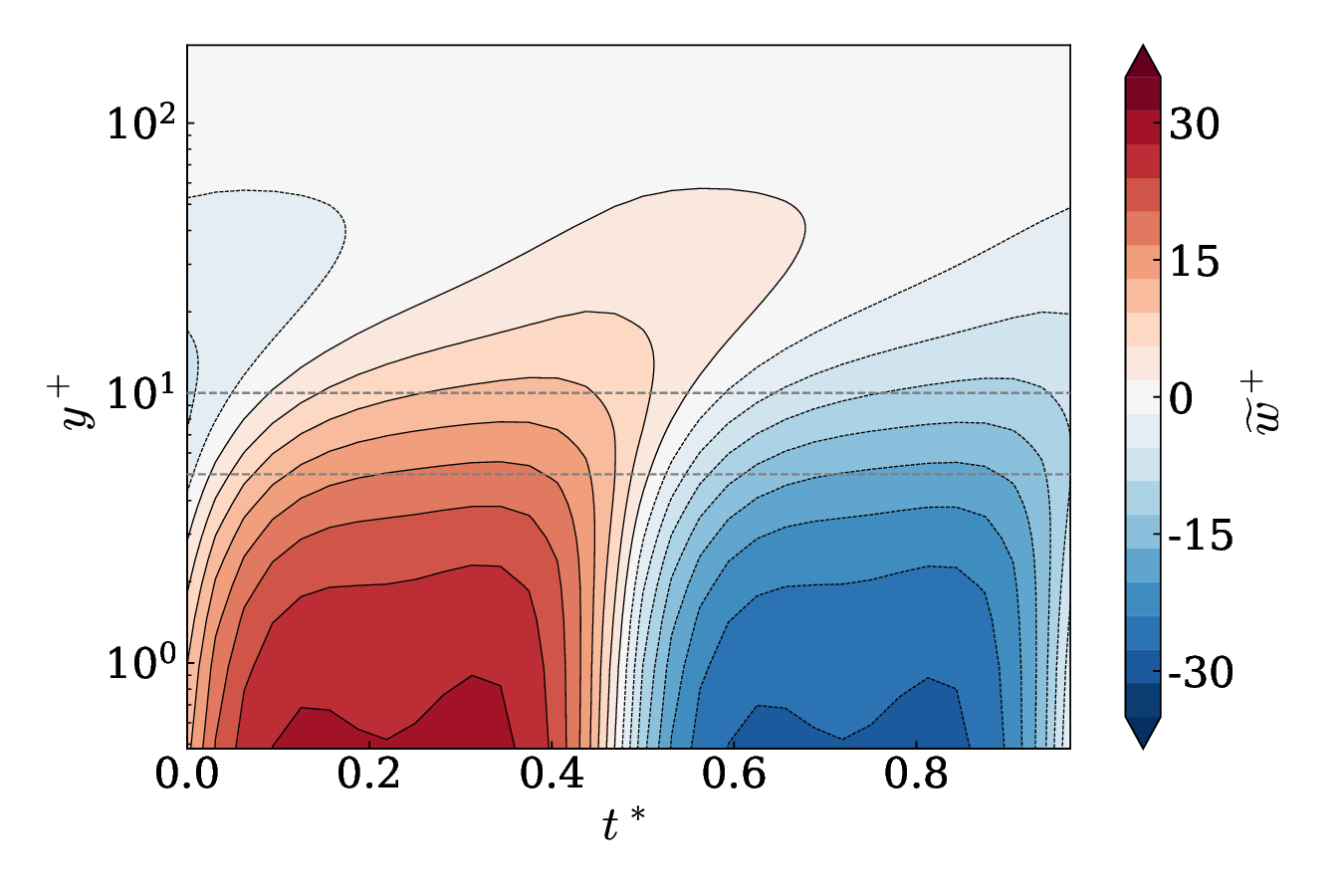}
		\vspace{0.2cm}
		\centerline{(\textit{a})}
	\end{minipage}
	\hfill
	\begin{minipage}{0.48\textwidth}
		\centering
		\includegraphics[width=\textwidth]{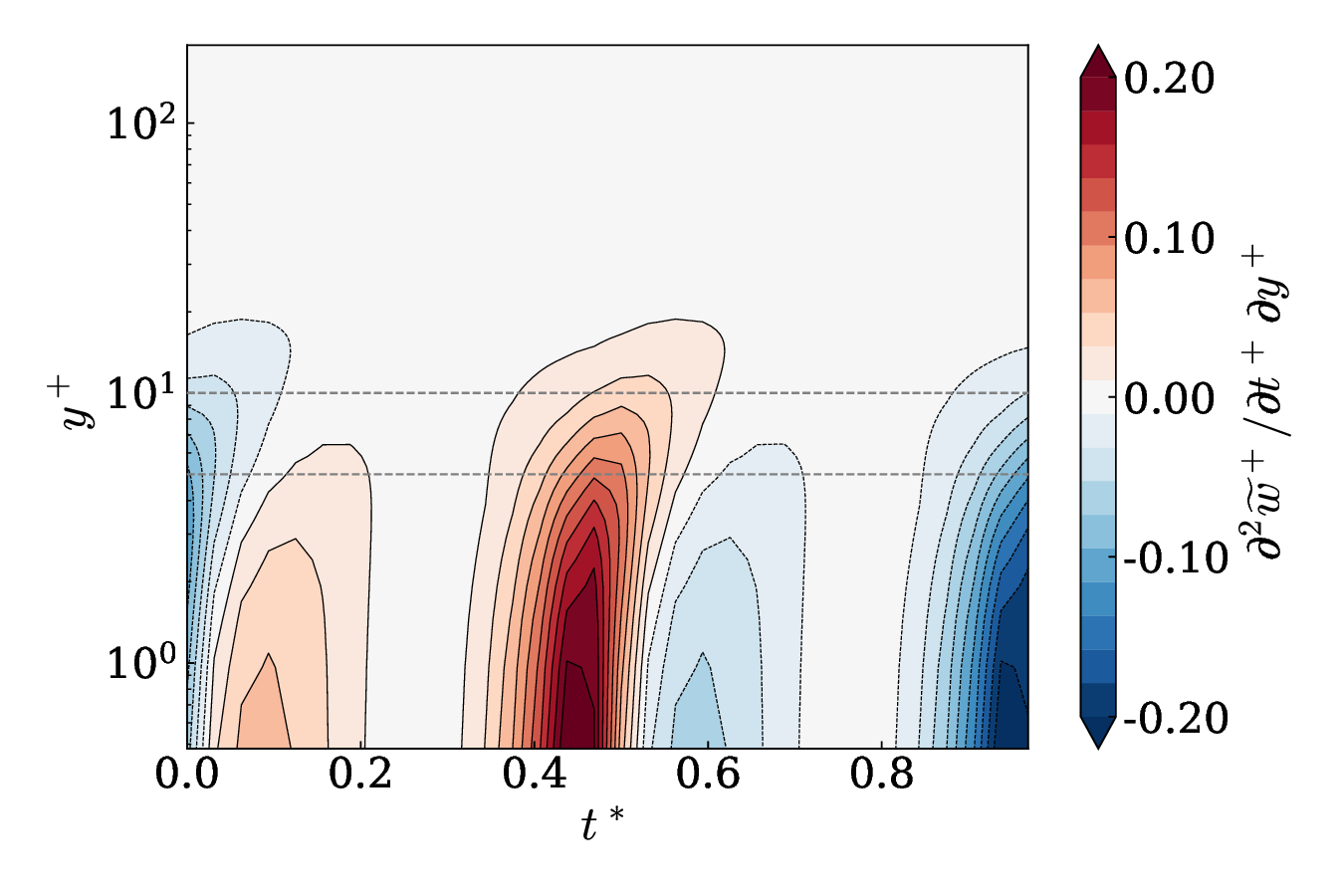}
		\vspace{0.2cm}
		\centerline{(\textit{b})}
	\end{minipage}
	\\[0.4cm]
	\begin{minipage}{0.6\textwidth}
		\centering
		\includegraphics[width=\textwidth]{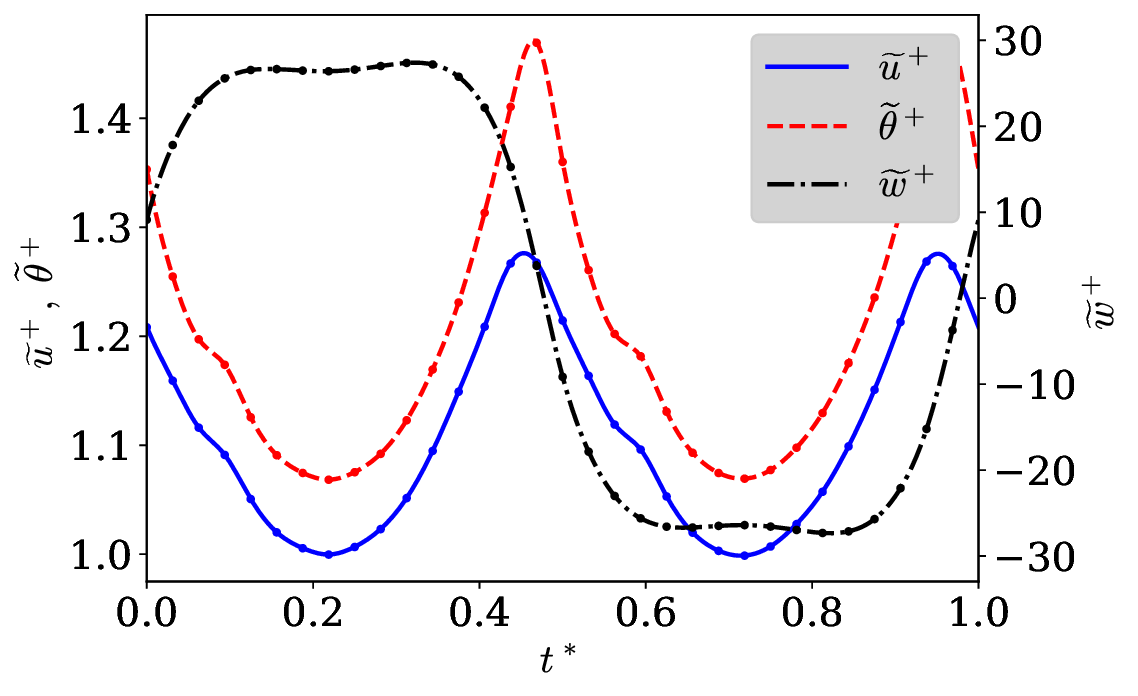}
		\vspace{0.2cm}
		\centerline{(\textit{c})}
	\end{minipage}
	\caption{Phase-resolved quantities for the optimal waveform at $T^+=350$, $W^+=30$. (\textit{a}) Phase-resolved map ($y^+$ vs $t^*$) of the spanwise velocity $\widetilde{w}^+$. (\textit{b}) Phase-resolved map of the Stokes-strain rate $\partial^2\widetilde{w}^+/\partial t^+\,\partial y^+$, the wall-normal derivative of the spanwise acceleration, on the same axes; the derivative is taken in viscous time $t^+$ whilst the abscissa remains the cycle phase $t^*$. (\textit{c}) Phase-averaged wall quantities $\widetilde{u}^+$, $\widetilde{\theta}^+$ and $\widetilde{w}^+$ at $y^+ \approx 1$.}
	\label{fig:stokes_wall}
\end{figure}

The phase-resolved evolution of the wall quantities and the spanwise Stokes layer for the optimal waveform configuration ($T^+=350$, $W^+=30$) is presented in figure~\ref{fig:stokes_wall}. The two-dimensional map of the phase-resolved spanwise velocity $\widetilde{w}^+$ (figure~\ref{fig:stokes_wall}\textit{a}) illustrates the Stokes layer structure characteristic of the quasi-plateau waveform: extended phases of nearly uniform amplitude separated by rapid reversal intervals during which the spanwise strain $\partial\widetilde{w}/\partial y$ penetrates into the buffer layer. The companion map of the Stokes-strain rate $\partial^2\widetilde{w}^+/\partial t^+\,\partial y^+$ (figure~\ref{fig:stokes_wall}\textit{b}) makes plain that this forcing is concentrated almost entirely within the two short reversal intervals and the near-wall region $y^+ \lesssim 10$, where it attains magnitudes of order $0.2$, whilst remaining negligible throughout the intervening plateaux. This mixed derivative, the rate of change of the Stokes strain, is the temporal gate of the duty-cycle modulation introduced in \S\,\ref{sec:methodology:governing}, partitioning each cycle into brief reversal phases of intense spanwise straining and extended lingering phases in which the strain rate is small; it is a partition to which the budget analysis repeatedly returns. The temporal traces of $\widetilde{u}^+$, $\widetilde{\theta}^+$, and $\widetilde{w}^+$ at $y^+ \approx 1$ (figure~\ref{fig:stokes_wall}\textit{c}) reveal a pronounced asymmetry between the velocity and temperature responses, which constitutes the phase-resolved manifestation of the differential wall-gradient enhancement identified in \S\,\ref{sec:results:time_avg}. During the suppression phase immediately following the Stokes-strain reversal ($t^* \lesssim 0.2$), both $\widetilde{u}^+$ and $\widetilde{\theta}^+$ decrease, consistent with the suppression of near-wall turbulent transport in the wake of the reversal. Beyond $t^* \approx 0.2$, as turbulent structures recommence, both quantities recover; however, $\widetilde{\theta}^+$ exhibits a markedly more pronounced recovery than $\widetilde{u}^+$ and remains above it at all phases; the two are directly comparable in the common wall scaling of \S\,\ref{sec:methodology:governing}, in which the unactuated velocity and temperature profiles nearly coincide at this height at $\Pr = 1$, so that the persistent excess of $\widetilde{\theta}^+$ indicates that the wall-temperature signature of the dissimilarity is maintained throughout the actuation cycle rather than confined to any particular reversal phase.

\begin{figure}
	\centering
	\begin{minipage}{0.48\textwidth}
		\centering
		\includegraphics[width=\textwidth]{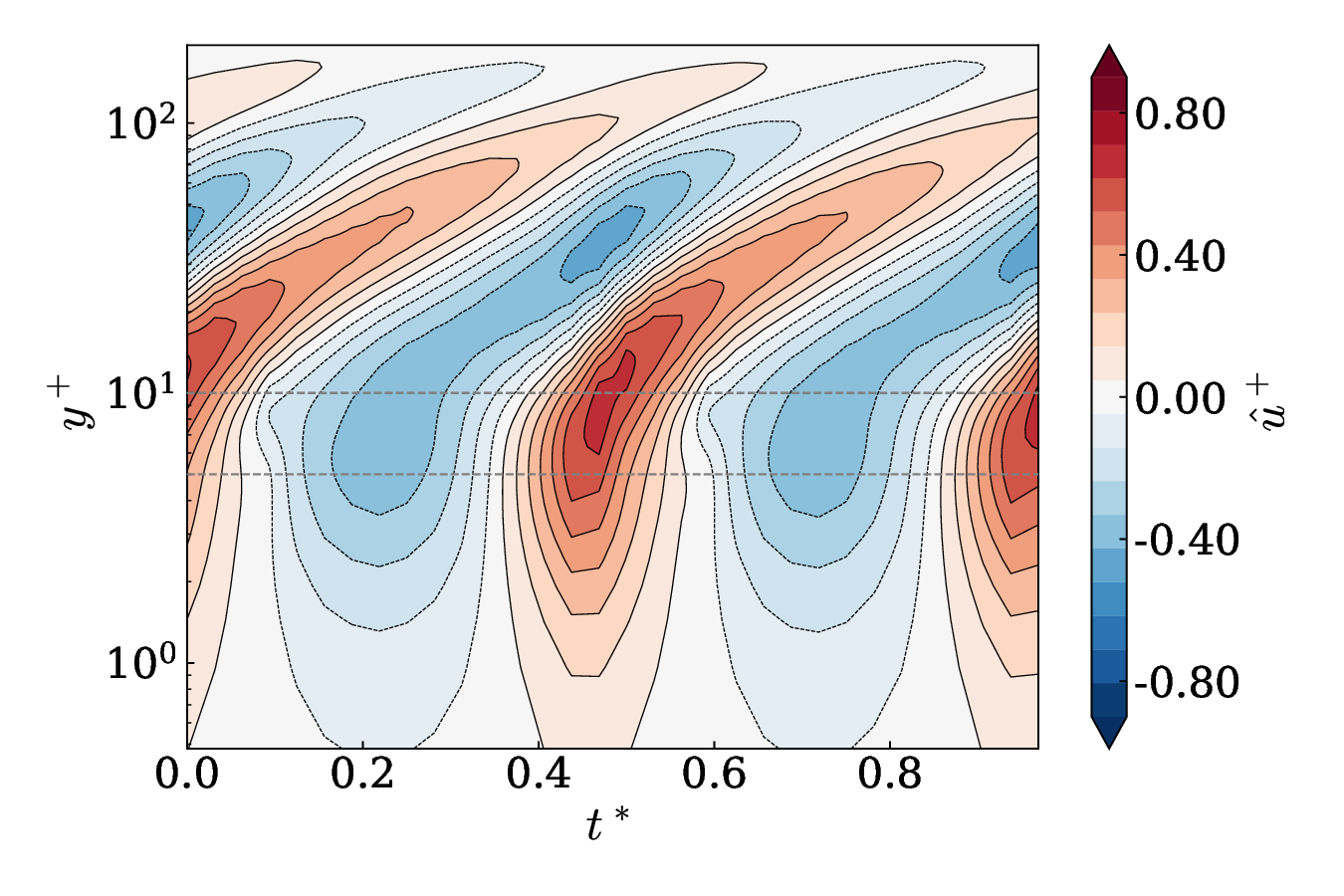}
		\vspace{0.2cm}
		\centerline{(\textit{a})}
	\end{minipage}
	\hfill
	\begin{minipage}{0.48\textwidth}
		\centering
		\includegraphics[width=\textwidth]{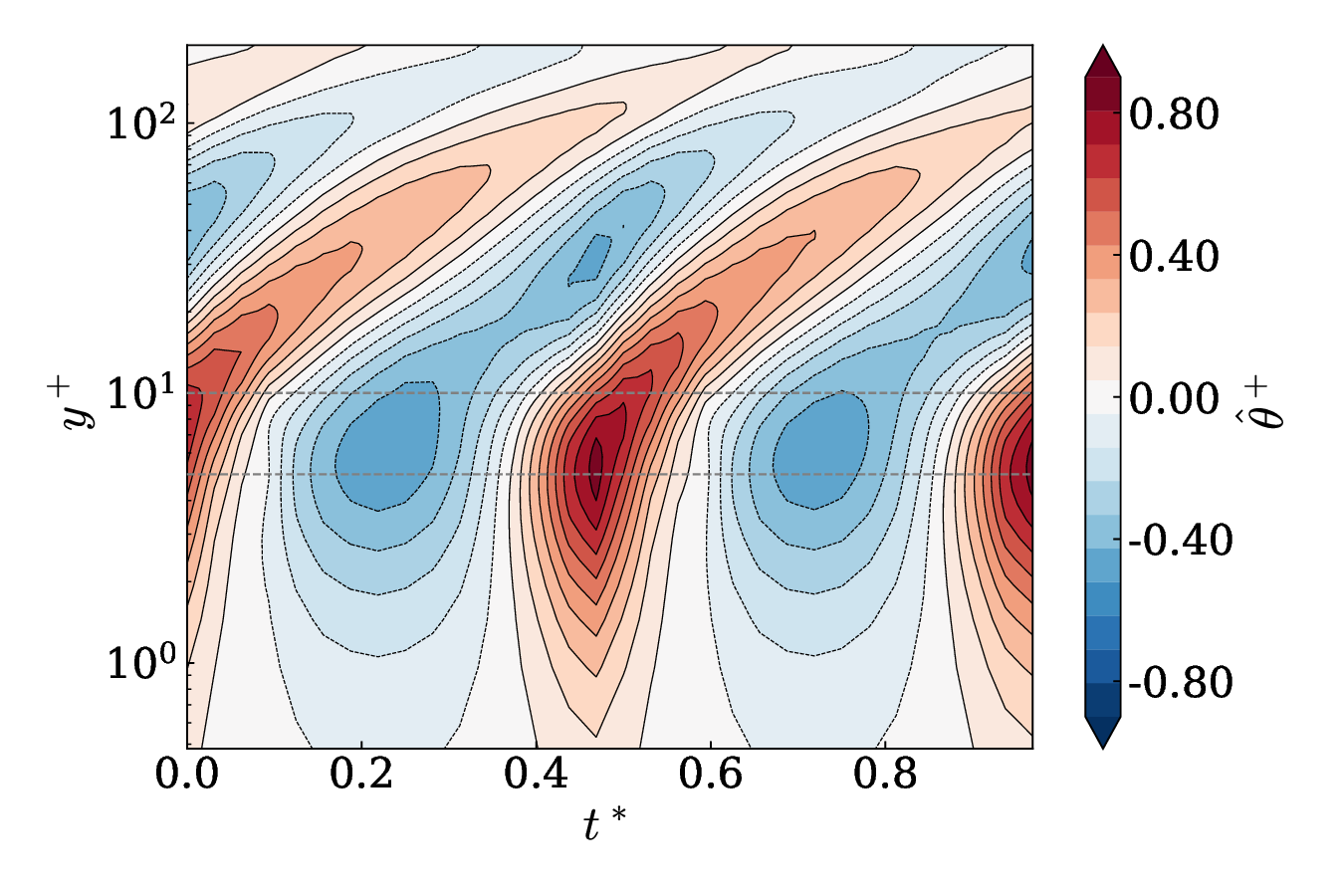}
		\vspace{0.2cm}
		\centerline{(\textit{b})}
	\end{minipage}
	\caption{Phase-resolved maps ($y^+$ vs $t^*$) for the $T^+=350$ case, plotted on a common colour axis. (\textit{a}) Coherent streamwise velocity $\hat{u}^+$. (\textit{b}) Coherent temperature $\hat{\theta}^+$.}
	\label{fig:coherent_relative}
\end{figure}

The coherent components $\hat{u}^+ = \widetilde{u}^+ - \overline{u}^+$ and $\hat{\theta}^+ = \widetilde{\theta}^+ - \overline{\theta}^+$ are presented as two-dimensional maps in figure~\ref{fig:coherent_relative}\textit{a} and \textit{b}, respectively. These panels convey the wall-normal extent of the phase-locked modulation: positive values indicate phases at which the coherent field exceeds its time-averaged counterpart, whilst negative values indicate phases of suppression. Three successive stages can be identified within each actuation cycle: (i) progressive weakening of the suppression as the buffer-layer Stokes strain $\partial\widetilde{w}/\partial y$ decays by viscous diffusion throughout the plateau; (ii) recommencement of the self-sustaining process (SSP) during the latter portion of the plateau as this strain approaches zero \citep{touber_near-wall_2012,agostini_turbulence_2015}; and (iii) pronounced amplification of streak intensity immediately prior to the subsequent Stokes strain reversal. This three-stage behaviour is a consequence of the long actuation period ($T^+=350$): each half-cycle affords a quiescent interval of the order of the characteristic SSP regeneration timescale ($\tau_\text{SSP}^+ \approx 100$--$200$; see \citealp{hamilton_regeneration_1995, jimenez_near-wall_2013}), permitting substantial turbulence recovery within each plateau phase, a window that the shorter periods of the drag-reduction regime do not afford. The asymmetry is quantitative as well as visual: measured as the excursion above its cycle minimum, normalised by that minimum, the coherent temperature amplitude recovers by approximately $40\%$ over the cycle against approximately $30\%$ for the streamwise velocity, a relative amplification greater by one-third, in keeping with the deeper $\hat{\theta}^+$ extrema of figure~\ref{fig:coherent_relative}\textit{b}. This asymmetry is consistent with the differential wall-gradient enhancement documented in \S\,\ref{sec:results:time_avg} and anticipates the production asymmetry analysed in \S\,\ref{sec:results:production}. In the conceptual timeline of figure~\ref{fig:mechanism_schematic}, the forcing stage established here, the wall velocity together with the gate against which every subsequent quantity is read, is condensed into panel (\textit{a}).

\subsection{Phase modulation of stochastic fluctuations}
\label{sec:results:stochastic}

The time-mean characterisation of \S\,\ref{sec:results:time_avg} established that the diagonal pressure-strain channel is compensatory in the time mean: the dissimilarity must therefore originate in the phase-resolved organisation of the actuation cycle, of which the time-mean budgets represent only the time-averaged residue. Although the turbulent fluxes $-\overline{u^{\dprime}v^{\dprime}}^+$ and $-\overline{\theta^{\dprime}v^{\dprime}}^+$ are the quantities directly governing drag and heat transfer, the streamwise variances $\overline{u^{\dprime}u^{\dprime}}^+$ and $\overline{\theta^{\dprime}\theta^{\dprime}}^+$ are not secondary to the present analysis: they constitute the energetic reservoir from which the wall-normal fluxes are sustained, and the phase-resolved perspective is necessary to identify at which stages of the actuation cycle the variance amplification occurs and whether the scalar and velocity fields amplify in phase or with a lag, as any such differential amplification must ultimately translate into the asymmetric flux enhancement documented in \S\,\ref{sec:results:time_avg}.

\begin{figure}
	\centering
	\begin{minipage}{0.48\textwidth}
		\centering
		\includegraphics[width=\textwidth]{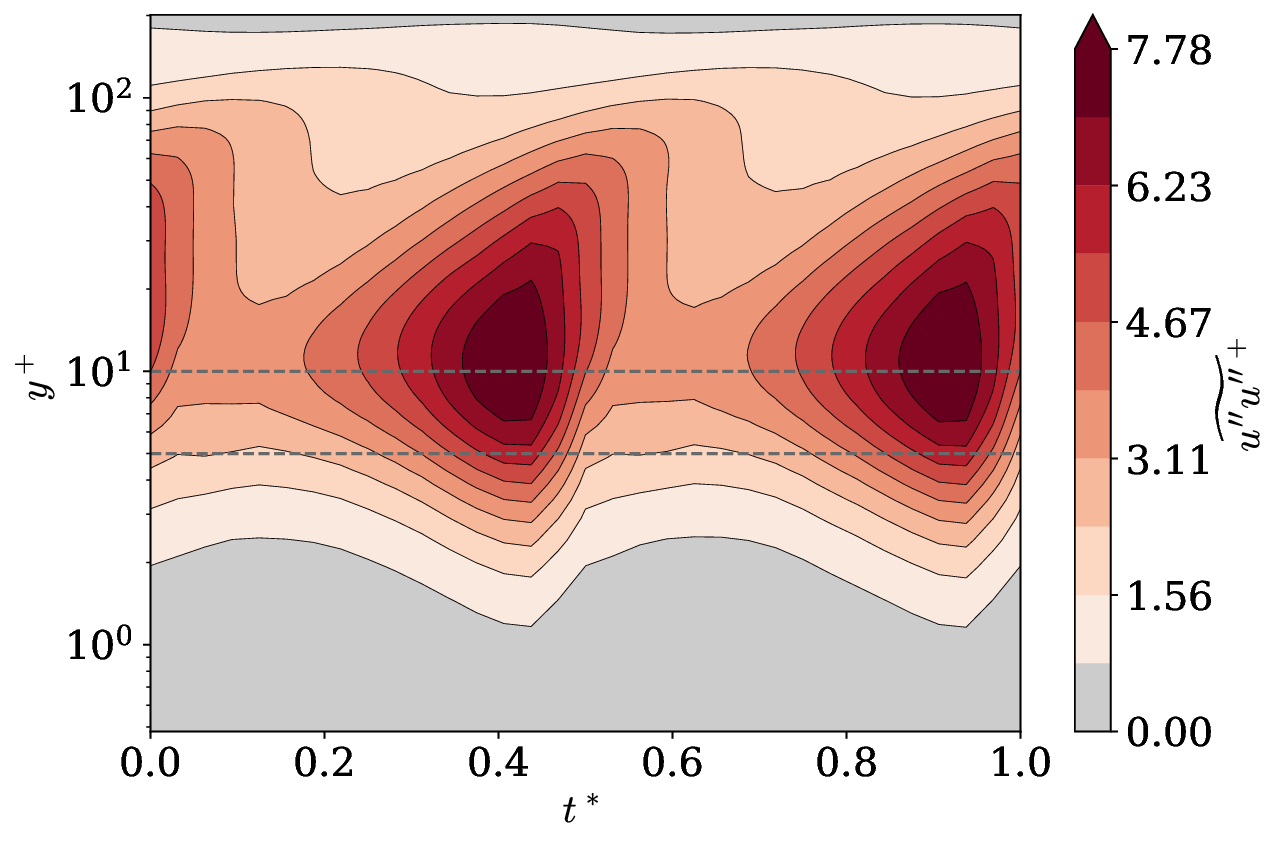}
		\vspace{0.2cm}
		\centerline{(\textit{a})}
	\end{minipage}
	\hfill
	\begin{minipage}{0.48\textwidth}
		\centering
		\includegraphics[width=\textwidth]{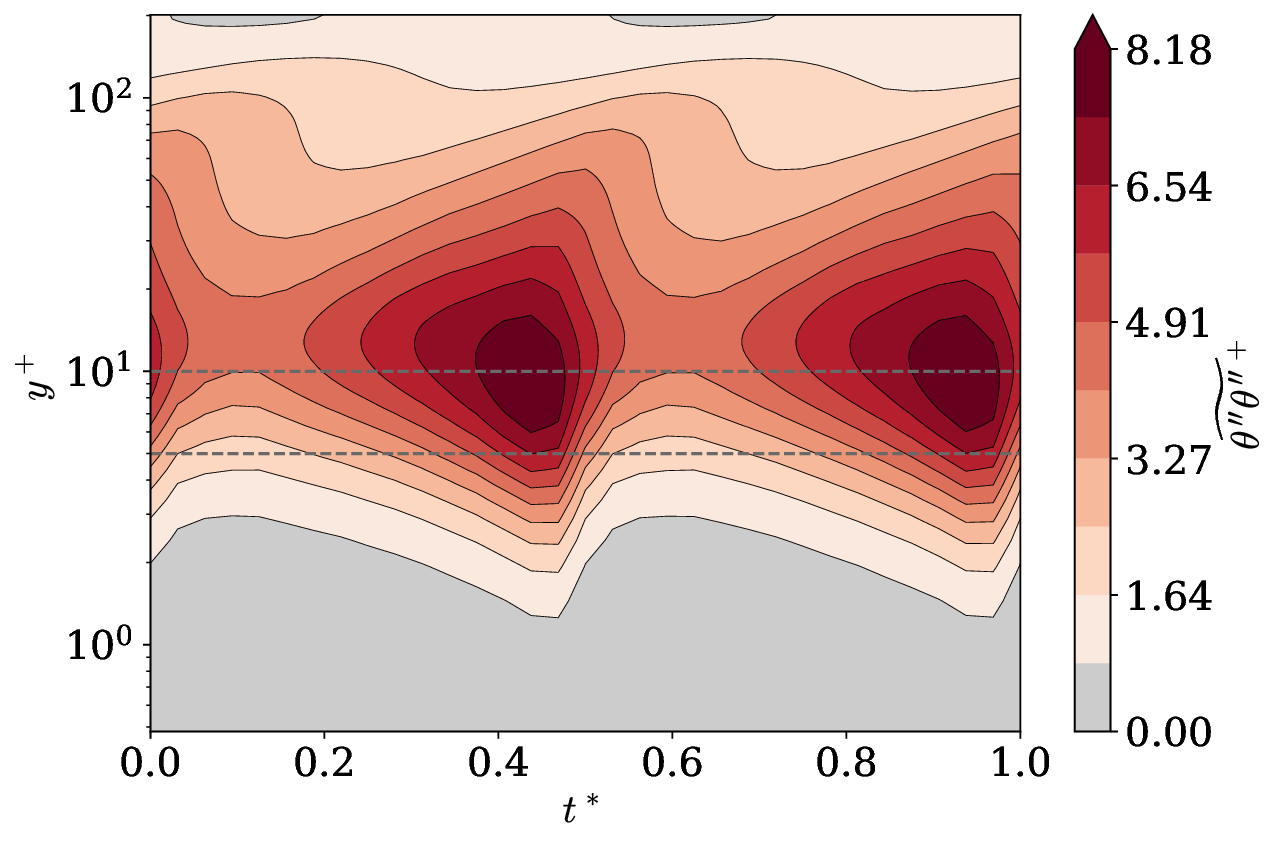}
		\vspace{0.2cm}
		\centerline{(\textit{b})}
	\end{minipage}
	\caption{Phase-resolved maps ($y^+$ vs $t^*$) of the stochastic streamwise velocity variance and temperature variance at $T^+=350$, $W^+=30$. (\textit{a}) Streamwise velocity variance $\uupp^+$. (\textit{b}) Temperature variance $\ttpp^+$.}
	\label{fig:uu_thth_phase}
\end{figure}

The phase-resolved distributions of the streamwise velocity variance $\uupp^+$ and the temperature variance $\ttpp^+$ are presented in figure~\ref{fig:uu_thth_phase}. Both quantities attain their maximum values in the vicinity of $y^+ \approx 10$, and their temporal evolution mirrors the three-stage pattern identified for the coherent fields in \S\,\ref{sec:results:coherent}: suppressed levels are maintained during the suppression phase immediately following the reversal, when the buffer-layer Stokes strain $\partial\widetilde{w}/\partial y$ is at its strongest and continues to disrupt the near-wall turbulence, followed by progressive recovery as this strain decays by viscous diffusion. This correspondence between coherent and stochastic responses establishes that the stochastic fluctuations are modulated by the same Stokes-layer dynamics that drives the coherent fields. The two variance maps do not, however, exhibit any pronounced visual asymmetry between the scalar and velocity responses: both fields show comparable suppression and recovery, and the differential amplification required to explain the flux asymmetry cannot be read directly from the variance maps themselves.

\begin{figure}
	\centering
	\includegraphics[width=0.75\textwidth]{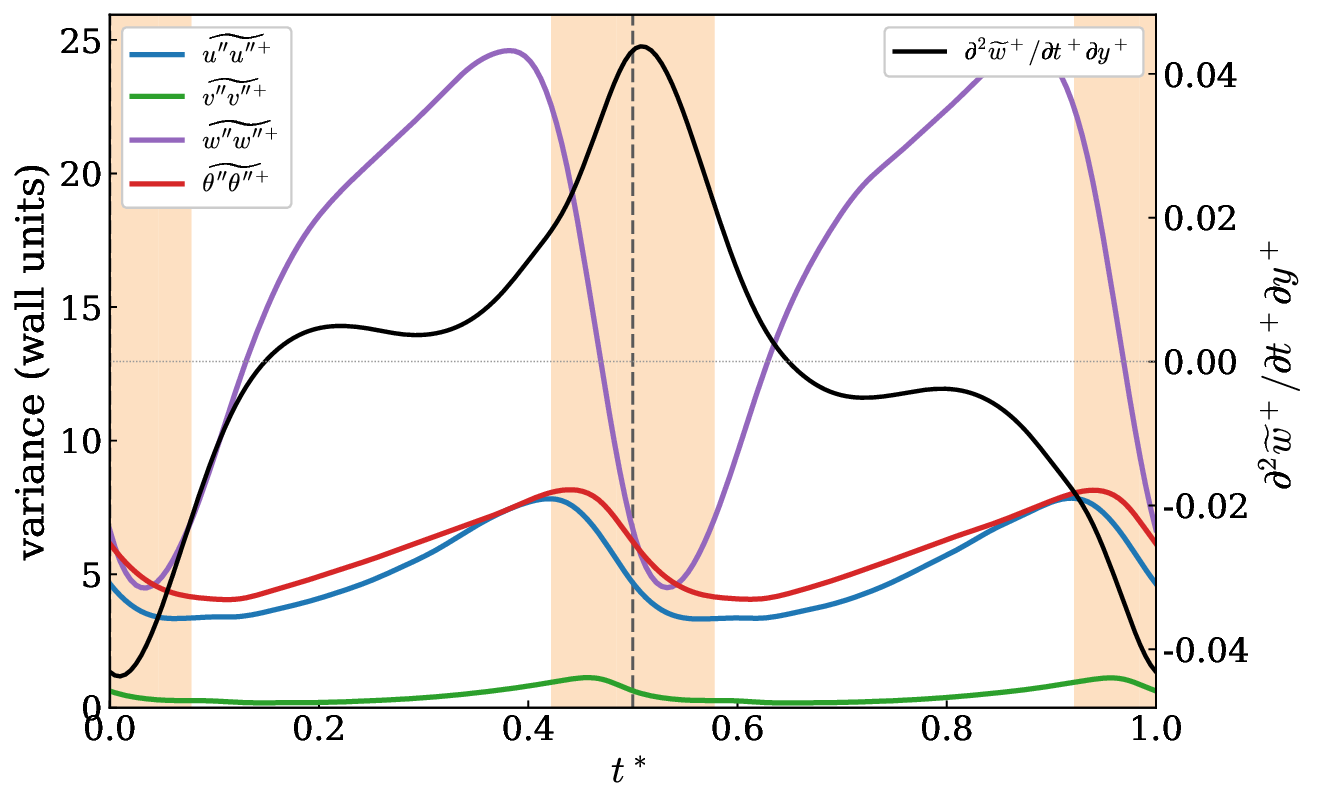}
	\caption{Velocity and temperature variances and the temporal gate at $y^+ = 10$ over the full actuation period for $T^+=350$, $W^+=30$. Left axis: the phase-resolved variances $\widetilde{u^{\dprime}u^{\dprime}}^+$, $\widetilde{v^{\dprime}v^{\dprime}}^+$, $\widetilde{w^{\dprime}w^{\dprime}}^+$ and $\widetilde{\theta^{\dprime}\theta^{\dprime}}^+$. Right axis: the Stokes-strain rate $\partial^2\widetilde{w}^+/\partial t^+\,\partial y^+$. The shaded windows mark the reversals, the phases over which $|\partial^2\widetilde{w}^+/\partial t^+\,\partial y^+|$ exceeds half its cycle maximum; the unshaded windows are the plateau, or lingering, phases over which it is small. The dashed vertical lines mark the two phases of maximum $|\partial^2\widetilde{w}^+/\partial t^+\,\partial y^+|$. This partition, and these reference lines, are retained in all subsequent phase-resolved figures.}
	\label{fig:variance_regions}
\end{figure}

The three velocity variances and the temperature variance are gathered at the single buffer-layer height $y^+ = 10$ in figure~\ref{fig:variance_regions}, set against the Stokes-strain rate $\partial^2\widetilde{w}^+/\partial t^+\,\partial y^+$, which renders explicit the partition into reversal and lingering phases introduced in \S\,\ref{sec:results:coherent}. Against this gate the evolution divides into three successive intervals. In the early plateau, whilst the gate decays from its reversal maximum, the spanwise variance $\widetilde{w^{\dprime}w^{\dprime}}^+$ stands apart: recharged almost immediately by the imposed Stokes shear, it grows to more than five times its post-reversal minimum over the plateau, whilst $\uupp$, $\widetilde{\theta^{\dprime}\theta^{\dprime}}^+$ and $\widetilde{v^{\dprime}v^{\dprime}}^+$ linger near their minima. Through the remainder of the plateau, the gate now small and nearly constant, all four variances grow together as the self-sustaining process recommences. As the gate magnitude rises towards the following reversal, the variances turn downward in a staggered sequence: $\widetilde{w^{\dprime}w^{\dprime}}^+$ first ($t^* \approx 0.38$, well before the reversal window opens at $t^* \approx 0.44$), its Stokes-shear source having faded with the slowly decaying strain, then $\uupp$ ($t^* \approx 0.42$), and last $\widetilde{\theta^{\dprime}\theta^{\dprime}}^+$ and $\widetilde{v^{\dprime}v^{\dprime}}^+$, nearly together ($t^* \approx 0.44$ and $0.46$), the wall-normal variance, small though it remains, thus peaking at essentially the same phase as the thermal variance. The streamwise and thermal variances accordingly track one another closely in amplitude, both suppressed at the reversal and recovering through the plateau, although not exactly in phase: $\uupp$ ceases to grow before $\widetilde{\theta^{\dprime}\theta^{\dprime}}^+$, which is sustained some two hundredths of the period longer into the reversal. This offset lies at the limit of the $32$-phase resolution; it is, however, systematic, recurring at both reversals, and constitutes the first legible trace of the dissimilarity: a hint that the streamwise field is arrested at the reversal by an agent that spares the scalar. Beyond this small lag and the modest differential suppression documented in \S\,\ref{sec:results:time_avg}, no amplitude asymmetry is disclosed at the variance level; the differential structure must be sought one level deeper, in the production terms $\widetilde{P}_{uu}$ and $\widetilde{P}_{\theta\theta}$ examined in the following subsection, where the same lag reappears, magnified, in the productions, at an amplitude no longer limited by the phase resolution. The staggered downturns of the four variances described here are collected in panel (\textit{c}) of the timeline figure~\ref{fig:mechanism_schematic}, the triangles there marking the sequence in which the components turn.

\subsection{Phase modulation of the variance production}
\label{sec:results:production}
\begin{figure}
	\centering
	\begin{minipage}{0.48\textwidth}
		\centering
		\includegraphics[width=\textwidth]{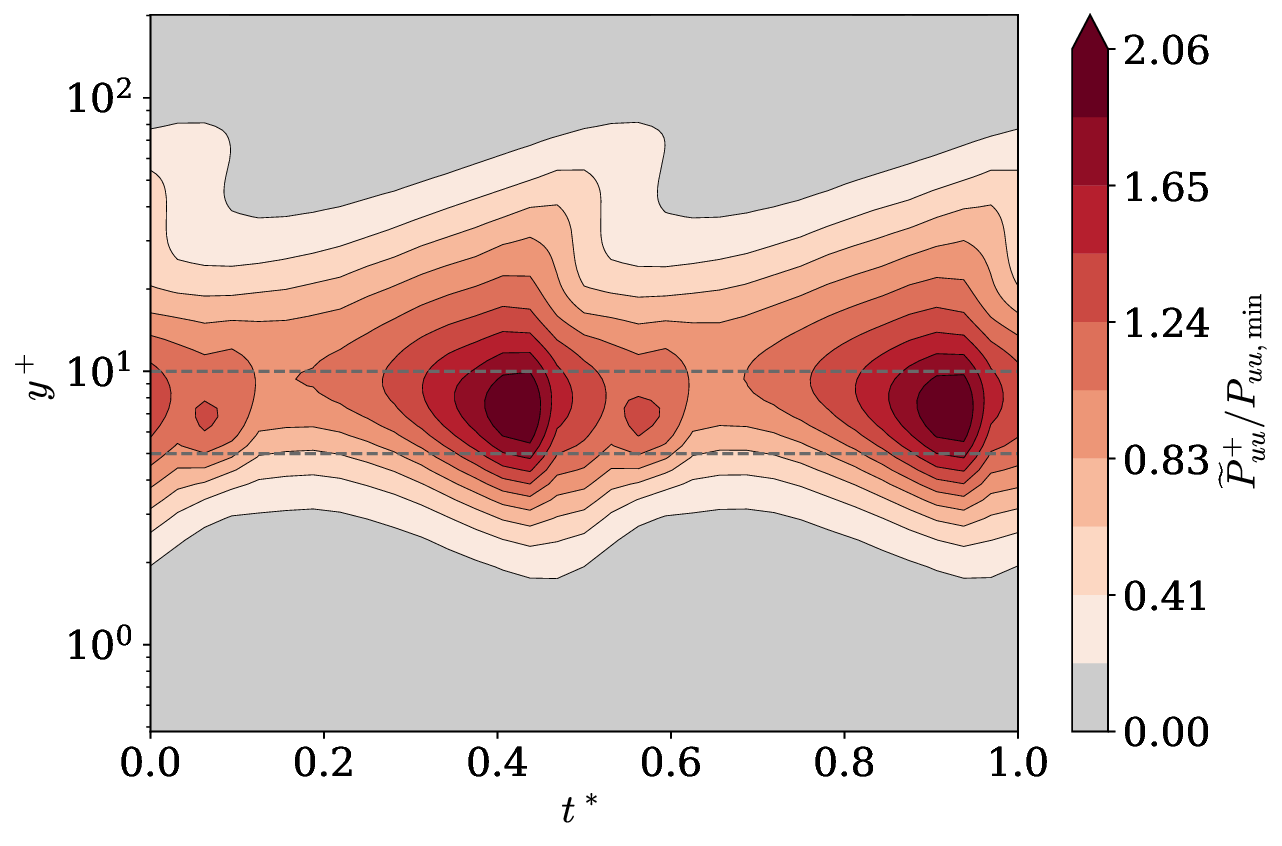}
		\vspace{0.2cm}
		\centerline{(\textit{a})}
	\end{minipage}
	\hfill
	\begin{minipage}{0.48\textwidth}
		\centering
		\includegraphics[width=\textwidth]{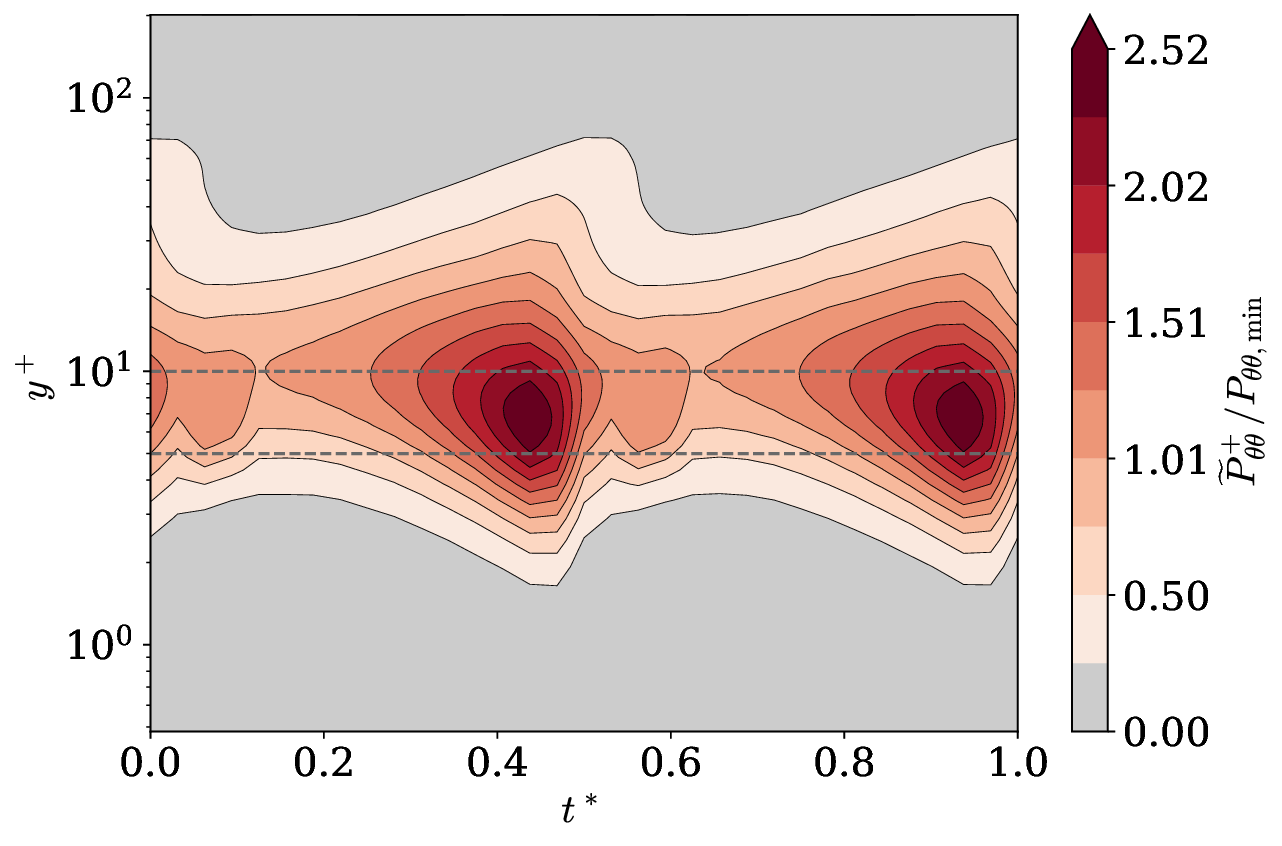}
		\vspace{0.2cm}
		\centerline{(\textit{b})}
	\end{minipage}
	\hfill
	\begin{minipage}{0.52\textwidth}
		\centering
		\includegraphics[width=\textwidth]{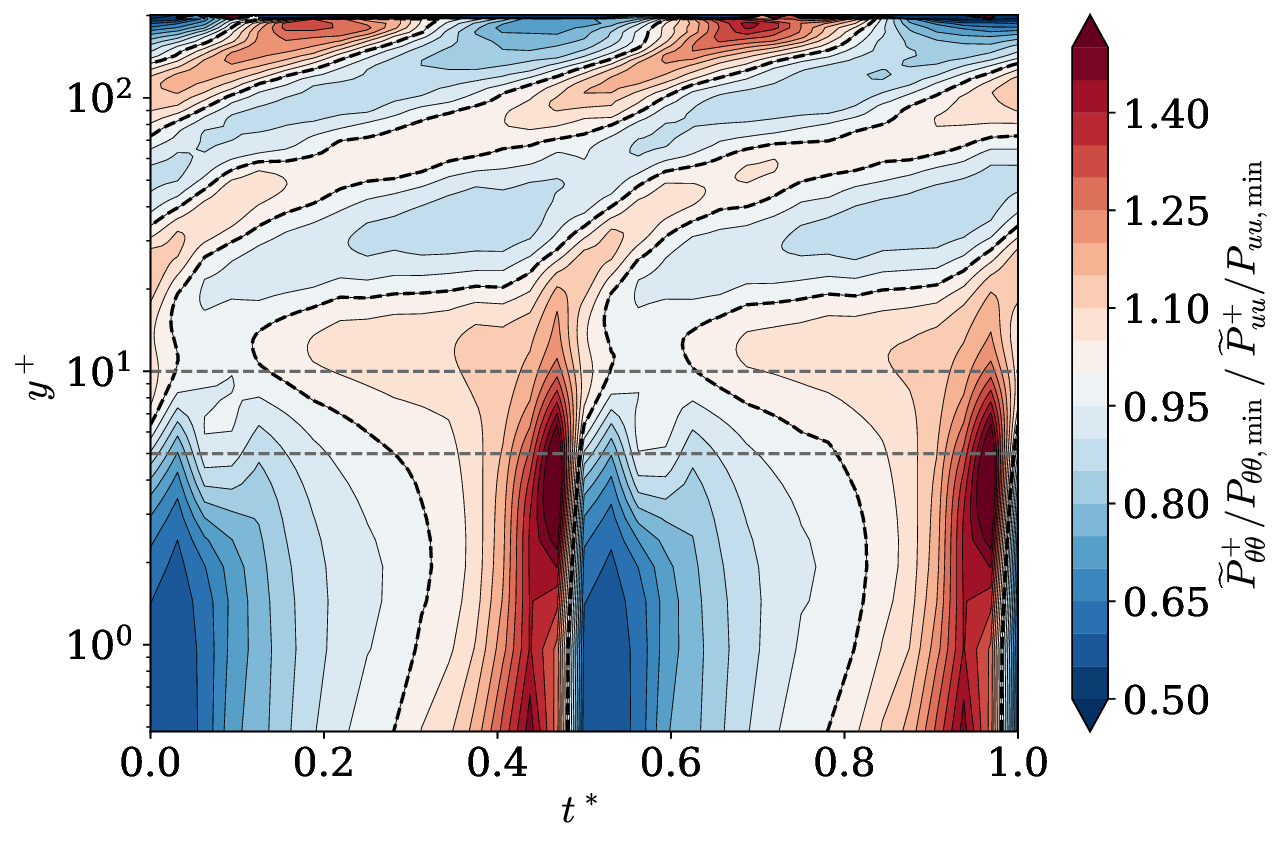}
		\vspace{0.2cm}
		\centerline{(\textit{c})}
	\end{minipage}
	\caption{Phase-resolved production terms at $T^+=350$, $W^+=30$, normalised by their respective phase-minimum values at $y^+ \approx 10$. (\textit{a}) Normalised velocity production $\widetilde{P}_{uu} / \widetilde{P}_{uu,\text{min}}$. (\textit{b}) Normalised temperature production $\widetilde{P}_{\theta\theta} / \widetilde{P}_{\theta\theta,\text{min}}$. (\textit{c}) Ratio of normalised productions: values above unity indicate that thermal production is relatively more amplified at that phase.}
	\label{fig:production_norm}
\end{figure}

The variance maps having disclosed no decisive amplitude asymmetry between the two fields, and the $y^+ = 10$ traces only a marginal lag, attention is directed to the production terms, which couple the variances to the wall-normal fluxes through the mean gradients and are the immediate candidates to carry the differential response. They are examined in figure~\ref{fig:production_norm}, each normalised by its own phase-minimum value at $y^+ \approx 10$; the two minima fall at the same phase, so that the fields share a common suppression phase and their relative amplification across the cycle may be read directly.

The ratio of the two normalised productions (figure~\ref{fig:production_norm}\textit{c}) is far from uniform: it carries a sharply localised spike at the Stokes-strain reversal, rising to some $30\%$ above unity, and relaxes to a milder excess of order $6\%$ through the plateau. The reversal spike is the more arresting feature, and panels (\textit{a}) and (\textit{b}) disclose its character: both productions grow together as the turbulence recovers; $\widetilde{P}_{uu}$, however, reaches its maximum and declines \emph{before} $\widetilde{P}_{\theta\theta}$, opening a window in which the scalar production still climbs whilst the velocity production already falls. The spike is thus the signature of a \emph{lag}, and a telling one: it is the streamwise production, and it alone, that is curtailed, and precisely at the reversal. As each production is the product of a flux and a mean gradient, $\widetilde{P}_{uu} = -2\widetilde{u^{\dprime}v^{\dprime}}\,\partial\widetilde{u}/\partial y$, the curtailment must be the work of one of two agents: a collapse of the momentum flux, or a collapse of the mean gradient. Each is examined in turn, the elimination of one being the first step towards the mechanism. The milder plateau excess is deferred until the spike is resolved.

\begin{figure}
	\centering
	\begin{minipage}{0.48\textwidth}
		\centering
		\includegraphics[width=\textwidth]{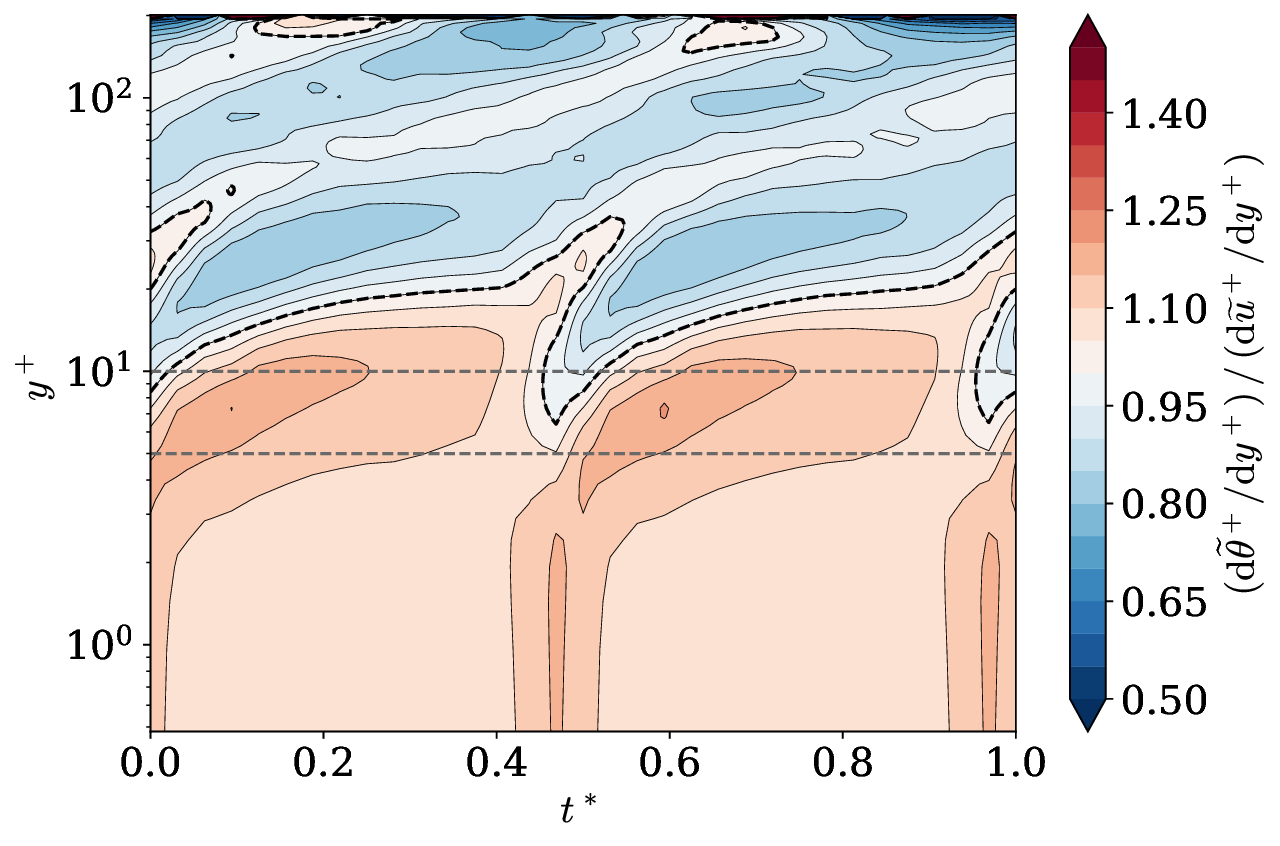}
		\vspace{0.2cm}
		\centerline{(\textit{a})}
	\end{minipage}
	\hfill
	\begin{minipage}{0.48\textwidth}
		\centering
		\includegraphics[width=\textwidth]{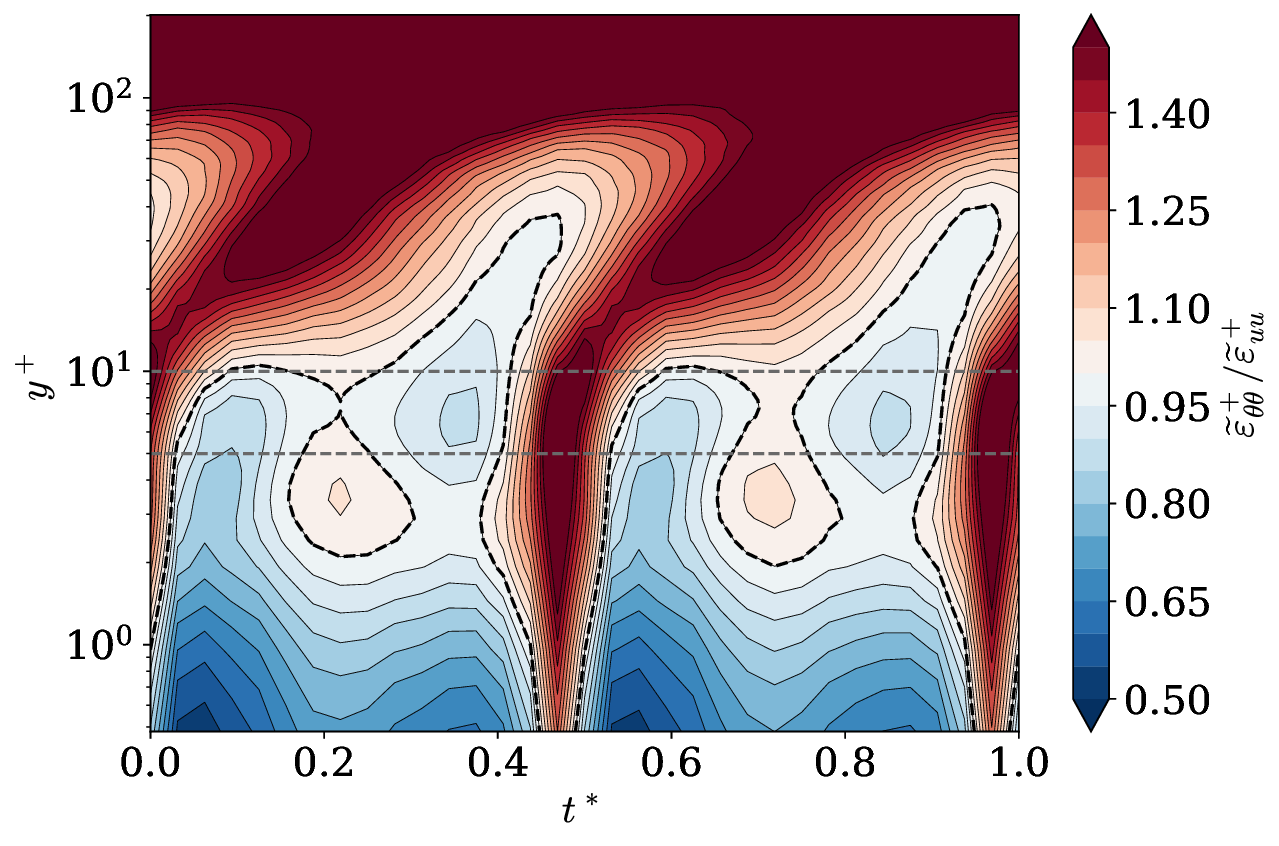}
		\vspace{0.2cm}
		\centerline{(\textit{b})}
	\end{minipage}
	\caption{Phase-resolved maps in the $(y^+,\,t^*)$ plane at $T^+=350$; $y^+$ is defined using the unactuated friction velocity $u_{\tau,0}$ and the horizontal lines mark $y^+=5$ and $y^+=10$. (\textit{a}) Mean-gradient ratio $(\partial\widetilde{\theta}^+/\partial y^+) / (\partial\widetilde{u}^+/\partial y^+)$. (\textit{b}) Dissipation ratio $\widetilde{\varepsilon}_{\theta\theta}^+ / \widetilde{\varepsilon}_{uu}^+$. The dashed isolines mark unit ratio in both panels.}
	\label{fig:budget_decomp}
\end{figure}

The two agents are separated by writing the production ratio $\widetilde{P}_{\theta\theta}/\widetilde{P}_{uu}$ as the product of a flux ratio $\widetilde{\theta^{\dprime}v^{\dprime}}/\widetilde{u^{\dprime}v^{\dprime}}$ and a mean-gradient ratio $(\partial\widetilde{\theta}/\partial y) / (\partial\widetilde{u}/\partial y)$. The gradient is the more readily dismissed, and is taken first (figure~\ref{fig:budget_decomp}\textit{a}). At $y^+ \approx 10$ the gradient ratio remains close to unity throughout the cycle, carrying a slight, steady plateau excess and \emph{no} reversal feature: the two mean gradients are modulated almost in step. The gradient is therefore exonerated, and the sharp reversal feature must be carried by the flux: it is the momentum flux $-\widetilde{u^{\dprime}v^{\dprime}}$, and not the mean gradient, that collapses at the reversal, the scalar flux being spared. The gradient's small plateau excess is not without consequence, however: as both fluxes are produced on the \emph{same} wall-normal fluctuation $\widetilde{v^{\dprime}v^{\dprime}}$, it is this residual gradient asymmetry that permits the scalar production to remain slightly the higher through the plateau, and so seeds the milder $6\%$ dominance whose mechanism is assembled in \S\,\ref{sec:results:synthesis}.

One accomplice remains to be eliminated: a dissipation asymmetry, which could either cancel the production asymmetry before it reaches the variances or itself masquerade as the cause. The phase-resolved dissipation ratio $\widetilde{\varepsilon}_{\theta\theta}^+ / \widetilde{\varepsilon}_{uu}^+$ is mapped in figure~\ref{fig:budget_decomp}(\textit{b}). The natural baseline for it is the unactuated value of the same ratio, which over the buffer-layer band $y^+\in[5,10]$ is $1.32$, already above unity as the scalar variance, and hence its dissipation, exceeds the streamwise even without actuation. Through the plateau the actuated ratio falls below this baseline, so no independent dissipation asymmetry sustains the $6\%$ dominance there. At the reversal it rises above the baseline, although only as a downstream echo of the production lag, the elevated scalar variance driving a correspondingly higher scalar dissipation; acting as a sink, this slightly \emph{attenuates} the spike rather than creating it. Dissipation is therefore a consequence, not an independent agent, and the production asymmetry passes intact to the variances.

\begin{figure}
	\centering
	\begin{minipage}{0.48\textwidth}
		\centering
		\includegraphics[width=\textwidth]{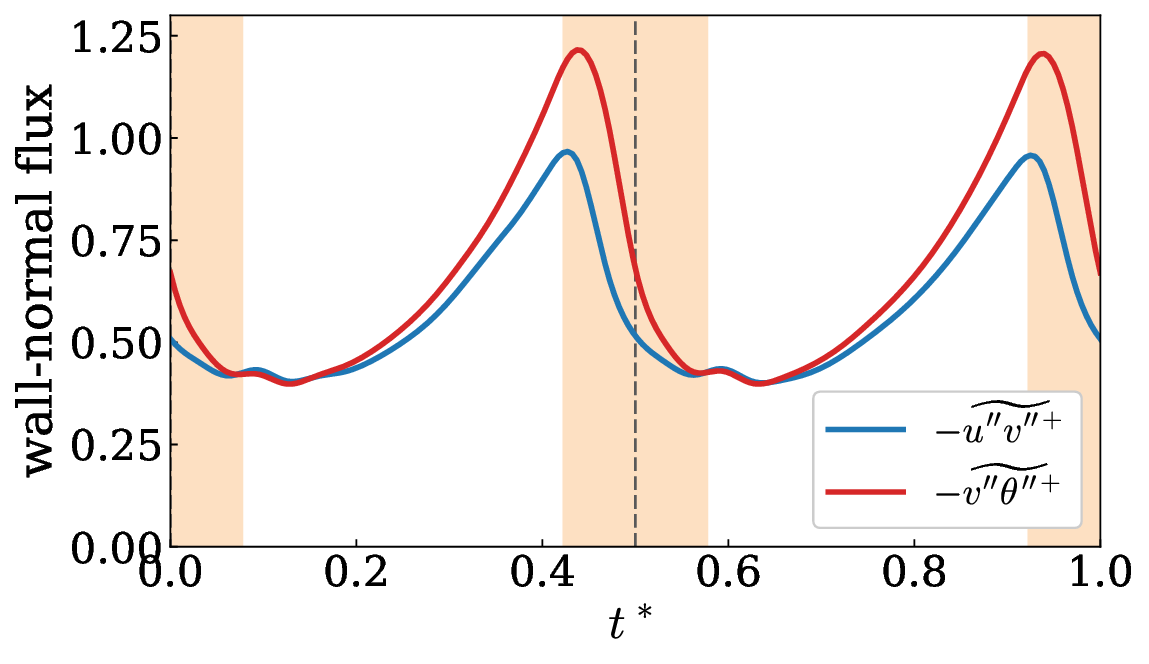}
		\centerline{(\textit{a})}
	\end{minipage}
	\hfill
	\begin{minipage}{0.48\textwidth}
		\centering
		\includegraphics[width=\textwidth]{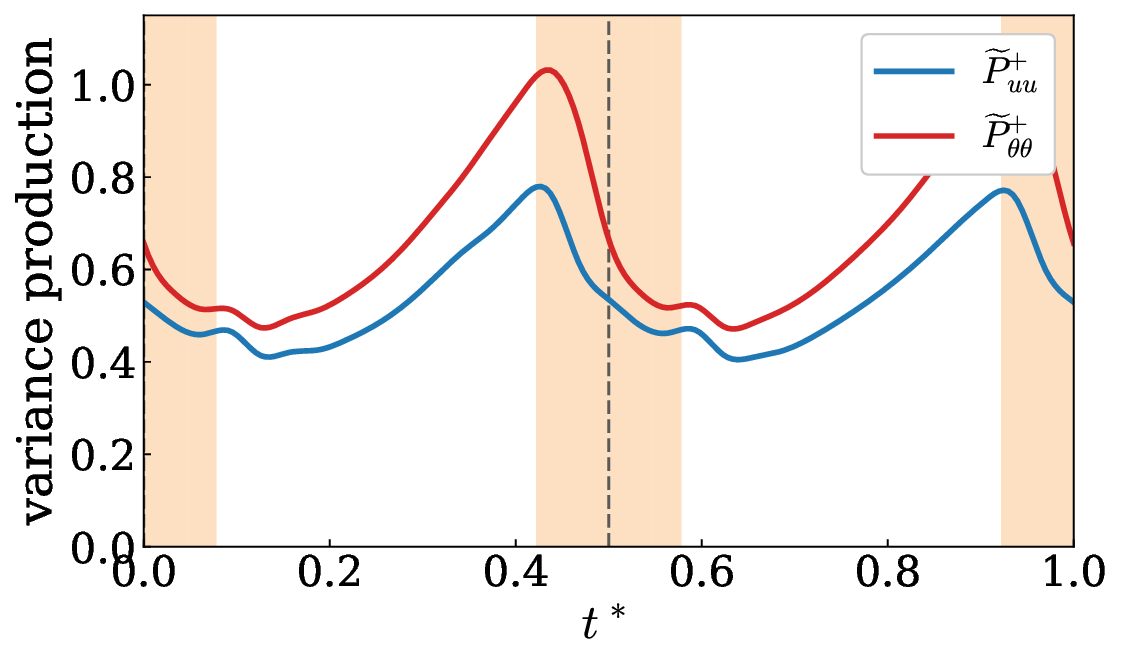}
		\centerline{(\textit{b})}
	\end{minipage}\\[2mm]
	\begin{minipage}{0.48\textwidth}
		\centering
		\includegraphics[width=\textwidth]{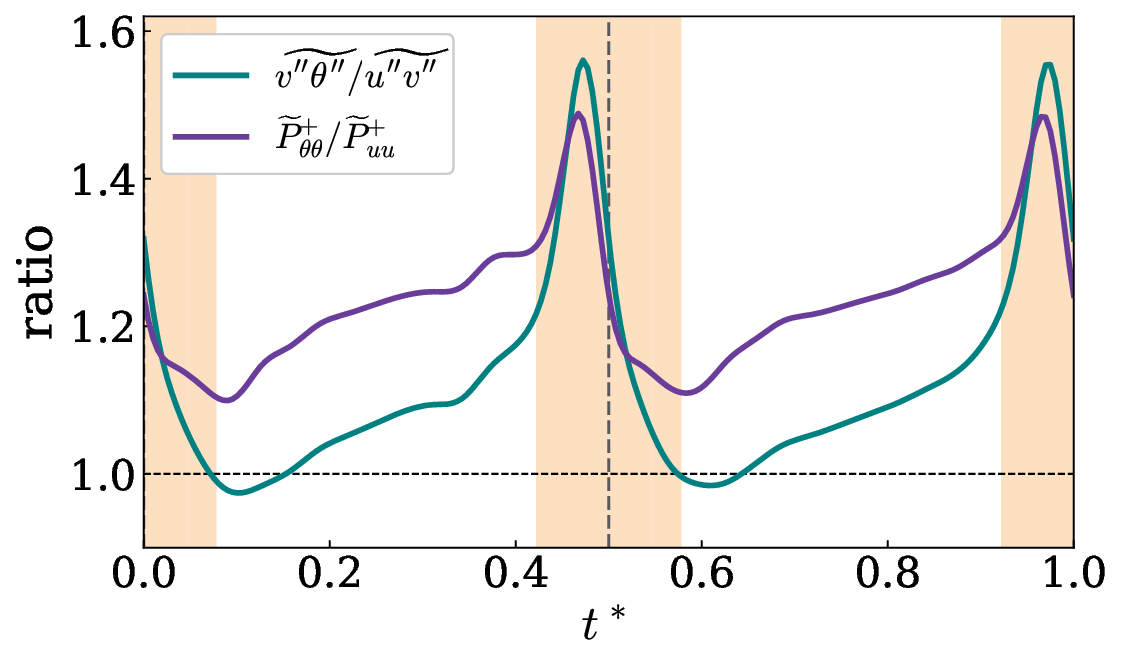}
		\centerline{(\textit{c})}
	\end{minipage}
	\caption{Phase-resolved budget quantities at $y^+ = 10$ over the full actuation period for $T^+=350$, $W^+=30$, all in wall units; the reversal windows are shaded as in figure~\ref{fig:variance_regions} and the dashed vertical lines mark the two phases of maximum $|\partial^2\widetilde{w}^+/\partial t^+\,\partial y^+|$. (\textit{a}) Wall-normal flux magnitudes $-\widetilde{u^{\dprime}v^{\dprime}}^+$ (momentum) and $-\widetilde{v^{\dprime}\theta^{\dprime}}^+$ (scalar). (\textit{b}) Variance productions $\widetilde{P}_{uu}^+$ and $\widetilde{P}_{\theta\theta}^+$. (\textit{c}) Production ratio $\widetilde{P}_{\theta\theta}^+/\widetilde{P}_{uu}^+$ and flux ratio $\widetilde{v^{\dprime}\theta^{\dprime}}/\widetilde{u^{\dprime}v^{\dprime}}$.}
	\label{fig:variance_lag}
\end{figure}

The strands are gathered at a single buffer-layer height, $y^+ = 10$, in figure~\ref{fig:variance_lag}, where the flux collapse that the elimination implied is seen directly, the reversal windows shaded as defined in figure~\ref{fig:variance_regions}. The two wall-normal fluxes, of comparable magnitude in the cycle mean (the scalar flux the larger by some $15\%$), part sharply at the reversal (panel~\textit{a}): the momentum flux $-\widetilde{u^{\dprime}v^{\dprime}}$ collapses whilst the scalar flux $-\widetilde{v^{\dprime}\theta^{\dprime}}$ holds, their ratio reaching about $1.55$. The productions follow (panel~\textit{b}), sharing a peak before $\widetilde{P}_{uu}$ decays the faster, whilst the surviving flux continues to feed the scalar production. Panel~(\textit{c}) sets their raw ratio $\widetilde{P}_{\theta\theta}/\widetilde{P}_{uu}$, as distinct from the phase-minimum-normalised ratio of figure~\ref{fig:production_norm}\textit{c}, beside the flux ratio $\widetilde{v^{\dprime}\theta^{\dprime}}/\widetilde{u^{\dprime}v^{\dprime}}$: the production ratio stands above unity throughout, the flux ratio essentially so, dipping some $2\%$ below unity as the fluxes trough after each reversal; the two spike together at each reversal, the production ratio rising to about $1.49$ and the flux ratio to about $1.55$, their near-coincidence there confirming that the mean-gradient ratio remains close to unity and that the reversal spike is carried by the flux. The immediate cause of the dissimilarity is thus isolated: a selective collapse of the momentum flux at the reversal, propagated through production to the variance and onward to the integrated transport. In the timeline of figure~\ref{fig:mechanism_schematic}, this stage of the argument occupies panels (\textit{e}) and (\textit{d}), in which the parting of the fluxes and the ensuing production lag are displayed against the gate. One question alone remains: what collapses that flux at the reversal, acting upon the momentum field and not upon the scalar? The lone structural difference between the two budgets names the suspect, and the following subsection places it on trial.

\subsection{The diagonal pressure-strain redistribution and the Stokes-strain rate}
\label{sec:results:pressure_strain}
\begin{figure}
	\centering
	\begin{minipage}{0.48\textwidth}
		\centering
		\includegraphics[width=\textwidth]{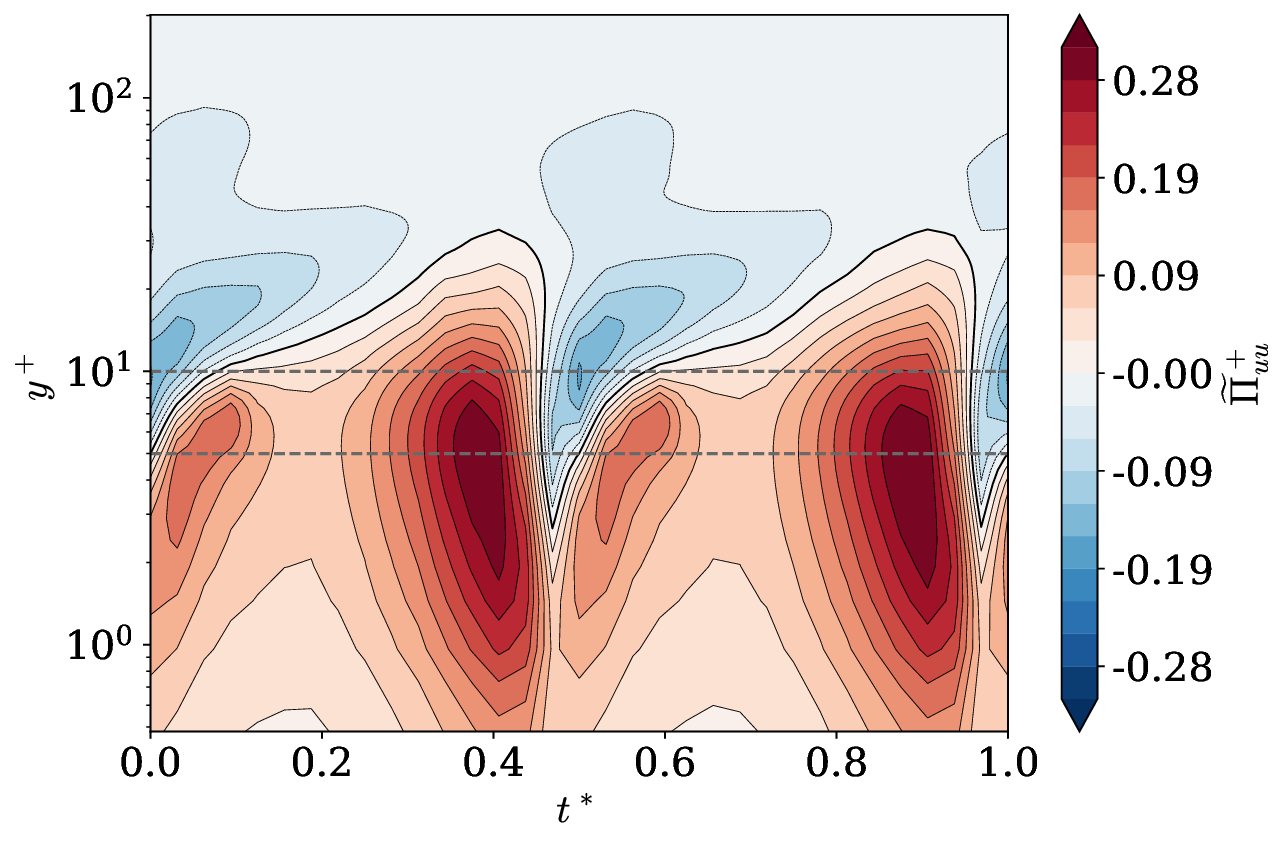}
		\vspace{0.2cm}
		\centerline{(\textit{a})}
	\end{minipage}
	\hfill
	\begin{minipage}{0.48\textwidth}
		\centering
		\includegraphics[width=\textwidth]{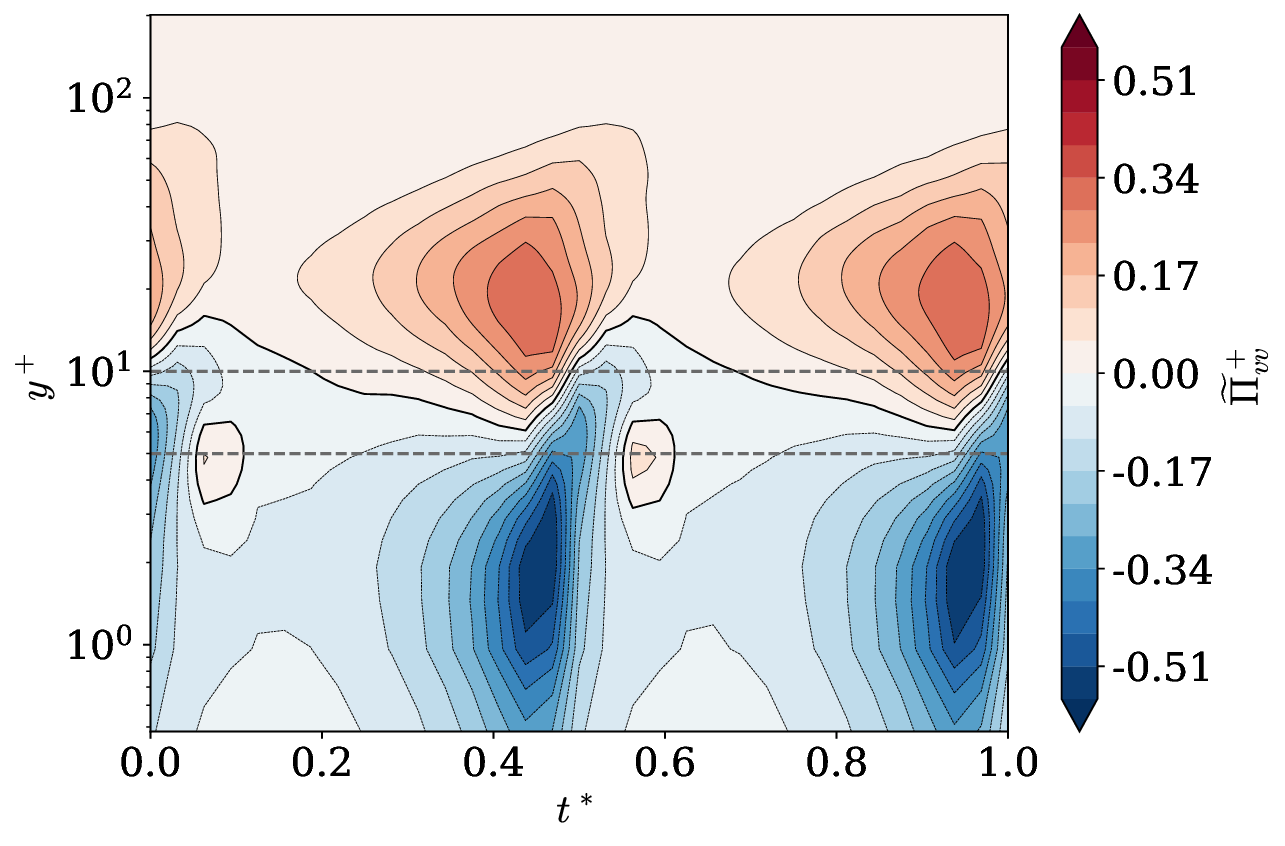}
		\vspace{0.2cm}
		\centerline{(\textit{b})}
	\end{minipage}
	\hfill
	\begin{minipage}{0.48\textwidth}
		\centering
		\includegraphics[width=\textwidth]{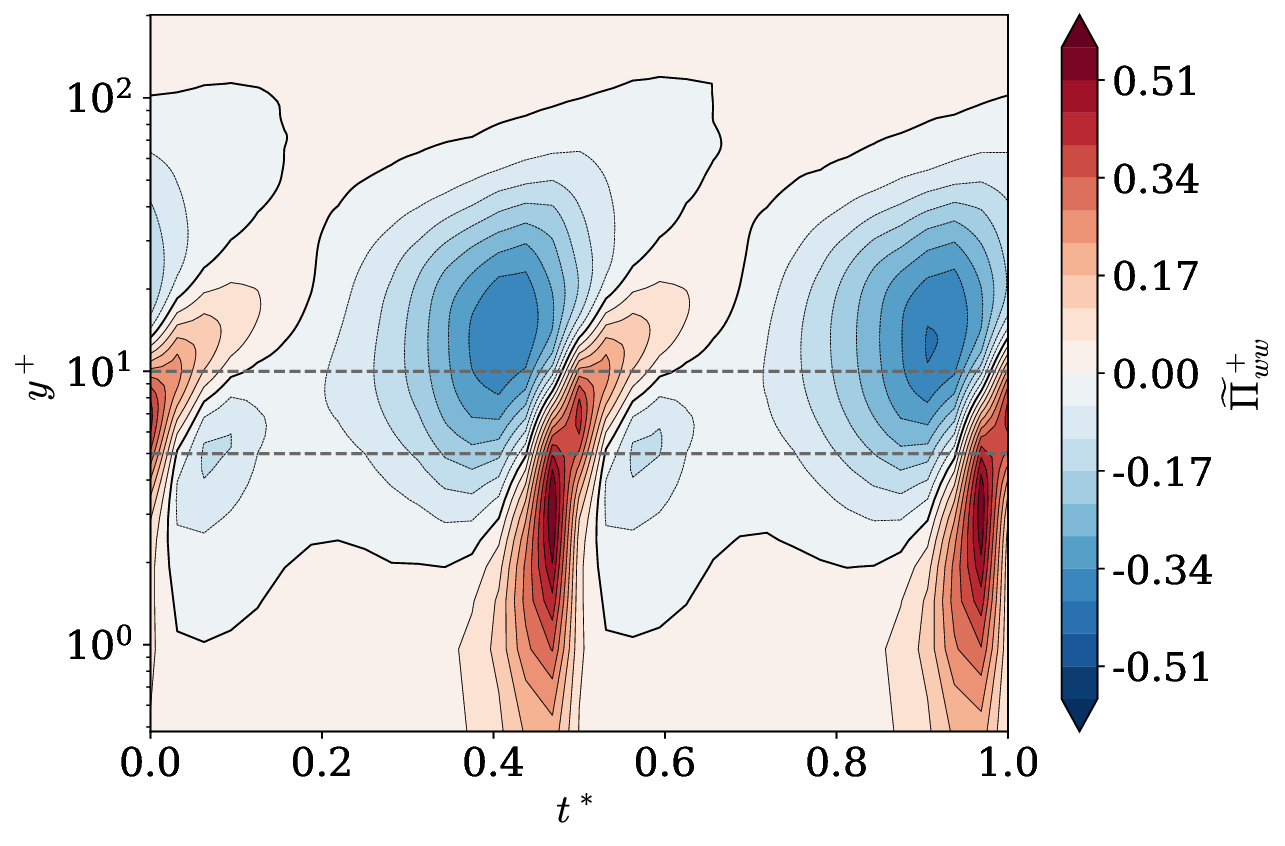}
		\vspace{0.2cm}
		\centerline{(\textit{c})}
	\end{minipage}
	\caption{Phase-resolved diagonal pressure-strain redistribution terms at $T^+=350$. (\textit{a}) $\Pi_{uu}^+$, (\textit{b}) $\Pi_{vv}^+$, (\textit{c}) $\Pi_{ww}^+$. Negative values indicate energy drained from the corresponding component; positive values indicate energy received via inter-component redistribution.}
	\label{fig:pressure_strain}
\end{figure}

The production asymmetry identified in \S\,\ref{sec:results:production}, a $30\%$ spike at the reversal and a sustained $6\%$ dominance during the plateau, indicates a structural difference in the dynamical equations governing $\uupp$ and $\widetilde{\theta^{\dprime}\theta^{\dprime}}$. The suspect named at the close of \S\,\ref{sec:results:production} is the diagonal pressure-strain redistribution, defined in equation~\eqref{eq:pi_diag}, which enters the velocity variance budget~\eqref{eq:budget_uu} and has no analogue in the scalar variance equation~\eqref{eq:budget_thth}. Its phase-resolved behaviour is presented in figure~\ref{fig:pressure_strain} and is examined against the partition into reversal and lingering phases established in \S\,\ref{sec:results:coherent} (figure~\ref{fig:variance_regions}). The plateau is considered first, then the sequencing of the redistribution amongst the three components as the reversal approaches, and the reversal itself; the cause of the reversal-phase behaviour is then isolated by means of a rapid/slow decomposition of the fluctuating pressure.

During the plateau, $\Pi_{ww}$ (figure~\ref{fig:pressure_strain}\textit{c}) is predominantly negative, indicating that spanwise energy accumulated through the imposed Stokes shear is progressively drained and redistributed to $\uupp$ and $\widetilde{v^{\dprime}v^{\dprime}}$. Correspondingly, both $\Pi_{uu}$ (figure~\ref{fig:pressure_strain}\textit{a}) and $\Pi_{vv}$ (figure~\ref{fig:pressure_strain}\textit{b}) are positive throughout this interval. The positive $\Pi_{uu}$ acts as a source for $\uupp$ and thereby sustains $\widetilde{P}_{uu}$; as no analogous pressure-redistribution term appears in the scalar variance equation, this source is absent for the scalar field, the diagonal mechanism acting during the plateau in the direction opposite to the observed $6\%$ thermal dominance. The positive $\Pi_{vv}$ simultaneously sustains the wall-normal velocity fluctuations $\widetilde{v^{\dprime}v^{\dprime}}$, which govern the turbulent transport of both momentum and heat; the extent to which this sustained wall-normal amplification translates symmetrically or asymmetrically into the two turbulent fluxes is examined in \S\,\ref{sec:results:off_diagonal}.

\begin{figure}
	\centering
	\begin{minipage}{0.48\textwidth}
		\centering
		\includegraphics[width=\textwidth]{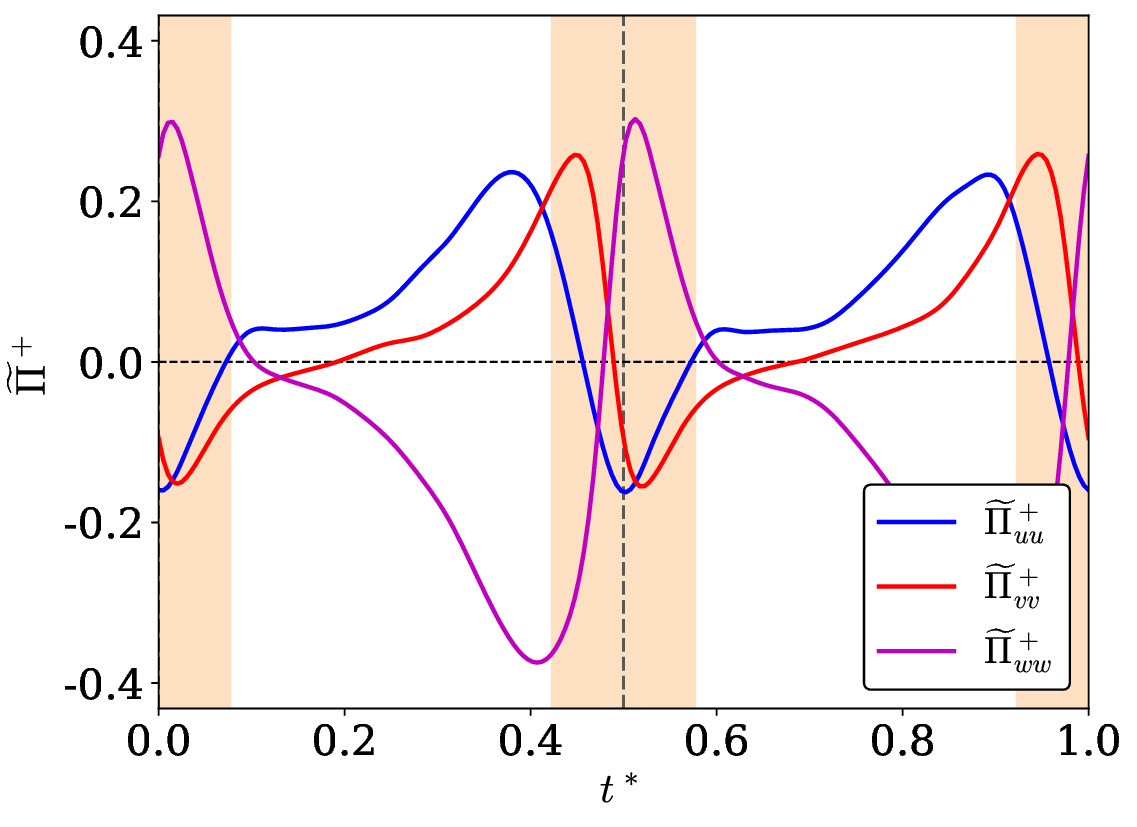}
		\vspace{0.2cm}
		\centerline{(\textit{a})}
	\end{minipage}
	\hfill
	\begin{minipage}{0.48\textwidth}
		\centering
		\includegraphics[width=\textwidth]{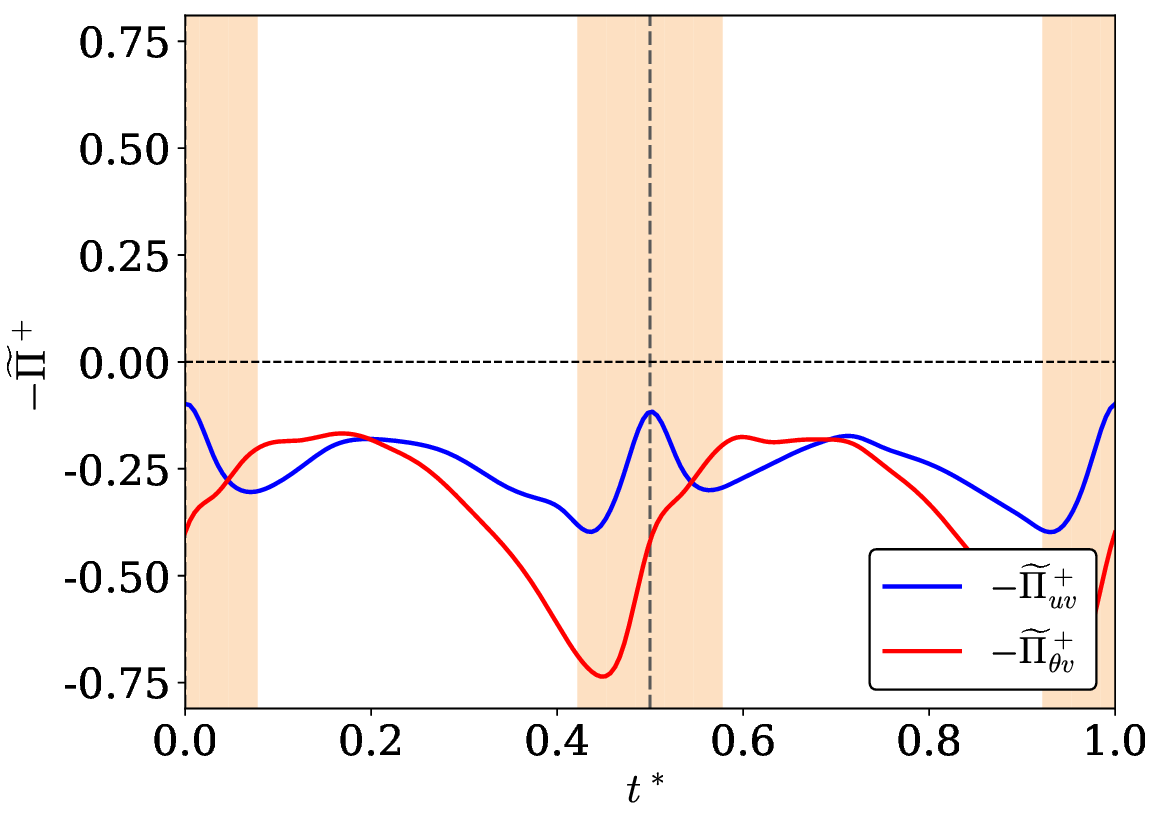}
		\vspace{0.2cm}
		\centerline{(\textit{b})}
	\end{minipage}
	\caption{Phase-resolved pressure-strain terms extracted at $y^+ = 10$ for $T^+=350$. (\textit{a}) Diagonal components $\widetilde{\Pi}_{uu}^+$, $\widetilde{\Pi}_{vv}^+$, and $\widetilde{\Pi}_{ww}^+$. (\textit{b}) Off-diagonal pressure-strain terms $-\widetilde{\Pi}_{uv}^+$ and $-\widetilde{\Pi}_{v\theta}^+$. The dashed horizontal line marks zero; the reversal windows are shaded and the dashed vertical lines mark the maxima of $|\partial^2\widetilde{w}^+/\partial t^+\,\partial y^+|$, both as defined in figure~\ref{fig:variance_regions}.}
	\label{fig:pi_1d}
\end{figure}

As the actuation cycle approaches the reversal, the temporal sequencing of the redistribution amongst the three components becomes material. The $t^*$-distributions of $\Pi_{uu}$, $\Pi_{vv}$, and $\Pi_{ww}$ extracted at $y^+ = 10$ are presented in figure~\ref{fig:pi_1d}\textit{a}, which renders the phase relationships more legibly than the two-dimensional fields. Although $|\Pi_{ww}|$ and $\Pi_{uu}$ grow concurrently during the plateau, as the spanwise variance charged by the Stokes shear is drained into the streamwise component, the two quantities do not peak simultaneously: $\Pi_{uu}$ attains its maximum at $t^* \approx 0.40$ and begins to decline, whilst $|\Pi_{ww}|$ continues to grow for a brief interval thereafter; the downturn of the streamwise variance at $t^* \approx 0.42$, noted in \S\,\ref{sec:results:stochastic}, follows immediately upon the decline of its pressure-strain source. This lag reflects the different physical quantities to which each term responds within the Stokes layer. The streamwise pressure-strain $\Pi_{uu}$ is governed primarily by the Stokes strain $\partial\widetilde{w}/\partial y$, which controls the disruption and regeneration of near-wall streaks through the tilting and stretching of quasi-streamwise vortices \citep{agostini_turbulence_2015}; the spanwise term $\Pi_{ww}$, in contrast, responds more directly to the spanwise velocity magnitude $\widetilde{w}$, which retains a large value even as the strain $\partial\widetilde{w}/\partial y$ has already peaked and begun to decline, the wall-normal diffusion of the spanwise motion causing the gradient to peak before the velocity. $\Pi_{uu}$ therefore peaks and declines before $|\Pi_{ww}|$ attains its maximum. At the culmination of the plateau activity ($t^* \approx 0.40$) the redistribution favours the streamwise component by a factor of about two, $\Pi_{uu}^+ \approx 0.24$ against $\Pi_{vv}^+ \approx 0.11$. The magnitude of this source is placed in perspective by the unactuated flow, in which the diagonal pressure-strain at the same height is a weak drain, $\Pi_{uu,0}^+ = -0.039$ (the slightly larger value $-0.045$ quoted in \S\,\ref{sec:results:time_avg} refers to the unactuated production peak at $y^+ \approx 12$): actuation converts a weak sink into a source of some six times its magnitude. As every unactuated diagonal term is of comparably negligible size, the dashed zero line of figure~\ref{fig:pi_1d}\textit{a} may be read as the unactuated baseline.

A direct consequence of this sequencing follows from the trace-free condition $\Pi_{uu} + \Pi_{vv} + \Pi_{ww} = 0$, which holds pointwise in the phase-averaged sense at every $(y, t^*)$ as established in \S\,\ref{sec:methodology:budgets}. As $\Pi_{uu}$ has begun to decline whilst $|\Pi_{ww}|$ continues to grow, the energy still extracted from $\widetilde{w^{\dprime}w^{\dprime}}$ must be redirected entirely to $\Pi_{vv}$, which therefore exhibits its steepest rise and attains its own peak during this narrow pre-reversal window. This temporal sequencing pre-charges the wall-normal velocity fluctuations $\widetilde{v^{\dprime}v^{\dprime}}$ to their maximum extent precisely at the onset of the reversal, when the diagonal drain of $\Pi_{uu}$ is about to suppress the streamwise variance. At the reversal itself, $\Pi_{ww}$ becomes strongly positive over a short temporal window, indicating rapid energy injection into the spanwise component by the reversing Stokes layer. Simultaneously, $\Pi_{uu}$ transitions sharply to negative values in the buffer-layer region ($y^+ \gtrsim 10$); this reversal-phase sink is the transient counterpart of the time-mean near-wall source identified in \S\,\ref{sec:results:time_avg}. As an energy sink, the reversal-phase $\Pi_{uu}$ actively drains $\uupp$. As the temperature variance equation contains no pressure-strain analogue, the scalar field is entirely unaffected by this inter-component redistribution and evolves unimpeded. This structural asymmetry directly explains the production lag identified in \S\,\ref{sec:results:production}: the transition of $\Pi_{uu}$ to negative values drains $\uupp$, and through the streak--vortex regeneration cycle the depleted streamwise fluctuations weaken the quasi-streamwise vortices that carry the shear stress $-\widetilde{u^{\dprime}v^{\dprime}}$, thereby curtailing $\widetilde{P}_{uu} = -2\widetilde{u^{\dprime}v^{\dprime}}\,\partial\widetilde{u}/\partial y$, whilst $\widetilde{P}_{\theta\theta}$ continues to amplify, generating the $\approx 30\%$ spike in the production ratio. At the $\Pi_{uu}$ trough, which is attained at the reversal itself ($t^* \approx 0$), $\Pi_{uu}^+ \approx -0.16$, quantifying the drain that suppresses the streamwise variance at the reversal. The zero-crossings at $y^+ = 10$ further characterise these reversal dynamics. $\Pi_{uu}$ crosses zero at $t^* \approx 0.47$ (source to sink), whilst $\Pi_{ww}$ crosses zero at $t^* \approx 0.49$ (sink to source). $\Pi_{vv}$ does not cross zero until $t^* \approx 0.51$, lagging $\Pi_{uu}$ by $\Delta t^* \approx 0.04$. This delay has two superimposed origins. The first is algebraic: the trace-free condition permits $\Pi_{vv}$ to remain positive as long as $|\Pi_{uu}|$ exceeds $\Pi_{ww}$, a condition satisfied during the early reversal. The second is dynamical: whilst $\Pi_{uu}$ responds directly to the disruption of streak structures by the changing spanwise strain, $\Pi_{vv}$ is governed by quasi-streamwise vortices through return-to-isotropy tendencies, and these vortical structures possess a finite lifetime that extends beyond the streak disruption \citep{agostini_spanwise_2014}. The positive $\Pi_{vv}$ lobe therefore persists into the early reversal, maintaining elevated wall-normal velocity fluctuations at the onset of the diagonal drain of $\Pi_{uu}$.

The foregoing has established that the streamwise drain coincides with the reversal of the spanwise strain; coincidence, however, is not causation, and the question of cause remains open. The trigger may at least be named precisely: the quantity that distinguishes the reversal from the plateau is the gate itself, the Stokes-strain rate $\partial^2\widetilde{w}^+/\partial t^+\,\partial y^+$ of \S\,\ref{sec:results:coherent}. Whether this gate acts upon $\Pi_{uu}$ \emph{directly}, as the instantaneous linear response of the pressure field to the imposed shear, or only indirectly, through some slower turbulence-mediated relaxation of the residual anisotropy, is a question that admits a clean test. The fluctuating pressure decomposes into a rapid part, the linear response to the mean strain, and a slow part, the turbulence--turbulence return to isotropy \citep{rotta_statistische_1951,lumley_return_1977,pope_turbulent_2000}; if the streamwise drain is the rapid response to the reversing Stokes shear it will reside in the rapid pressure-strain and follow the gate in phase, whereas if it is a slow relaxation it will reside in the slow part and lag.

\begin{figure}
	\centering
	\begin{minipage}{0.7\textwidth}
		\centering
		\includegraphics[width=\textwidth]{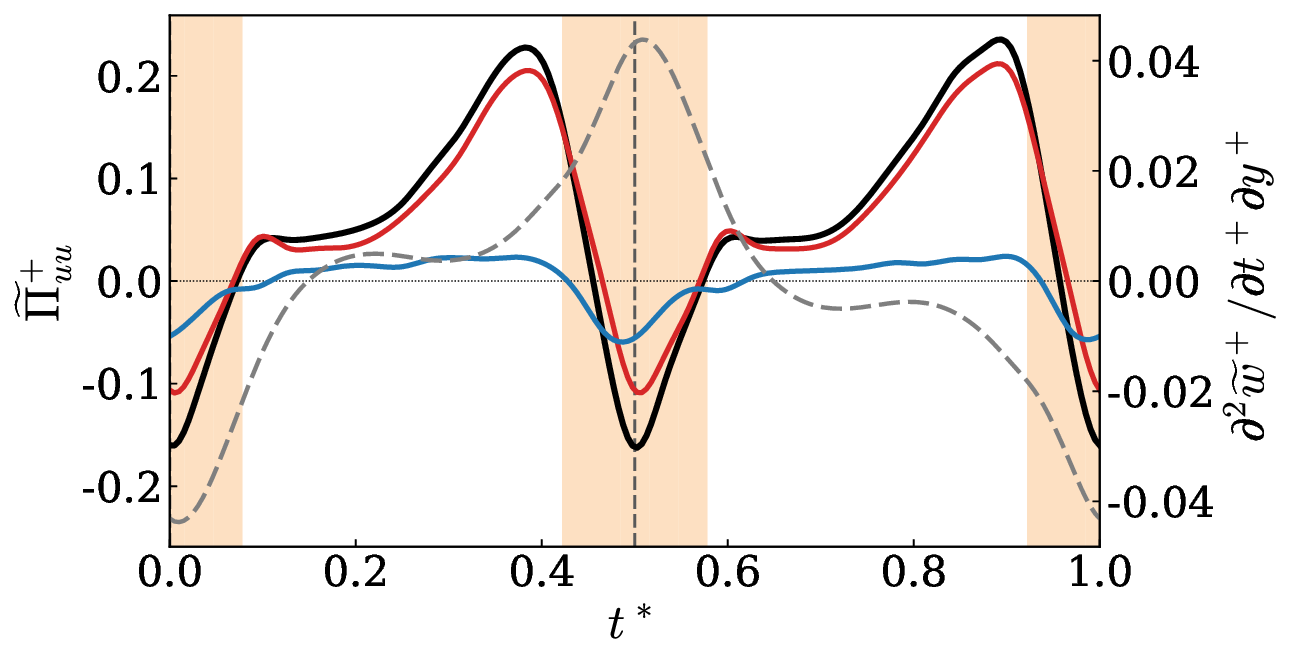}
		\centerline{(\textit{a})}
	\end{minipage}\\[2mm]
	\begin{minipage}{0.7\textwidth}
		\centering
		\includegraphics[width=\textwidth]{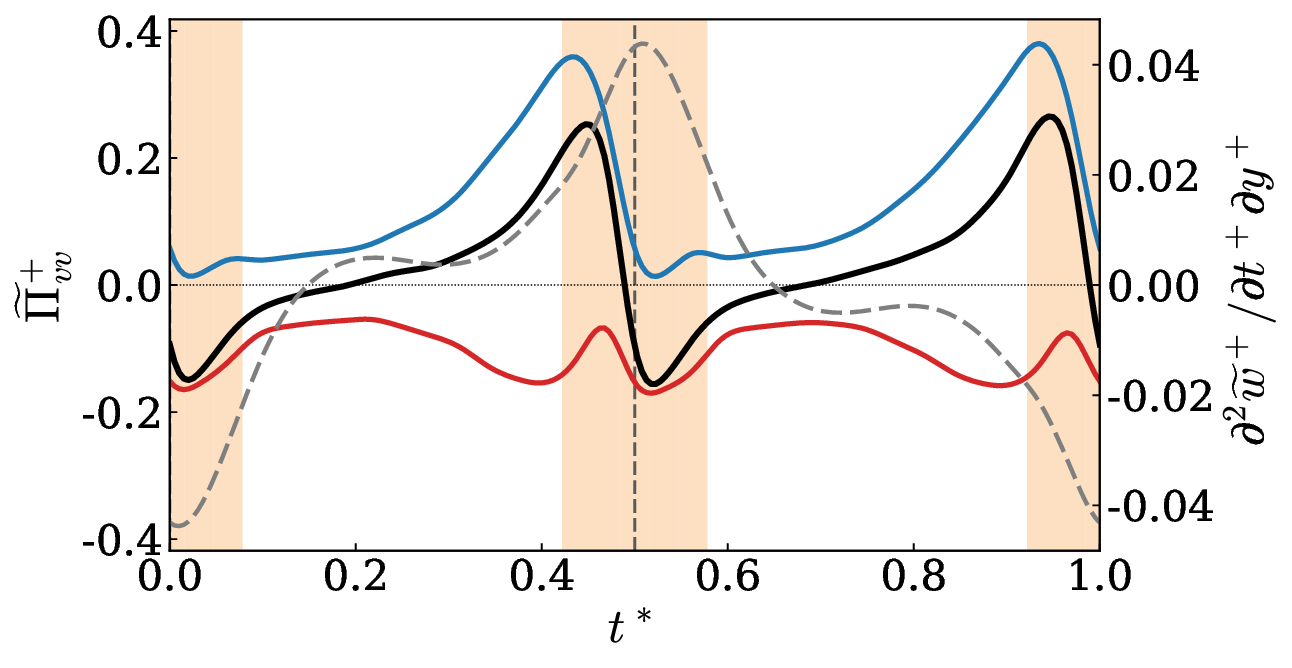}
		\centerline{(\textit{b})}
	\end{minipage}\\[2mm]
	\begin{minipage}{0.7\textwidth}
		\centering
		\includegraphics[width=\textwidth]{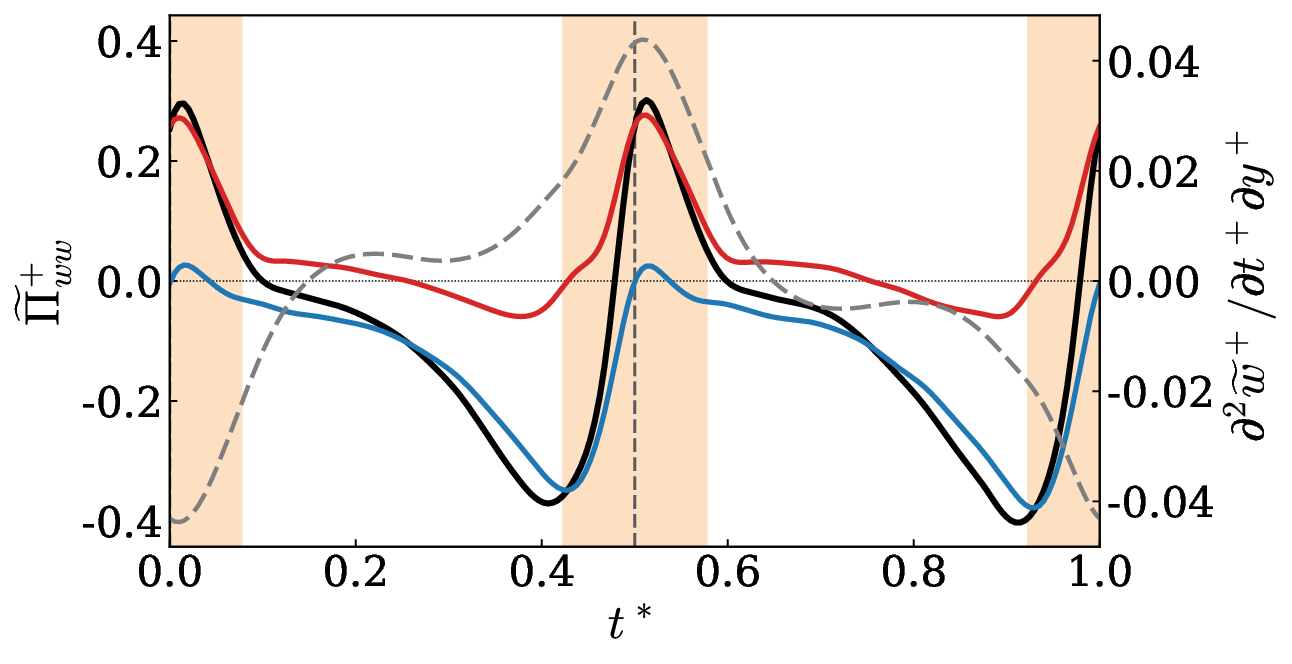}
		\centerline{(\textit{c})}
	\end{minipage}
	\caption{Rapid/slow decomposition of the diagonal pressure-strain at $y^+ = 10$ over the full actuation period ($T^+=350$, $W^+=30$), computed from the instantaneous DNS fields (appendix~\ref{app:rapid_slow}): total (black), rapid mean-shear part (red) and slow return-to-isotropy part (blue). The Stokes-strain rate $\partial^2\widetilde{w}^+/\partial t^+\partial y^+$ (grey dashed, right axis) marks the gate; the reversal windows are shaded and the dashed vertical lines mark the two phases of maximum $|\partial^2\widetilde{w}^+/\partial t^+\partial y^+|$, as in figure~\ref{fig:variance_regions}. (\textit{a}) $\Pi_{uu}$, (\textit{b}) $\Pi_{vv}$, (\textit{c}) $\Pi_{ww}$.}
	\label{fig:rapid_slow}
\end{figure}

\begin{figure}
	\centering
	\begin{minipage}{0.48\textwidth}
		\centering
		\includegraphics[width=\textwidth]{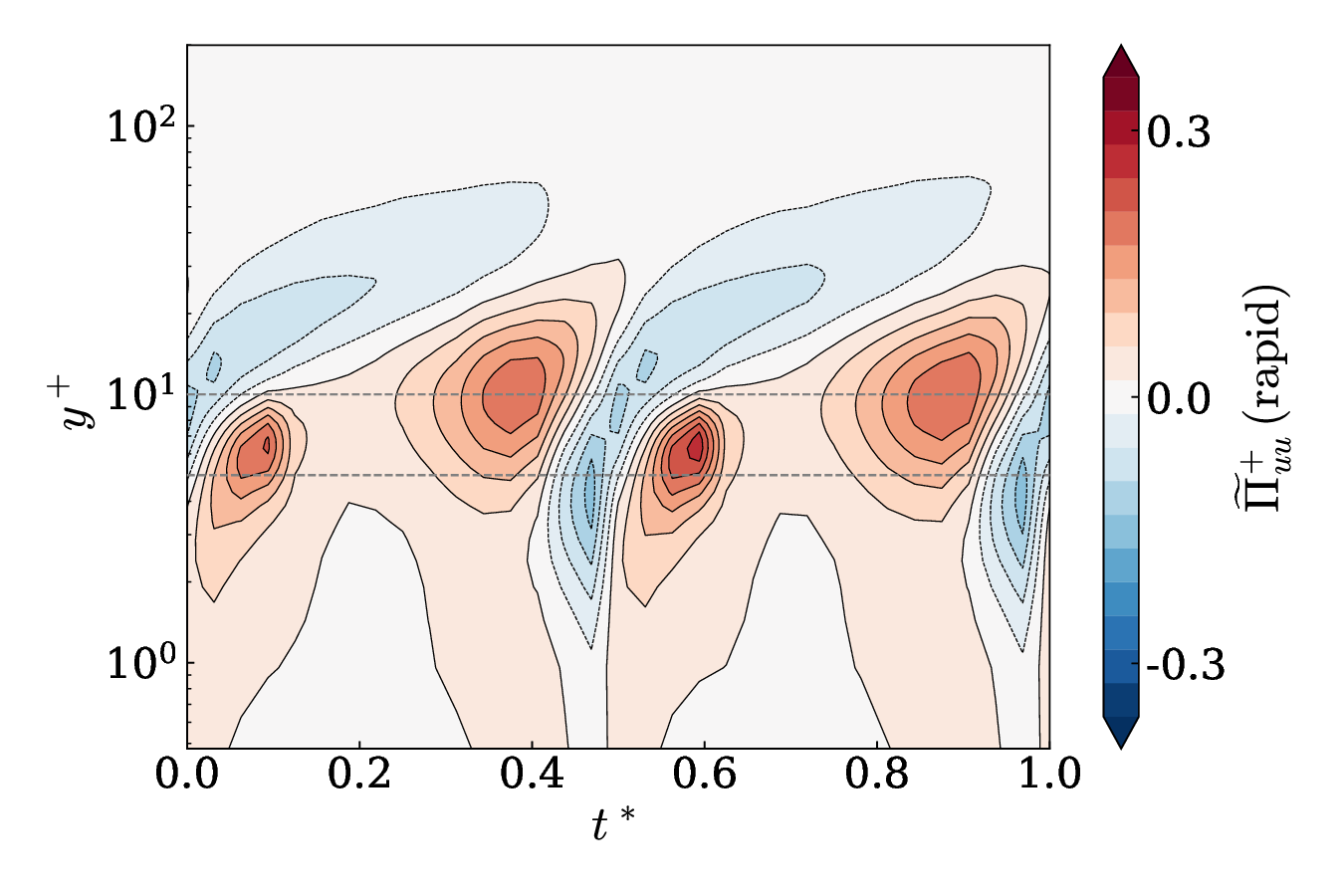}
		\centerline{(\textit{a})}
	\end{minipage}
	\hfill
	\begin{minipage}{0.48\textwidth}
		\centering
		\includegraphics[width=\textwidth]{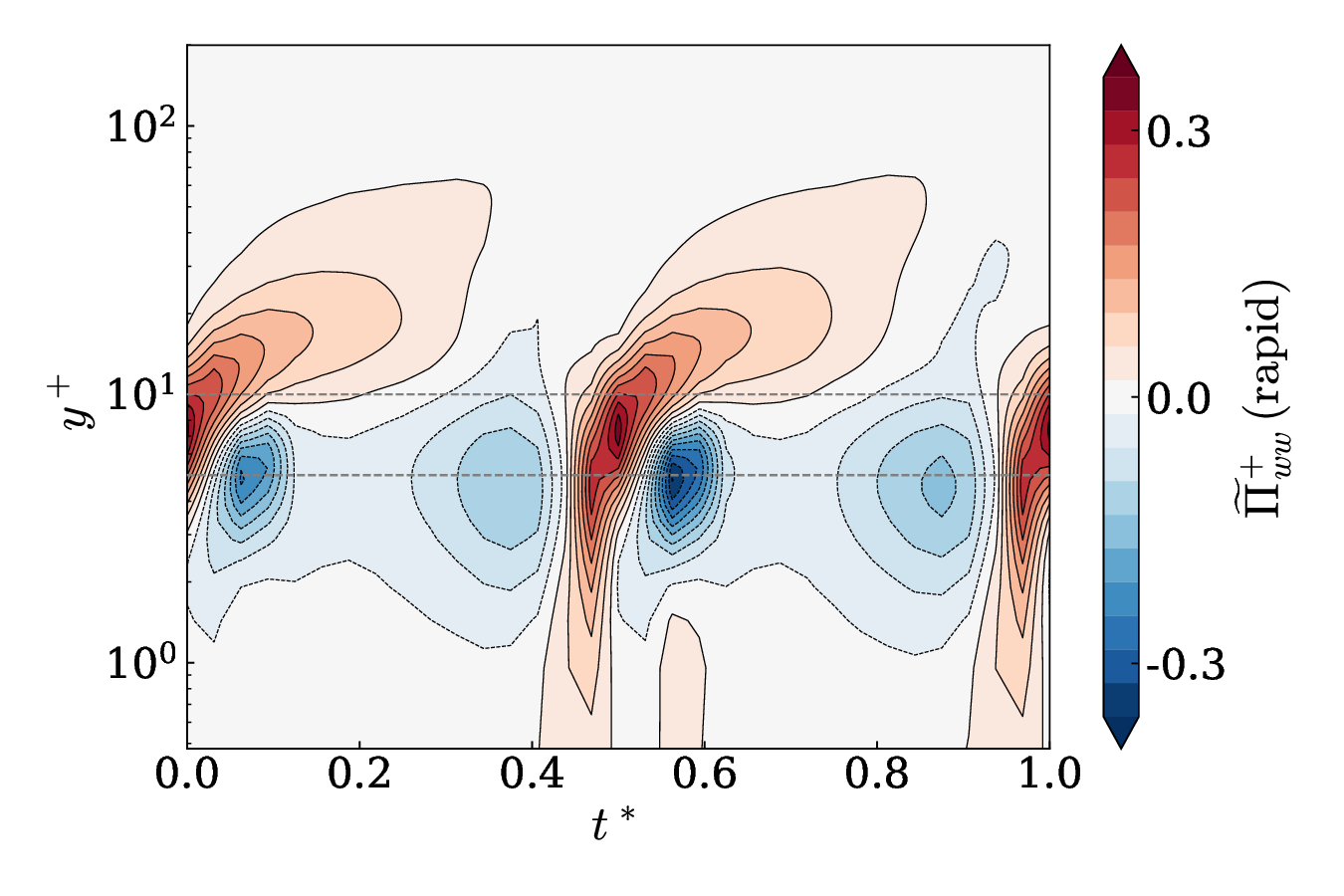}
		\centerline{(\textit{b})}
	\end{minipage}
	\caption{Phase--wall-normal maps of the rapid diagonal pressure-strain ($T^+=350$, $W^+=30$; wall units, computed from the instantaneous DNS fields, appendix~\ref{app:rapid_slow}). (\textit{a}) Rapid streamwise redistribution $\widetilde{\Pi}_{uu}^+$; (\textit{b}) rapid spanwise redistribution $\widetilde{\Pi}_{ww}^+$. The dashed horizontal lines mark $y^+ = 5$ and $10$.}
	\label{fig:rapid_slow_maps}
\end{figure}

The decomposition is here computed directly from the instantaneous DNS fields rather than modelled, the rapid pressure being obtained by solving its Poisson equation~\eqref{eq:poisson_pressure}, $\nabla^2 p^{(r)} = -2\,(\partial U_i/\partial x_j)\,(\partial u_j^{\dprime}/\partial x_i)$, with the imposed mean shear as the sole source and the slow part taken as the exact remainder $p^{\dprime}-p^{(r)}$ (appendix~\ref{app:rapid_slow}); the diagonal terms so obtained reproduce the phase-resolved totals to within a few per cent. The verdict is presented in figure~\ref{fig:rapid_slow}, in which each diagonal component at $y^+ = 10$ is displayed together with its rapid and slow parts and with the gate itself. The picture divides at a glance: in the streamwise and spanwise panels the rapid part shadows the total through the deep excursions of the reversal windows, whilst in the wall-normal panel the rapid part remains nearly flat and the cyclic modulation is carried by the slow part. The streamwise drain is thus overwhelmingly rapid (figure~\ref{fig:rapid_slow}\textit{a}): the rapid part carries some four-fifths of the cyclic variation of $\Pi_{uu}$ and attains its minimum in phase with the strain rate at the reversal, the slow part contributing only a smaller and smoother background. The spanwise charge is rapid likewise (figure~\ref{fig:rapid_slow}\textit{c}), its reversal-core maximum correlating with the gate at approximately $+0.84$; the streamwise drain and the spanwise charge are thereby revealed as the two halves of a single rapid, linear exchange $\widetilde{u^{\dprime}u^{\dprime}}\!\leftrightarrow\!\widetilde{w^{\dprime}w^{\dprime}}$, driven instantaneously by the reversing Stokes shear. The phase--wall-normal maps of the two rapid terms (figure~\ref{fig:rapid_slow_maps}) exhibit this exchange directly, as near-mirror patterns confined to the buffer layer and locked to the strain reversals.

The wall-normal component behaves in the opposite manner (figure~\ref{fig:rapid_slow}\textit{b}): its rapid part is a weak, quasi-steady shear-drain, whilst the positive pre-charge that sustains $\widetilde{v^{\dprime}v^{\dprime}}$ across the reversal resides in the \emph{slow} return to isotropy, decoupled from the strain rate. The lag of $\Pi_{vv}$ inferred above from the trace-free constraint is thereby assigned its physical identity: the wall-normal fluctuations are charged not by the forcing directly but by the slower redistribution of the spanwise energy the forcing has deposited, and so persist into the reversal on the turbulence timescale, prolonging the window over which the flux channel survives.

This settles the question of cause, and with it the reversal spike. The selective collapse of the momentum flux at the reversal is the rapid, linear pressure-strain response of the streamwise component to the reversing Stokes strain, a response that the passive scalar, possessing no pressure term and hence no rapid pressure-strain, cannot share, so that $\widetilde{P}_{\theta\theta}$ proceeds unimpeded whilst $\widetilde{P}_{uu}$ is curtailed. The selectivity of the collapse corroborates this attribution: the wall-normal fluctuation shared by the two fluxes persists through the early reversal upon the slow $\Pi_{vv}$ pre-charge (figure~\ref{fig:rapid_slow}\textit{b}), so that the loss of $-\widetilde{u^{\dprime}v^{\dprime}}$ reflects the depletion of its streamwise constituent, the component drained by $\Pi_{uu}$, rather than a wholesale destruction of the vortical motions, which would depress both fluxes alike. The structural argument advanced by \citet{hasegawa_dissimilar_2011}, that the divergence-free constraint introduces a pressure-mediated coupling from which the passive scalar is structurally exempt, is thereby converted from a structural admissibility into a demonstrated physical process: the tensor-rank asymmetry between the rapid pressure--velocity-gradient and pressure--scalar-gradient correlations (appendix~\ref{app:rapid_slow}) is realised, in the present flow, as a leading-order rapid drain of $\widetilde{u^{\dprime}u^{\dprime}}$ at every strain reversal with no scalar counterpart, the cyclical reversal of the sign of $\Pi_{uu}$ in the buffer layer being its time-resolved signature. There remains the milder plateau dominance. During the plateau $\Pi_{uu}$ is itself a source for $\uupp$, acting, if anything, against a thermal advantage, so the plateau excess cannot be of diagonal-drain origin; its seed was identified in \S\,\ref{sec:results:production} as the small residual mean-gradient asymmetry acting upon the shared wall-normal fluctuation that the diagonal $\Pi_{vv}$ sustains. The diagonal interplay established here, the reversal drain of $\Pi_{uu}$ and the persistent $\Pi_{vv}$ pre-charge, is condensed into panel (\textit{b}) of the timeline figure~\ref{fig:mechanism_schematic}. Before the two are assembled into a single mechanism, however, the most obvious competing explanation must be confronted and eliminated: that the pressure acts not through the diagonal redistribution at all, but directly and preferentially upon the turbulent fluxes themselves, draining the momentum flux more than the scalar. This possibility is examined in the following subsection.

\subsection{Exoneration of the off-diagonal pressure}
\label{sec:results:off_diagonal}

The competing explanation may be stated at once. The wall-normal fluxes possess their own magnitude budgets, in which the off-diagonal pressure-strain correlations $-\Pi_{uv}$ and $-\Pi_{v\theta}$ of equation~\eqref{eq:pi_offdiag} enter as sinks; were the pressure to drain the momentum flux more severely than the scalar flux, it would favour the latter directly, with no need of the diagonal mechanism at all. The flux-magnitude budgets (equations~\eqref{eq:budget_uv} and~\eqref{eq:budget_vth}), closed from the present DNS to within one to two per cent and presented in full in appendix~\ref{app:offdiag}, do not sustain this reading. The off-diagonal pressure-strain does drain the scalar flux somewhat more deeply than the momentum flux in absolute terms ($\Pi_{v\theta}/\Pi_{uv} \approx 1.4$ at $y^+ = 10$); this deeper sink is, however, attended by a correspondingly larger pressure-transport \emph{source} upon the scalar flux, the two nearly cancelling, so that the \emph{net} pressure action upon the two fluxes is all but indistinguishable; such difference as survives the cancellation is, in the cycle mean, of the sign that opposes the scalar flux ($\overline{\Delta} \approx -0.12$ integrated across the lower buffer layer; figure~\ref{fig:offdiag_phase} of appendix~\ref{app:offdiag}). When each correlation is normalised by the flux it drains, moreover, the pressure-strain is found to remove a near-constant fraction, close to $0.6$ in the $y^+ = 5$--$10$ band, of whichever flux is present: it is a return-to-isotropy sink that scales with the flux it acts upon, following the larger scalar flux rather than penalising the momentum flux. The off-diagonal pressure is accordingly exonerated; the deeper scalar sink it presents is a consequence, not a cause, of the larger scalar flux. The one off-diagonal source-term asymmetry that does favour the scalar is the differential production already encountered in \S\,\ref{sec:results:production}, the residual mean-gradient asymmetry $\partial\widetilde{\theta}/\partial y > \partial\widetilde{u}/\partial y$ acting upon the shared wall-normal fluctuation. It is modest, confined to the plateau (the off-diagonal flux productions being nearly equal at the reversals), and accounts for the milder $6\%$ plateau dominance; the reversal spike owes nothing to it. With the obvious alternative thus eliminated, the diagonal mechanism stands alone; the synthesis follows.

\subsection{Synthesis of the mechanism}
\label{sec:results:synthesis}

\begin{figure}
	\centering
	\includegraphics[trim={1.85cm 0cm 0cm 0cm}, clip, width=1.1\textwidth]{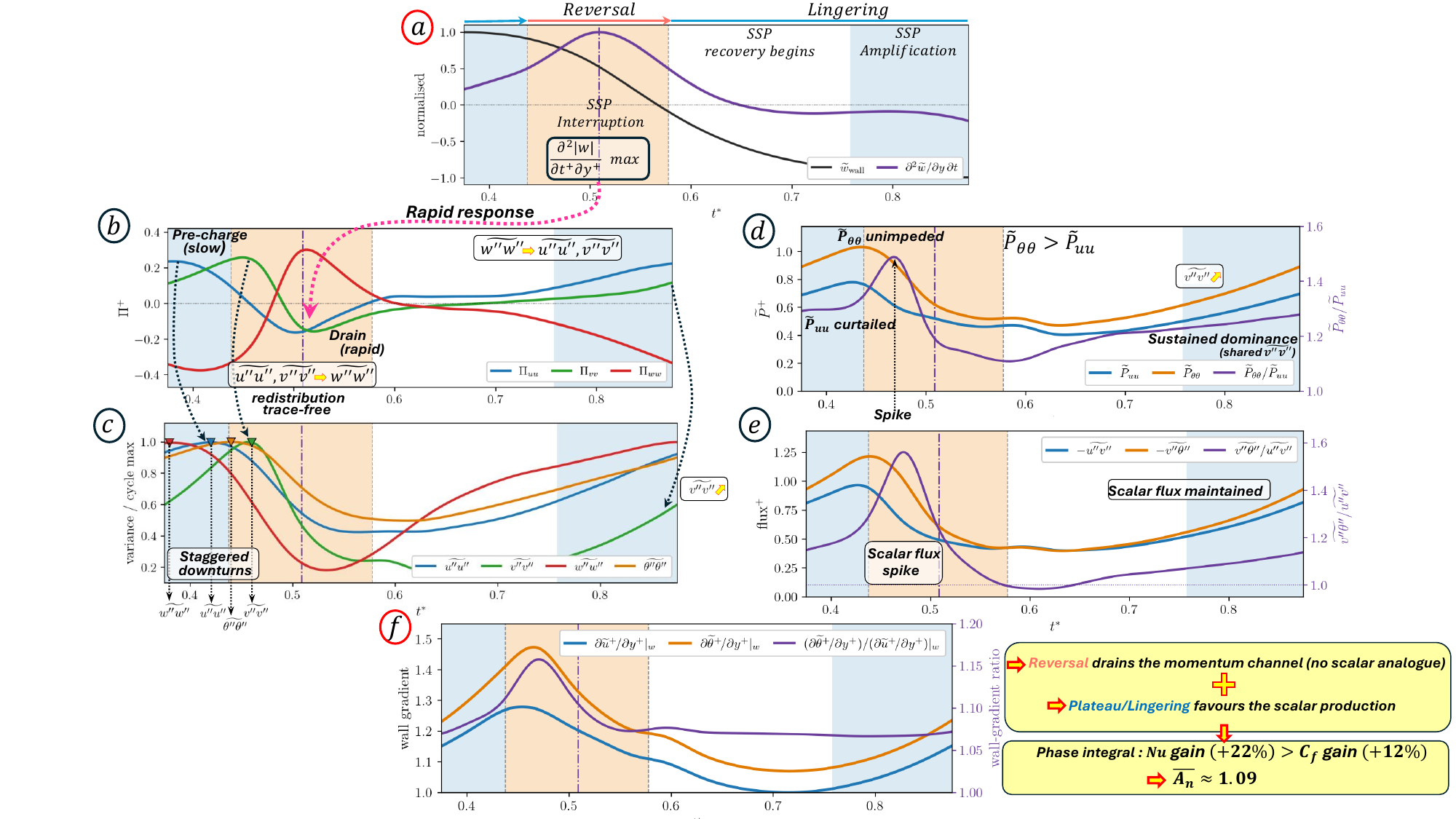}
	\caption{Phase timeline of the mechanism at $y^+=10$ over one half-cycle of the quasi-plateau actuation ($T^+=350$, $W^+=30$), drawn between successive extrema of the wall velocity. In every panel the reversal window is shaded in orange and the regenerative second half of the plateau in blue, its quiescent first half remaining unshaded; the dashed vertical lines mark the reversal-window edges and the dash-dot line the maximum of the Stokes-strain-rate gate. (\textit{a}) The wall velocity $\widetilde{w}_{\mathrm{wall}}$ and the gate $\partial^2\widetilde{w}^+/\partial y^+\,\partial t^+$, each normalised by its cycle maximum. (\textit{b}) The diagonal pressure-strain components $\Pi_{uu}^+$, $\Pi_{vv}^+$ and $\Pi_{ww}^+$. (\textit{c}) The stochastic variances, each normalised by its cycle maximum, the triangles marking the successive downturns. (\textit{d}) The productions $\widetilde{P}_{uu}^+$ and $\widetilde{P}_{\theta\theta}^+$ with their ratio (right axis). (\textit{e}) The wall-normal fluxes $-\widetilde{u^{\dprime}v^{\dprime}}^+$ and $-\widetilde{v^{\dprime}\theta^{\dprime}}^+$ with their ratio (right axis). (\textit{f}) The wall gradients of mean velocity and temperature, in unactuated wall units, with their ratio (right axis). The annotations name the successive stages of the self-sustaining process and the inter-component transfers, and the box at lower right summarises the two expressions of the mechanism together with their cycle-integrated imprint.}
	\label{fig:mechanism_schematic}
\end{figure}

The pieces may now be assembled into a single causal chain. The breaking of the Reynolds analogy is governed by the diagonal, inter-component pressure-strain redistribution, the channel that the divergence-free constraint ($\nabla\cdot\bm{u}=0$) opens in the velocity budget and denies to the passive scalar. Its decisive expression is the rapid drain of $\Pi_{uu}$ at the Stokes-strain reversal, the linear pressure-strain response of the streamwise component to the reversing spanwise shear established in \S\,\ref{sec:results:pressure_strain}, which curtails the streamwise production whilst the scalar production, lacking any such term, continues undiminished: this is the reversal spike. A second and milder expression operates through the plateau, where the same diagonal redistribution, now sustaining through $\Pi_{vv}$ the wall-normal fluctuation shared by both fluxes, permits the residual mean-gradient asymmetry to maintain the scalar production slightly the higher. The off-diagonal pressure, confronted as the obvious alternative, was found to favour neither flux, scrambling a near-constant fraction of each (\S\,\ref{sec:results:off_diagonal}, appendix~\ref{app:offdiag}), and plays no preferential part. The same diagonal channel leaves its trace upon the time-mean budgets of \S\,\ref{sec:results:time_avg} as the sign reversal of $\Pi_{uu}^+$ that opened the investigation, now recognised as the cycle-residue of the alternating reversal drain, dissipated locally and conferring no time-mean advantage of its own. The dissimilarity thus emerges as the time integral of these two diagonal contributions, dominated by the reversal spike, and the evidence is assembled, panel by panel, in figure~\ref{fig:mechanism_schematic}: the forcing and its gate (panel \textit{a}), the diagonal redistribution (\textit{b}), the variances (\textit{c}), the productions (\textit{d}), the fluxes (\textit{e}) and, as the final imprint, the wall gradients (\textit{f}).

The timeline illustrates how the mechanism unfolds across the actuation cycle. During the reversal phases (shaded orange), the Stokes-strain rate $\partial^2\widetilde{w}/\partial y\,\partial t$ is large (panel \textit{a}) and the self-sustaining process is interrupted. At the diagonal level (panel \textit{b}), $\Pi_{ww}$ is positive, energy being injected into the spanwise component by the reversing Stokes layer, whilst $\Pi_{uu}$ transitions to negative values and acts as a sink for the streamwise variance, a drain with no analogue in the temperature variance budget. As $\Pi_{vv}$ remains positive slightly beyond the $\Pi_{uu}$ sign change, pre-charged by the strain/velocity phase lag identified in figure~\ref{fig:pi_1d}\textit{a}, wall-normal fluctuations persist (panel \textit{c}) and sustain the turbulent flux channel (panel \textit{e}). At the production level (panel \textit{d}), $\widetilde{P}_{uu}$ and $\widetilde{P}_{\theta\theta}$ both decrease during the reversal; $\widetilde{P}_{uu}$ decays earlier owing to the $\Pi_{uu}$ drain, whilst the undrained $\widetilde{P}_{\theta\theta}$ continues; the production ratio $\widetilde{P}_{\theta\theta}/\widetilde{P}_{uu}$ rises to $\approx 1.49$, which, referred to the phase minima of the two productions, is the $\approx 30\%$ relative spike of \S\,\ref{sec:results:production}. During the lingering phases, the Stokes strain decays through diffusion (panel \textit{a}) and the self-sustaining process recommences progressively (panel \textit{c}). $\Pi_{ww}$ is now negative, redistributing spanwise energy to $\Pi_{uu} > 0$ and $\Pi_{vv} > 0$ (panel \textit{b}), which act as sources for $\uupp$ and $\widetilde{v^{\dprime}v^{\dprime}}$. In the second half of the plateau (shaded blue), the diagonal $\Pi_{vv}$ sustains the shared wall-normal fluctuation, upon which the residual mean-gradient asymmetry holds the scalar production slightly the higher (panel \textit{d}); the off-diagonal pressure that drains the two fluxes follows them proportionally and favours neither (appendix~\ref{app:offdiag}), the tensor-rank asymmetry between the pressure--scalar-gradient and pressure--velocity-gradient correlations (appendix~\ref{app:rapid_slow}) permitting, though not dictating, the residual differential.

It is instructive at this point to articulate the dual role fulfilled by the turbulent fluxes $-\widetilde{u^{\dprime}v^{\dprime}}$ and $-\widetilde{v^{\dprime}\theta^{\dprime}}$, the natural bridge between the buffer-layer dynamics characterised in the preceding subsections and the macroscopic indicator $\overline{A}_n$. Within the local budgets, the fluxes enter the production terms $\widetilde{P}_{uu} = -2\widetilde{u^{\dprime}v^{\dprime}}\,\partial\widetilde{u}/\partial y$ and $\widetilde{P}_{\theta\theta} = -2\widetilde{v^{\dprime}\theta^{\dprime}}\,\partial\widetilde{\theta}/\partial y$ of equations~(\ref{eq:budget_uu}) and~(\ref{eq:budget_thth}), and it is in this capacity that their differential phase-resolved behaviour, panel (\textit{e}) feeding panel (\textit{d}), constitutes the immediate cause of the production asymmetry documented in \S\,\ref{sec:results:production}. At the level of the integrated transport, the time-averaged magnitudes of the same fluxes determine directly the modifications of momentum and thermal transfer that underlie $\overline{A}_n$. No separate transport mechanism therefore need be posited to connect the turbulent micro-dynamics to the macroscopic dissimilarity indicator: the fluxes themselves serve as both the sites at which the pressure-mediated asymmetry acts and the carriers through which it reaches the wall.

Building upon this dual-level view, the manner in which the two channels operate together over each actuation cycle may now be summarised. The dissimilarity is not carried across the suppression trough by a stored thermal variance excess, the rapid scalar dissipation precluding any inter-phase storage of variance (appendix~\ref{app:offdiag}); the two channels are instead regenerated afresh within each cycle, the diagonal $\Pi_{uu}$ drain producing the reversal spike and the differential production, acting on the wall-normal fluctuation amplified by the diagonal $\Pi_{vv}$, producing the sustained plateau dominance, and the time-mean dissimilarity arises as the cycle integral of these two repeatedly reactivated contributions. The structural argument of \citet{hasegawa_dissimilar_2011}, originally advanced on the basis of the Fr\'{e}chet differential of the velocity and scalar response to wall transpiration, is thereby placed on a phase-resolved budget-level foundation: the diagonal pressure-strain components $\Pi_{uu}$ and $\Pi_{vv}$ are identified as the specific terms through which the structural asymmetry of the transport equations is dynamically expressed under SWO actuation, the off-diagonal pressure being found to favour neither flux. The direction and magnitude of the demonstrated production asymmetry, namely a sustained $6\%$ dominance during the plateau compounded by a $30\%$ spike at each reversal, are consistent with the global analogy factor $\overline{A}_n \approx 1.09$ reported by \citet{guerin_PBO_2026}. The imprint upon the integrated transport is registered directly at the wall (panel \textit{f}): the wall gradients of mean velocity and temperature, which measure the instantaneous friction and heat-transfer enhancements over the unactuated flow, both surge within the reversal window, the thermal gradient the harder, their ratio peaking at $\approx 1.16$ near the gate maximum and settling to $\approx 1.07$ through the plateau; the cycle mean of this ratio, $\approx 1.08$, approximates the global analogy factor, the dissimilarity being earned at each reversal. The chain of evidence, from the reversing strain of panel (\textit{a}) to the wall gradients of panel (\textit{f}), is thereby closed; although a formal cycle-integrated closure connecting the phase-resolved budget asymmetry to the absolute $\overline{A}_n$ is beyond the scope of the present analysis, the identification of the causal chain provides the physical foundation upon which such quantitative models may be constructed.

This mechanism also provides a resolution of the protrusion paradox identified by \citet{guerin_PBO_2026}, as manifested by the two configurations examined in the companion study, namely the quasi-plateau waveform at $T^+=350$, $W^+=30$ and the sinusoidal optimum at $T^+=325$, $W^+=40$. The governing parameter for DHT appears not to be the spatial penetration depth $\ell_{0.01}^+$ of the Stokes layer; a candidate governing parameter is rather the lingering fraction, defined here as the fraction of the actuation cycle over which the Stokes-strain rate satisfies $|\partial^2\widetilde{w}^+/\partial y^+\,\partial t^+| < 0.01$ in the buffer-layer region $y^+ \gtrsim 10$, during which the self-sustaining process can recommence and the pressure-strain asymmetry can be cyclically activated, the lingering fraction thereby coinciding with the duty-cycle ratio of the self-sustaining process under spanwise oscillation \citep{agostini_duty_cycle_2026}. The lingering fraction is not quantified in the present single-budget dataset; the quasi-plateau waveform is, however, observed to sustain extended quasi-steady phases at reduced amplitude, thereby attaining comparable dissimilarity to the sinusoidal optimum at $W^+=40$ through temporal organisation rather than spatial penetration. The two configurations of the companion study converge to moderate periods ($T^+ \in [325, 350]$) rather than the extended periods that would maximise protrusion height, suggesting that the temporal structure of the imposed spanwise shear, rather than its spatial extent, may constitute the primary control parameter for DHT enhancement; establishing the lingering fraction as a predictor capable of ranking the protrusion-paradox cases would, however, require a broader parametric study across waveform topologies, which is deferred to future work.

%% file: conclusion.tex
\section{Conclusion}
\label{sec:conclusion}

The present investigation has elucidated the physical mechanism responsible for dissimilar heat transfer under optimised spanwise wall oscillation waveforms through phase-resolved analysis of the stochastic variance transport budgets. The central result is the identification of the diagonal, inter-component pressure-strain redistribution, present in the velocity budgets and absent from the scalar equation, as the mechanism that generates the preferential thermal enhancement observed by \citet{guerin_PBO_2026}, characterised at the quasi-plateau optimum ($T^+ = 350$, $W^+ = 30$) by an analogy factor $\overline{A}_n = 1.087$ in which the Nusselt-number gain ($+22.3\%$) exceeds the skin-friction gain ($+12.4\%$).

The redistribution acts through two temporally sequenced expressions. At each Stokes-strain reversal, $\Pi_{uu}$ drains streamwise fluctuation energy through the inter-component transfer enforced by the divergence-free constraint; no analogous term exists in the temperature variance budget, so the velocity production begins decaying earlier in the phase than its thermal counterpart, generating a production ratio spike of approximately $30\%$ at the reversal. During the extended plateau, the diagonal $\Pi_{vv}$ sustains the wall-normal fluctuation common to both fluxes, upon which a small mean-gradient asymmetry maintains the scalar production some $6\%$ the higher. The off-diagonal pressure is exonerated: the larger pressure-transport source received by the scalar flux and the deeper pressure-strain sink that drains it nearly cancel, the pressure-strain scrambling a near-constant fraction of each flux as a return-to-isotropy sink that follows, rather than penalises, the larger scalar flux (appendix~\ref{app:offdiag}). In the time-mean budgets of \S\,\ref{sec:results:time_avg}, the reversal drain survives only as the sign reversal of $\Pi_{uu}^+$, from a modest unactuated drain to a substantial near-wall source that is locally dissipated and confers no net advantage in the mean. The two expressions are regenerated afresh within each cycle as the self-sustaining process recommences during the plateau, the rapid scalar dissipation precluding any inter-phase storage of variance (appendix~\ref{app:offdiag}).

The analysis provides in addition a resolution of the protrusion paradox identified by \citet{guerin_PBO_2026}: the DHT performance is proposed to be governed not by the spatial penetration depth $\ell_{0.01}^+$ of the Stokes layer but by the lingering fraction of the actuation cycle, the duty-cycle ratio of the self-sustaining process under spanwise oscillation \citep{agostini_duty_cycle_2026}, as defined in \S\,\ref{sec:results:synthesis}. The quasi-plateau waveform attains a dissimilarity comparable to that of the sinusoidal optimum at $W^+=40$ through temporal organisation rather than spatial penetration, establishing waveform topology as a genuine design variable for DHT enhancement, distinct from a strengthening of the actuation amplitude; a parametric study across a wider range of waveform topologies would be required to establish the lingering fraction as a universal predictor.

The mechanism is structural in nature, as it originates from the presence of the pressure in the Navier--Stokes equations and its absence from the passive scalar transport equation, and is therefore expected to persist at higher Reynolds numbers provided the waveform maintains sufficient lingering phases for the asymmetric regeneration cycle to operate. At Prandtl numbers different from unity, the molecular diffusivity asymmetry would superimpose upon the pressure-mediated asymmetry; the structural feature itself is independent of $\Pr$, though the quantitative scaling of $\overline{A}_n$ remains to be established. Nor is the framework specific to temporal oscillation: any spanwise forcing that alternates Stokes-layer reversals with extended recovery phases, including spatially travelling waves, may plausibly activate the same channel. The mechanism is, moreover, demonstrated under a purely predetermined, open-loop actuation requiring no state feedback: the structural route to dissimilarity established by \citet{hasegawa_dissimilar_2011} through suboptimal control laws requiring full-field information is thereby shown to be accessible through the class of wall-information-only, predetermined actuation identified by \citet{kasagi_control_2012} as the configuration of greatest practical relevance, of which the quasi-plateau waveform constitutes a concrete realisation.

%% file: appendix.tex
\appendix

\section{Tensor-rank asymmetry of the rapid pressure-strain and pressure-scalar-gradient correlations}
\label{app:rapid_slow}

It is instructive to examine briefly the mathematical origin of the asymmetry between the pressure-strain tensor components $\Pi_{ij}$ and the pressure-temperature-gradient correlation $\Pi_{v\theta}$ that motivates the budget analysis of \S\,\ref{sec:results}. The fluctuating pressure satisfies the Poisson equation
\begin{equation}
	\nabla^2 p^{\dprime} = -2\,\frac{\partial U_i}{\partial x_j}\,\frac{\partial u_j^{\dprime}}{\partial x_i} + \text{(slow terms)},
	\label{eq:poisson_pressure}
\end{equation}
the source being driven by the mean velocity gradient \citep{rotta_statistische_1951,lumley_return_1977}. The Green's function solution thereof, expressed in Fourier space, introduces a kernel of the form $\kappa_i\kappa_j/|\bm{\kappa}|^2$, the trace-free residual of which is the transverse projector $\mathcal{P}_{ij}(\bm{\kappa}) = \delta_{ij} - \kappa_i\kappa_j/|\bm{\kappa}|^2$. Upon contraction with $\partial u_i^{\dprime}/\partial x_j$, which is itself subject to the solenoidal constraint $\partial u_i^{\dprime}/\partial x_i = 0$, the rapid pressure-strain correlation acquires the structure
\begin{equation}
	\Pi_{ij}^{(r)} = M_{ijkl}\,\frac{\partial U_k}{\partial x_l},
	\label{eq:rapid_pi_velocity}
\end{equation}
in which $M_{ijkl}$ is a fourth-order spectral integral involving the projector and the turbulence-kinetic-energy spectrum. The coupling between the rapid pressure-strain and the mean shear is therefore fourth-order in tensor rank, a structure formalised, in the modelling literature, in the SSG family of closures \citep{speziale_modelling_1991}.

The case of the rapid pressure-temperature-gradient correlation is, on this point, qualitatively different. The scalar gradient $\partial\theta^{\dprime}/\partial x_j$ enters in place of the second strain component; as the passive scalar is unconstrained by any divergence relation, no solenoidal projector acts on this gradient. The corresponding rapid contribution acquires the form
\begin{equation}
	\Pi_{j\theta}^{(r)} = G_{jk}\,\frac{\partial \Theta}{\partial x_k},
	\label{eq:rapid_pi_scalar}
\end{equation}
in which the kernel $G_{jk}$ is a second-order spectral integral. The asymmetry between the two correlations is therefore one of tensor rank: the rapid pressure-strain couples to the mean shear through a fourth-order kernel, whereas the rapid pressure-scalar-gradient correlation couples to the mean scalar gradient through a second-order kernel \citep{pope_turbulent_2000}. The two correlations are accordingly free to evolve in dissimilar fashion under a common forcing, and this freedom underpins the dissimilar response of the velocity and scalar variance budgets documented in \S\,\ref{sec:results}.

The rapid/slow decomposition employed in \S\,\ref{sec:results:pressure_strain} is computed directly from the instantaneous DNS fields rather than modelled. For the present channel flow the mean field reduces to the streamwise profile $U(y)$ and the spanwise Stokes layer $W(y,t)$, so that the rapid-pressure source of~\eqref{eq:poisson_pressure} takes the explicit form $\nabla^2 p^{(r)} = -2\left[(\partial U/\partial y)\,\partial v^{\dprime}/\partial x + (\partial W/\partial y)\,\partial v^{\dprime}/\partial z\right]$. This Poisson problem is solved for each realisation by Fourier transformation in the homogeneous $x$ and $z$ directions and a tridiagonal inversion on the stretched wall-normal grid, subject to homogeneous Neumann conditions at the walls, the slow pressure being recovered as the exact remainder $p^{(s)} = p^{\dprime} - p^{(r)}$. The diagonal rapid and slow pressure-strain components then follow as $\Pi_{ii}^{(r)} = 2\,\overline{p^{(r)}\,\partial u_i^{\dprime}/\partial x_i}$ and $\Pi_{ii}^{(s)} = \Pi_{ii} - \Pi_{ii}^{(r)}$ (no summation over $i$). The procedure was verified against a manufactured solution of the Poisson equation, for which the relative error is below $0.5\%$, and against the phase-resolved totals reconstructed from the statistical database, which it reproduces to within a few per cent at $y^+ = 10$; the rapid and slow parts are, as required, each separately trace-free.

\section{Off-diagonal flux-magnitude budgets}
\label{app:offdiag}

This appendix details the flux-magnitude budgets on which the exoneration of the off-diagonal pressure in \S\,\ref{sec:results:off_diagonal} rests.

The two off-diagonal fluxes $\widetilde{u^{\dprime}v^{\dprime}}^+$ and $\widetilde{v^{\dprime}\theta^{\dprime}}^+$, which carry the pressure-strain correlations~\eqref{eq:pi_offdiag} and which set the production terms of the variance budgets, obey the phase-resolved transport equations
\begin{equation}
	\begin{split}
	\frac{\partial \widetilde{u^{\dprime}v^{\dprime}}^+}{\partial t^+} ={}&
	\underbrace{-\,\widetilde{v^{\dprime}v^{\dprime}}^+ \frac{\partial \widetilde{u}^+}{\partial y^+}}_{P_{uv}^+}
	+ \underbrace{\widetilde{p^{\dprime +}\!\left( \frac{\partial u^{\dprime +}}{\partial y^+} + \frac{\partial v^{\dprime +}}{\partial x^+} \right)}}_{\Pi_{uv}^+}
	- \underbrace{\frac{\partial \widetilde{p^{\dprime}u^{\dprime}}^+}{\partial y^+}}_{T_{p,uv}^+} \\
	&- \underbrace{\frac{\partial \widetilde{u^{\dprime}v^{\dprime}v^{\dprime}}^+}{\partial y^+}}_{T_{t,uv}^+}
	+ \underbrace{\frac{\partial^2 \widetilde{u^{\dprime}v^{\dprime}}^+}{\partial {y^+}^{2}}}_{D_{\nu,uv}^+}
	- \underbrace{2\,\widetilde{\frac{\partial u^{\dprime +}}{\partial x_k^+} \frac{\partial v^{\dprime +}}{\partial x_k^+}}}_{\varepsilon_{uv}^+}
	\end{split}
	\label{eq:budget_uv}
\end{equation}
\begin{equation}
	\begin{split}
	\frac{\partial \widetilde{v^{\dprime}\theta^{\dprime}}^+}{\partial t^+} ={}&
	\underbrace{-\,\widetilde{v^{\dprime}v^{\dprime}}^+ \frac{\partial \widetilde{\theta}^+}{\partial y^+}}_{P_{v\theta}^+}
	+ \underbrace{\widetilde{p^{\dprime +} \frac{\partial \theta^{\dprime +}}{\partial y^+}}}_{\Pi_{v\theta}^+}
	- \underbrace{\frac{\partial \widetilde{p^{\dprime}\theta^{\dprime}}^+}{\partial y^+}}_{T_{p,v\theta}^+} \\
	&- \underbrace{\frac{\partial \widetilde{v^{\dprime}v^{\dprime}\theta^{\dprime}}^+}{\partial y^+}}_{T_{t,v\theta}^+}
	+ \underbrace{\tfrac{1}{2}\!\left(1+\Pr^{-1}\right) \frac{\partial^2 \widetilde{v^{\dprime}\theta^{\dprime}}^+}{\partial {y^+}^{2}}}_{D_{\nu,v\theta}^+}
	- \underbrace{\left(1+\Pr^{-1}\right)\widetilde{\frac{\partial v^{\dprime +}}{\partial x_k^+} \frac{\partial \theta^{\dprime +}}{\partial x_k^+}}}_{\varepsilon_{v\theta}^+}
	\end{split}
	\label{eq:budget_vth}
\end{equation}
where $P$ is the production, $\Pi_{uv}$ and $\Pi_{v\theta}$ the off-diagonal pressure-strain correlations~\eqref{eq:pi_offdiag}, $T_p$ a pressure-transport term, $T_t$ the turbulent transport, $D_\nu$ the molecular diffusion and $\varepsilon$ the pseudo-dissipation. In contrast with the diagonal variance budgets~\eqref{eq:budget_uu}--\eqref{eq:budget_thth}, a pressure-transport term $T_p$ appears explicitly here, the divergences $\partial\widetilde{p^{\dprime}u^{\dprime}}^+/\partial y^+$ and $\partial\widetilde{p^{\dprime}\theta^{\dprime}}^+/\partial y^+$ not vanishing under the streamwise and spanwise homogeneity; the molecular diffusion of the heat flux carries the coefficient $\tfrac{1}{2}(1+\Pr^{-1})$, equal to unity at $\Pr=1$. The corresponding time-mean balances follow by setting the left-hand-side time-derivative to zero and replacing the phase average $\widetilde{(\cdot)}$ by the time-mean $\overline{(\cdot)}$. Both fluxes being negative, the budgets for the flux magnitudes $-\widetilde{u^{\dprime}v^{\dprime}}^+$ and $-\widetilde{v^{\dprime}\theta^{\dprime}}^+$ mapped throughout the actuation cycle below follow by reversing the sign of every term, whereupon the production $-P$ acts as a source and the pressure-strain $-\Pi_{uv},\,-\Pi_{v\theta}$ as sinks.

\begin{figure}
	\centering
	\begin{minipage}{0.48\textwidth}
			\centering
			\includegraphics[width=\textwidth]{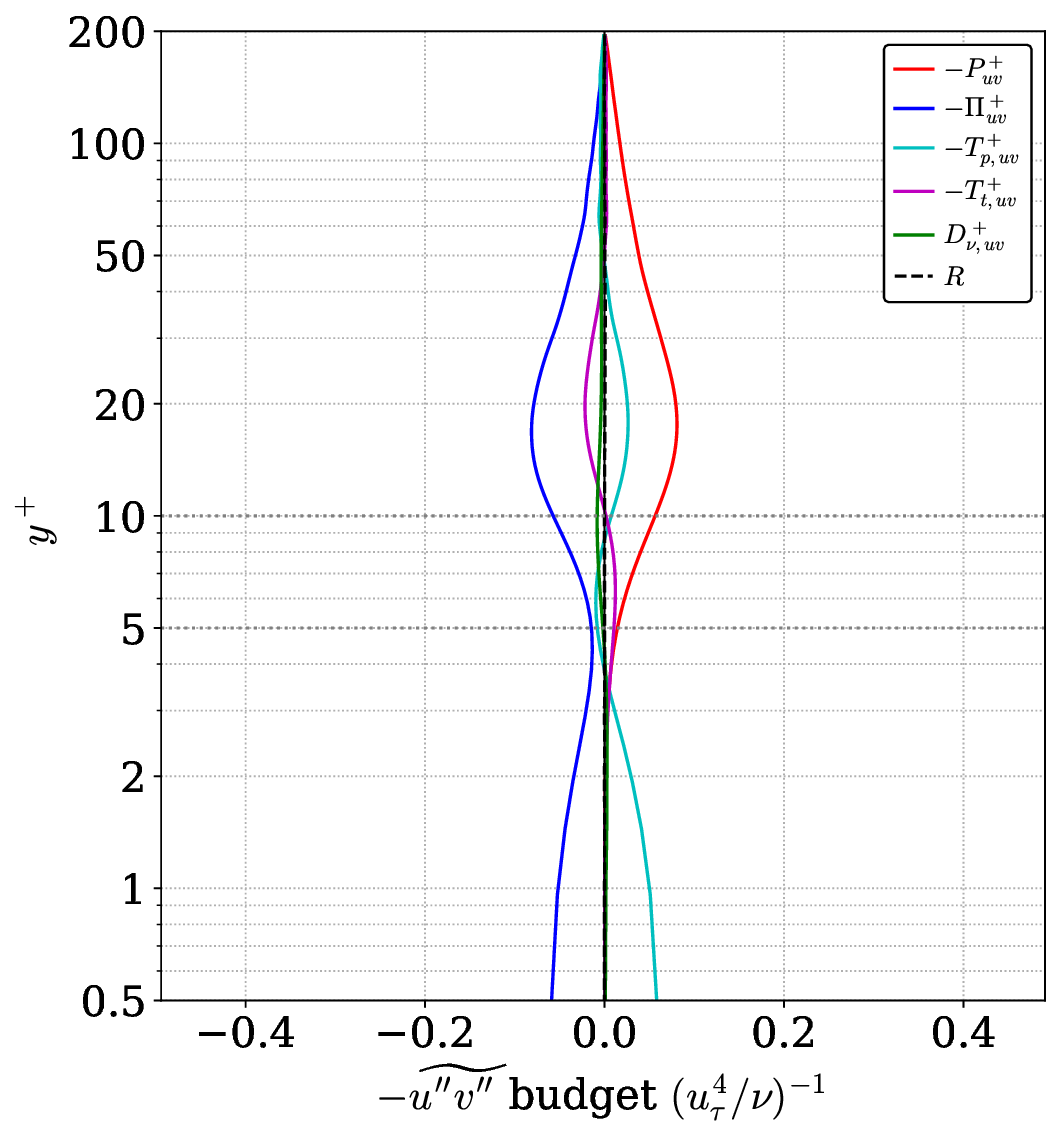}
			\vspace{0.2cm}
			\centerline{(\textit{a})}
		\end{minipage}
		\hfill
		\begin{minipage}{0.48\textwidth}
			\centering
			\includegraphics[width=\textwidth]{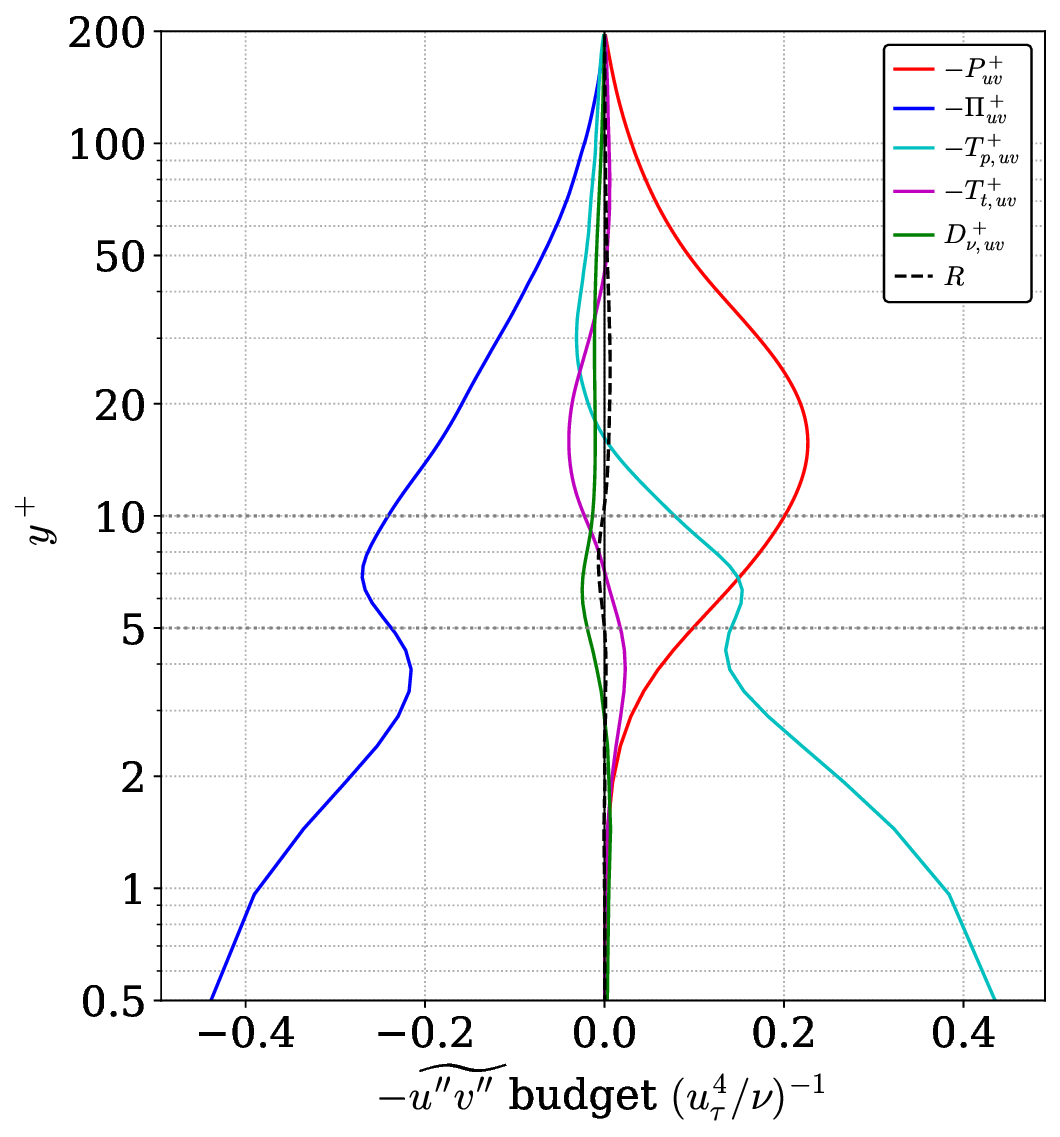}
			\vspace{0.2cm}
			\centerline{(\textit{b})}
		\end{minipage}
		\\[0.4cm]
		\begin{minipage}{0.48\textwidth}
			\centering
			\includegraphics[width=\textwidth]{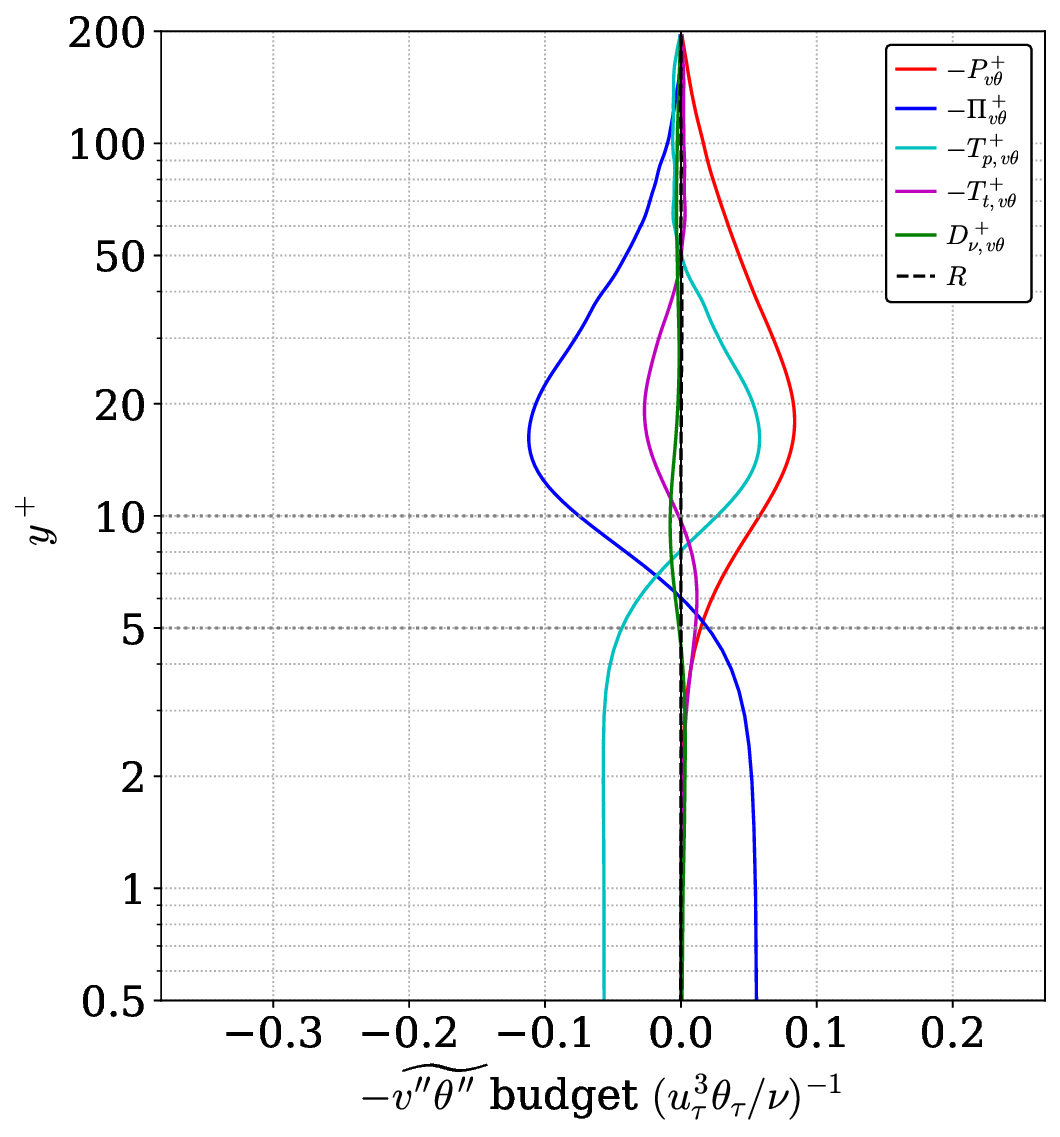}
			\vspace{0.2cm}
			\centerline{(\textit{c})}
		\end{minipage}
		\hfill
		\begin{minipage}{0.48\textwidth}
			\centering
			\includegraphics[width=\textwidth]{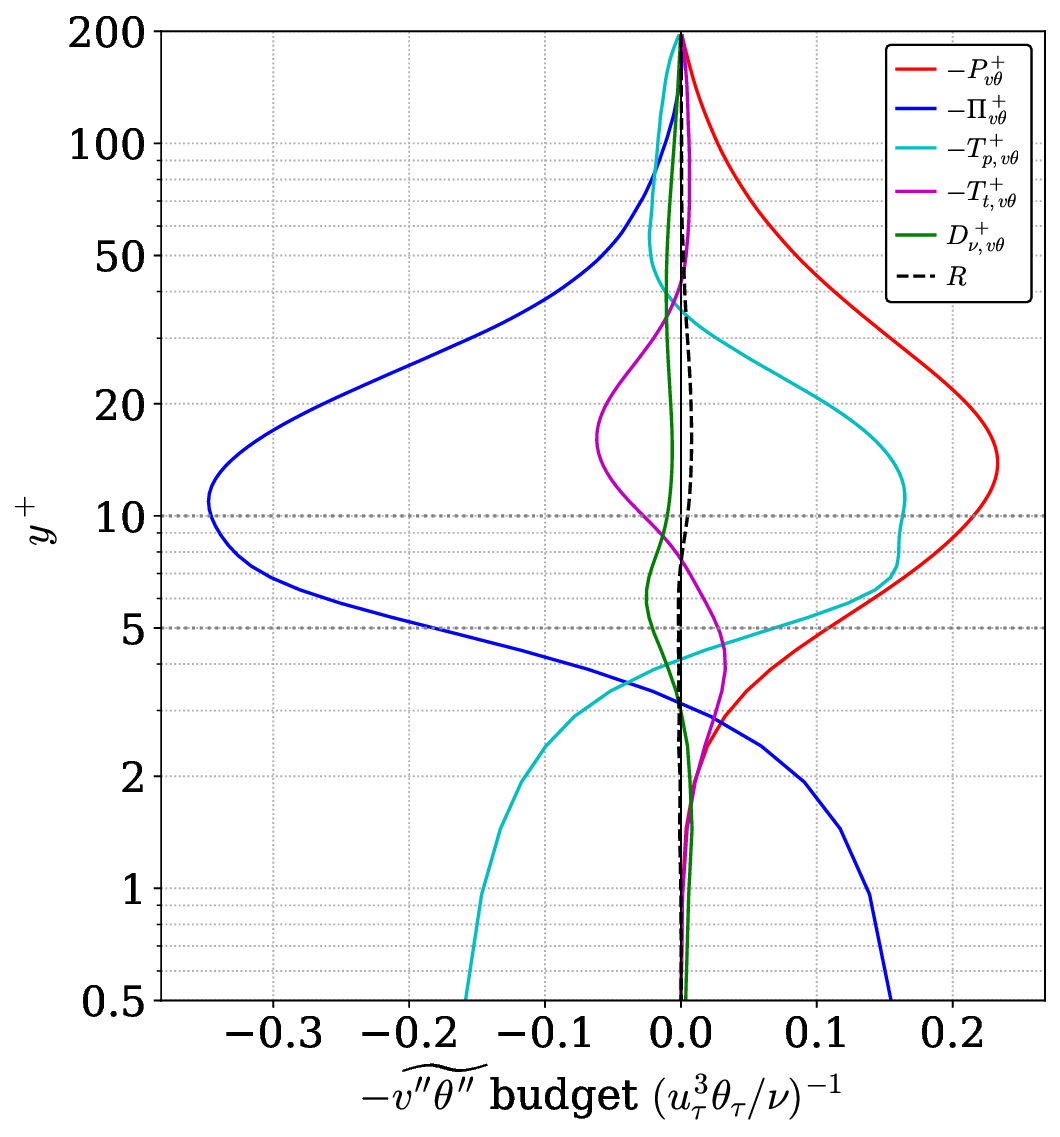}
			\vspace{0.2cm}
			\centerline{(\textit{d})}
	\end{minipage}
	\caption{Time-mean budgets of the turbulent flux magnitudes for the unactuated baseline (left column) and the actuated case ($T^+=350$, $W^+=30$; right column), computed from the rebuilt database with solver-consistent compact derivatives. (\textit{a}, \textit{b}) Budget of $-\overline{u^{\dprime}v^{\dprime}}^+$, normalised by $\nu/u_\tau^4$; (\textit{c}, \textit{d}) budget of $-\overline{v^{\dprime}\theta^{\dprime}}^+$, normalised by $\nu/(u_\tau^3\theta_\tau)$. Terms shown are production $-P^+$, pressure-strain $-\Pi^+$, pressure transport $-T_p^+$, turbulent transport $-T_t^+$, molecular diffusion $D_\nu^+$, and the residual $R$; panels share $x$-limits row-wise.}
	\label{fig:flux_budget}
\end{figure}

The production terms $\widetilde{P}_{uu} = -2\widetilde{u^{\dprime}v^{\dprime}}\,\partial\widetilde{u}/\partial y$ and $\widetilde{P}_{\theta\theta} = -2\widetilde{\theta^{\dprime}v^{\dprime}}\,\partial\widetilde{\theta}/\partial y$ depend on both the mean gradients and the turbulent fluxes $\widetilde{u^{\dprime}v^{\dprime}}$ and $\widetilde{\theta^{\dprime}v^{\dprime}}$; the budgets of these fluxes, in turn, contain the off-diagonal pressure-strain terms $\Pi_{uv}$ and $\Pi_{v\theta}$ as source or sink contributions. The complete causal chain is therefore $-\Pi_{uv} \to -\widetilde{u^{\dprime}v^{\dprime}} \to \widetilde{P}_{uu} \to \uupp$, with the scalar analogue $-\Pi_{v\theta} \to -\widetilde{\theta^{\dprime}v^{\dprime}} \to \widetilde{P}_{\theta\theta} \to \widetilde{\theta^{\dprime}\theta^{\dprime}}$: any differential action of the pressure field on the two fluxes propagates, through production, to the variances, and ultimately to the observed dissimilarity in the integrated quantities $C_f$ and $\Nu$. The budgets for the flux magnitudes $-\widetilde{u^{\dprime}v^{\dprime}}$ and $-\widetilde{\theta^{\dprime}v^{\dprime}}$ are presented in figure~\ref{fig:flux_budget}. These budgets close to within $1$--$2\%$ of the summed term magnitudes in wall units and identify the production and the pressure transport as the source terms that sustain each flux magnitude, the off-diagonal pressure-strain correlations $-\Pi_{uv}$ and $-\Pi_{v\theta}$ (equation~(\ref{eq:pi_offdiag})) entering as the leading sinks, negative throughout the buffer layer; actuation amplifies every term by a factor of three to four without altering this structure. Although the pressure-transport source is approximately twice as large for the scalar flux ($\widetilde{T}_{p,v\theta}^+\approx0.17$ against $\widetilde{T}_{p,uv}^+\approx0.08$ at $y^+=10$), it is set against a correspondingly deeper scalar pressure-strain sink ($-\Pi_{v\theta}^+\approx-0.36$ against $-\Pi_{uv}^+\approx-0.25$), the two productions meanwhile remaining comparable. Whether these two larger scalar pressure sub-terms combine to a net pressure preference for either flux is examined phase by phase in figure~\ref{fig:offdiag_phase} below.

\begin{figure}
	\centering
	\begin{minipage}{0.48\textwidth}
		\centering
		\includegraphics[width=\textwidth]{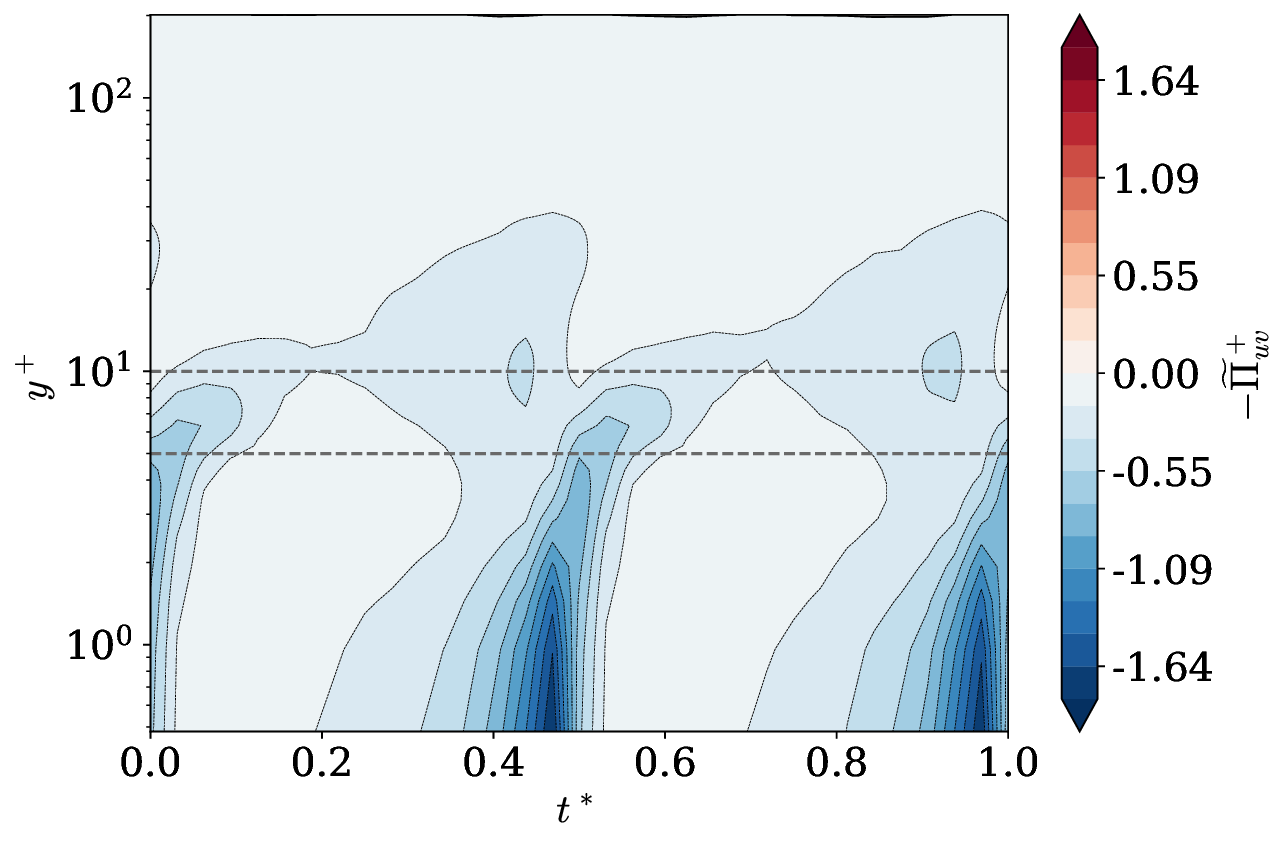}
		\vspace{0.2cm}
		\centerline{(\textit{a})}
	\end{minipage}
	\hfill
	\begin{minipage}{0.48\textwidth}
		\centering
		\includegraphics[width=\textwidth]{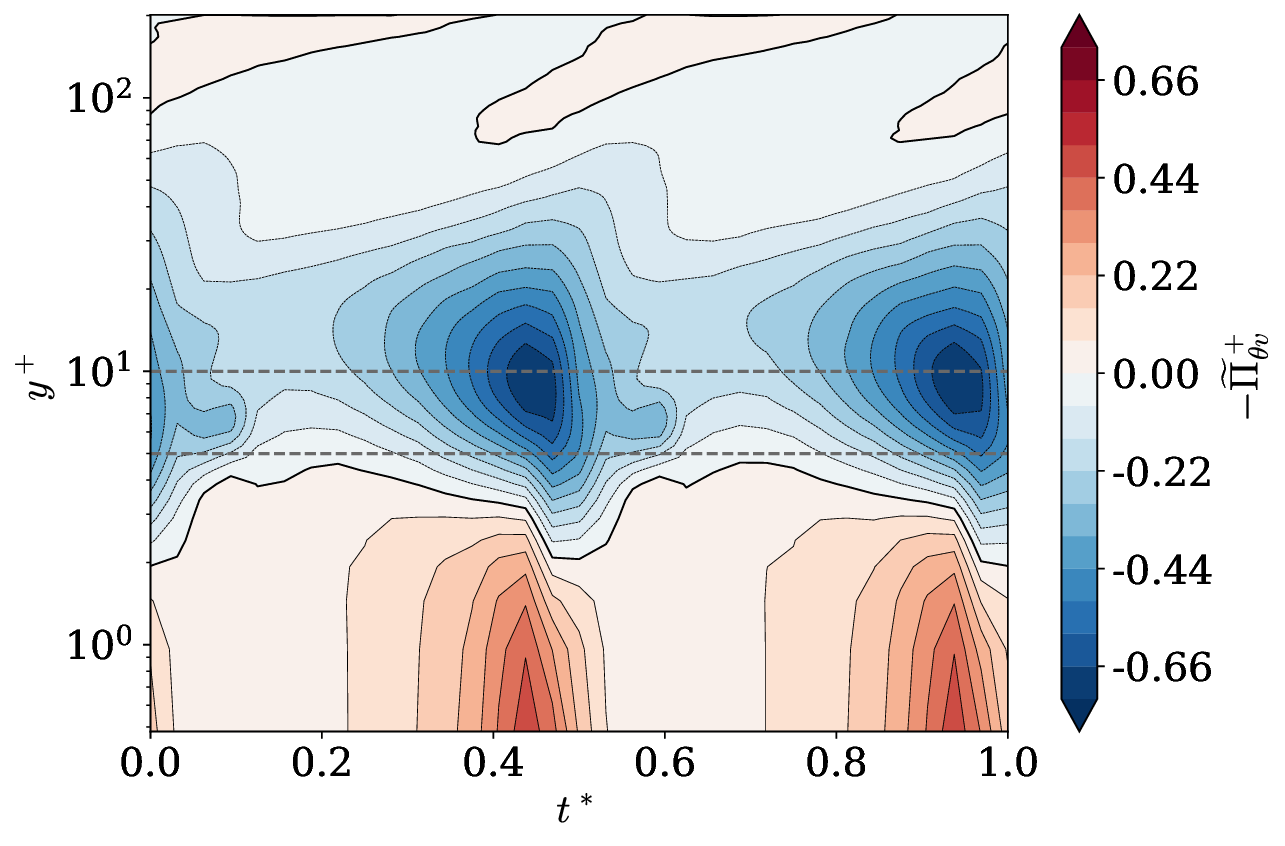}
		\vspace{0.2cm}
		\centerline{(\textit{b})}
	\end{minipage}
	\begin{minipage}{0.48\textwidth}
		\centering
		\includegraphics[width=\textwidth]{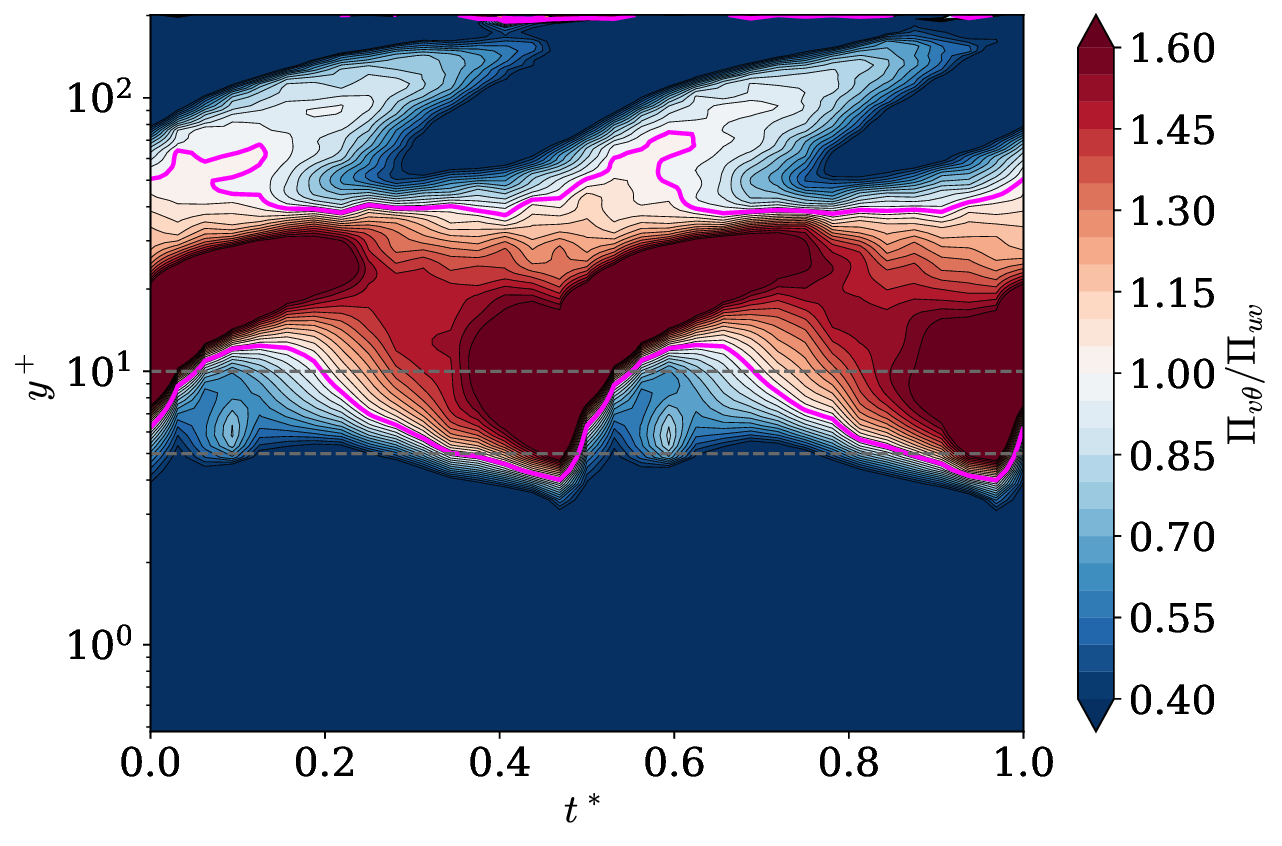}
		\vspace{0.2cm}
		\centerline{(\textit{c})}
	\end{minipage}
	\hfill
	\begin{minipage}{0.48\textwidth}
		\centering
		\includegraphics[width=\textwidth]{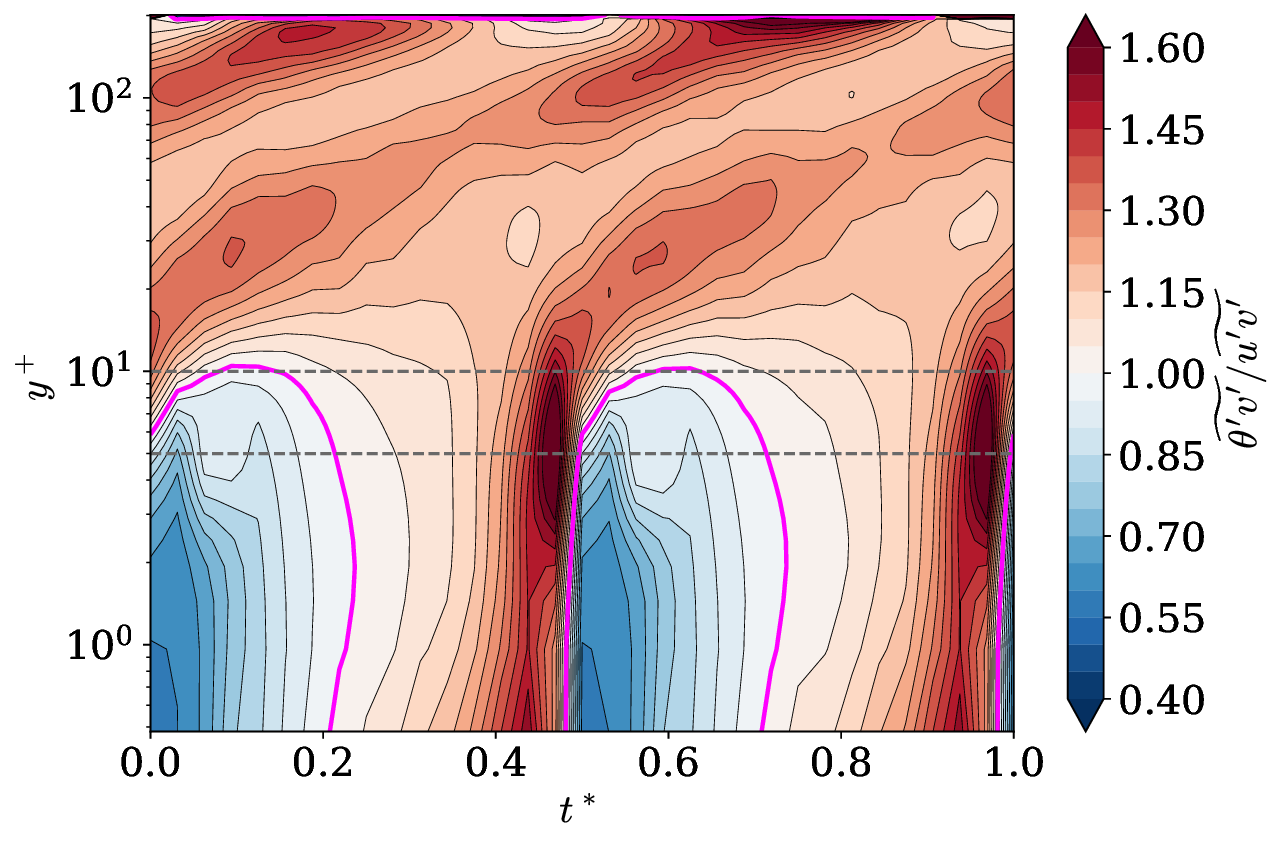}
		\vspace{0.2cm}
		\centerline{(\textit{d})}
	\end{minipage}
	\caption{Off-diagonal pressure terms and turbulent flux ratio at $T^+=350$, $W^+=30$. (\textit{a}) $-\Pi_{uv}$, the pressure-strain term in the $-\widetilde{u^{\dprime}v^{\dprime}}$ budget. (\textit{b}) $-\Pi_{v\theta}$, the pressure-strain term in the $-\widetilde{\theta^{\dprime}v^{\dprime}}$ budget; the positive values at $y^+\lesssim3$ are a viscous-sublayer feature outside the region of interest. (\textit{c}) Ratio $\Pi_{v\theta}/\Pi_{uv}$: values exceeding unity indicate a stronger pressure-strain drain of the scalar flux. (\textit{d}) Flux ratio $\widetilde{\theta^{\dprime}v^{\dprime}}/\widetilde{u^{\dprime}v^{\dprime}}$: values above unity indicate preferential wall-normal scalar flux enhancement. The magenta isoline marks the boundary where the ratio exceeds unity.}
	\label{fig:off_diag_pi}
\end{figure}

The phase-resolved fields of $-\Pi_{uv}$ and $-\Pi_{v\theta}$ are presented in figure~\ref{fig:off_diag_pi}\textit{a} and \textit{b}. Both quantities are predominantly negative in the buffer-layer region ($y^+ \approx 5$--$20$), confirming that the off-diagonal pressure-strain drains both flux magnitudes throughout the actuation cycle. In the unactuated flow, both $-\Pi_{uv}^+$ and $-\Pi_{v\theta}^+$ are negligible throughout the buffer layer; the pressure-strain activity observed under actuation is therefore attributable entirely to the interaction of the stochastic pressure field with the Stokes-layer-induced turbulence. The ratio $\Pi_{v\theta}/\Pi_{uv}$ (figure~\ref{fig:off_diag_pi}\textit{c}) is not, however, constant in time: it exceeds unity through much of the plateau and attains approximately $1.4$ in the cycle mean at $y^+=10$, signifying that the pressure-strain drains the scalar flux more strongly than the momentum flux. The net effect of the pressure on each flux combines this strain sink with the pressure-transport source of figure~\ref{fig:flux_budget}; the two oppose and nearly cancel, so that the maps of the net pressure contribution $-(\Pi+T_p)^+$ are nearly indistinguishable between the two fluxes (figure~\ref{fig:offdiag_phase}\textit{a},\textit{b}); their difference, $\Delta = -(\Pi+T_p)_{v\theta}^+ + (\Pi+T_p)_{uv}^+$, is negative through the plateau phases and in the cycle mean ($\overline{\Delta}\approx-0.12$), reverting to positive only transiently at the reversals (figure~\ref{fig:offdiag_phase}\textit{c},\textit{d}). On balance the off-diagonal pressure-strain is therefore a near-proportional return-to-isotropy sink that does \emph{not} favour the scalar flux (its marginally deeper action upon the scalar, in absolute terms, following the larger scalar flux that it scrambles), the excess of $-\widetilde{\theta^{\dprime}v^{\dprime}}$ over $-\widetilde{u^{\dprime}v^{\dprime}}$ originating not in the off-diagonal pressure but in the flux-budget production asymmetry $\widetilde{P}_{v\theta}>\widetilde{P}_{uv}$ (the mean-gradient asymmetry acting on the shared wall-normal fluctuation).

The temporal structure of this asymmetry is resolved more explicitly by the $t^*$-distributions extracted at $y^+ = 10$ (figure~\ref{fig:pi_1d}\textit{b}). The pressure-strain drains both fluxes throughout the plateau, the scalar sink $-\Pi_{v\theta}$ exceeding the momentum sink $-\Pi_{uv}$ in magnitude over its latter half, in accordance with the ratio $\Pi_{v\theta}/\Pi_{uv} > 1$ of figure~\ref{fig:off_diag_pi}\textit{c}. The net maintenance of each flux is set, however, by the balance of this sink against the production and pressure-transport sources of figure~\ref{fig:flux_budget}: the larger scalar pressure-transport source does not overcome the deeper scalar pressure-strain sink, so that the net off-diagonal pressure does not favour the scalar flux through the plateau and in the cycle mean (figure~\ref{fig:offdiag_phase}). The differential production, the mean-gradient asymmetry acting on the shared wall-normal fluctuation, is the only source-term asymmetry \emph{within the off-diagonal flux budget} that favours the scalar flux, and a modest one confined largely to the plateau, the off-diagonal flux productions being nearly equal at the reversals ($\widetilde{P}_{v\theta}/\widetilde{P}_{uv}\approx1$, rising to $\approx1.1$ in the plateau); the resulting flux excess (figure~\ref{fig:off_diag_pi}\textit{d}) is thus of production rather than off-diagonal-pressure origin (the production differential, integrated across the lower buffer layer, remains positive at $\approx+0.07$ in the cycle mean; figure~\ref{fig:offdiag_phase}\textit{d}), the net flux level reflecting the balance of this favouring production against the non-favouring, flux-proportional off-diagonal pressure sink and the remaining transport and dissipation terms. This off-diagonal route is, moreover, a subsidiary, plateau-confined contribution to the favouring: the larger surviving scalar flux $-\widetilde{v^{\dprime}\theta^{\dprime}}$ it sustains feeds the diagonal channel's variance-production lag $\widetilde{P}_{\theta\theta}/\widetilde{P}_{uu}$ (\S\ref{sec:results:production}), which dominates at the reversal. The characterisation of the off-diagonal pressure-strain as an opposition warrants a qualification, however. When each correlation is normalised by its own flux magnitude, the pressure-strain is found to scramble a near-constant fraction of each flux, $\Pi_{uv}/(-\widetilde{u^{\dprime}v^{\dprime}}) \approx \Pi_{v\theta}/(-\widetilde{\theta^{\dprime}v^{\dprime}}) \approx 0.6$ in the $y^+ = 5$--$10$ band, identifying it as a return-to-isotropy sink that scales with whatever flux is present rather than as a term that selectively penalises the scalar field. On this reading the deeper scalar pressure-strain sink, in absolute terms, is a consequence of the larger scalar flux (itself set by the production forcing through the mean-gradient asymmetry $\partial\widetilde{\theta}/\partial y > \partial\widetilde{u}/\partial y$ acting on the shared $\widetilde{v^{\dprime}v^{\dprime}}$) and not an independent agency acting against it; the proportionality holds most cleanly in the band average, the single station $y^+=10$ retaining a residual deeper scalar sink ($\approx 0.54$ against $0.43$).

This result may be contrasted with the earlier large-eddy simulations of \citet{fang_heat_2010}, who reported, for sinusoidal SWO at the substantially shorter period $T^+ \approx 104$ in the drag-reduction regime, that the quadrant contributions of $\overline{u^{\prime}v^{\prime}}$ and $\overline{\theta^{\prime}v^{\prime}}$ are modulated in concert by the actuation. The implication is that the differential flux enhancement documented in the present case is not a generic feature of the SWO mechanism, but emerges specifically when the actuation period is sufficient to accommodate extended plateau phases during which the self-sustaining process recommences and the differential production can act over a substantial fraction of the cycle; in the short-period, drag-reducing regime, the suppression of turbulence by the persistent Stokes strain precludes the activation of this differential channel, the analogy being thereby preserved.

The asymmetry in the response of the two correlations is admitted by a structural difference established through the rapid/slow decomposition of the fluctuating pressure employed in \S\,\ref{sec:results:pressure_strain} and developed further in appendix~\ref{app:rapid_slow}. The solenoidal projection couples the rapid pressure--velocity-gradient correlation to the mean field through a fourth-order kernel $M_{ijkl}$, whereas the rapid pressure--scalar-gradient correlation couples through a second-order kernel $G_{jk}$ (appendix~\ref{app:rapid_slow}), so that the two correlations are structurally free to respond in dissimilar fashion to a common forcing. This tensor-rank asymmetry does not, however, render either correlation magnitude-suppressed; $\Pi_{uu}$ is itself a substantial near-wall source under the same solenoidal constraint, as documented in \S\,\ref{sec:results:time_avg}. The direction observed, namely the deeper pressure-strain sink and the larger pressure-transport source acting on the scalar flux relative to the momentum flux during the active phases of the cycle, is accordingly an empirical result of the present DNS, the tensor-rank asymmetry establishing only that such a differential response is admissible and not its sign or net effect.

The consequence of this differential maintenance is rendered explicit by the flux ratio $\widetilde{\theta^{\dprime}v^{\dprime}}/\widetilde{u^{\dprime}v^{\dprime}}$ (figure~\ref{fig:off_diag_pi}\textit{d}), which exceeds unity in the buffer-layer region during the high-transport phases, confirming that the scalar flux is preferentially maintained at higher levels than the momentum flux. This flux ratio is a robust feature of the data; through the flux-to-production chain it propagates directly to a production dominance $\widetilde{P}_{\theta\theta} > \widetilde{P}_{uu}$, and subsequently to a sustained excess of $\widetilde{\theta^{\dprime}\theta^{\dprime}}$ over $\uupp$, consistent with the time-averaged variance results of \S\,\ref{sec:results:time_avg}. At the reversal, the diagonal drain $\Pi_{uu}<0$ depletes $\uupp$ and, through the attendant weakening of the quasi-streamwise vortices that carry the shear stress, curtails $\widetilde{P}_{uu}$, whilst the scalar flux, subject to no analogous pressure-strain drain, evolves unimpeded and so maintains $\widetilde{P}_{\theta\theta}$ and the growth of $\widetilde{\theta^{\dprime}\theta^{\dprime}}$; the compounding of diagonal momentum suppression with this unimpeded scalar evolution amplifies the production ratio spike beyond what the suppression of $\widetilde{P}_{uu}$ alone would produce.

The foregoing establishes that the scalar flux is preferentially maintained during the high-transport phases; it does not, however, by itself settle the attribution of the dominance observed at plateau onset, when turbulent structures remain weak following the preceding reversal and the ratio $\Pi_{v\theta}/\Pi_{uv}$ is close to unity. Two explanations are then admissible. The first is that the elevated production ratio at plateau onset is actively maintained, by a differential-production mechanism that has already begun to operate. The second, hereafter the residual-inheritance hypothesis, is that the pronounced thermal variance excess generated at the preceding reversal persists through the suppression phase by passive dissipation, the plateau-onset dominance being thus inherited rather than maintained. The two are distinguished by a timescale analysis of the phase-resolved scalar variance budget, which is presented in the remainder of the present subsection.

To distinguish between the two hypotheses, the rate at which the scalar variance would decay by dissipation alone, in the absence of any production activity, is compared with the rate of variance suppression observed during the post-reversal suppression interval. The characteristic thermal dissipation timescale,
\begin{equation}
	\tau_\theta = \frac{\widetilde{\theta^{\dprime}\theta^{\dprime}}_{\max}}{\widetilde{\varepsilon_{\theta\theta}}},
\end{equation}
is evaluated at $y^+ \approx 10$ at the phase of peak scalar variance preceding the suppression interval. From the present DNS data, $\widetilde{\theta^{\dprime}\theta^{\dprime}}_{\max} = 8.16$ and $\widetilde{\varepsilon_{\theta\theta}} = 0.57$, both expressed in wall units, producing $\tau_\theta = 14.2\,t^+$ in viscous wall-time units, equivalent to $\tau_\theta \approx 0.041\,T$, that is, a dissipative timescale short relative to the actuation period. With the suppression interval spanning $\Delta t_{\text{suppression}} = 0.129\,T = 45.15\,t^+$, the timescale ratio is $\tau_\theta/\Delta t_{\text{suppression}} \approx 0.32$, the dissipation being thus appreciably more rapid than the suppression interval is long. Free decay alone would accordingly reduce the scalar variance to $\exp(-45.15/14.2) \approx 4.2\%$ of its peak, amounting to a near-complete destruction of the variance by plateau onset. The measured scalar variance at the valley is, however, $50\%$ of the peak, exceeding the free-decay prediction by a factor of approximately $12$. This excess is attributed to production, which strongly sustains the scalar variance against the rapid dissipation throughout the suppression interval. The scalar variance during the suppression interval is therefore governed primarily by production, the dissipation being rapid rather than negligible.

The present result eliminates the residual-inheritance hypothesis, passive inheritance being precluded when the dissipation is sufficiently rapid to destroy the scalar variance within some fourteen viscous time units, and directly constrains the attribution of the plateau-phase production dominance. The scalar variance level at plateau onset is determined primarily by production activity during the suppression phase, not by dissipative relaxation from the preceding spike. Consequently, the 6\% excess of $\widetilde{P}_{\theta\theta}$ over $\widetilde{P}_{uu}$ sustained during the second half of the plateau cannot be attributed to passive inheritance; it must be actively maintained by an ongoing mechanism. The flux-magnitude budgets identify this ongoing mechanism as the differential production that preferentially maintains the scalar flux; the off-diagonal pressure-strain, with the ratio $\Pi_{v\theta}/\Pi_{uv}$ exceeding unity during the active plateau (figure~\ref{fig:off_diag_pi}\textit{c}), drains the scalar flux more strongly than the momentum flux in absolute terms; scrambling a near-constant fraction of each flux, however, it does not itself confer the preference. The two contributions to the production dominance are thus distinguished: the modest excess during the early plateau ($t^* \lesssim 0.13$) reflects the streamwise velocity production $\widetilde{P}_{uu}$ remaining depressed, the streamwise variance drained at the reversal by the diagonal $\Pi_{uu}$ being actively regenerated by the self-sustaining process over its recovery time, which is long relative to the scalar dissipation time $\tau_\theta$; this active recovery effect is distinct from, and is not to be confused with, the passive scalar-variance inheritance precluded by the timescale analysis above; the sustained $6\%$ dominance during the active plateau ($t^* \gtrsim 0.13$) is actively maintained by the ongoing differential production acting on the amplified wall-normal fluctuation, the off-diagonal pressure-strain a flux-proportional sink that does not favour it.

\begin{figure}
	\centering
	\begin{minipage}{0.48\textwidth}
		\centering
		\includegraphics[width=\textwidth]{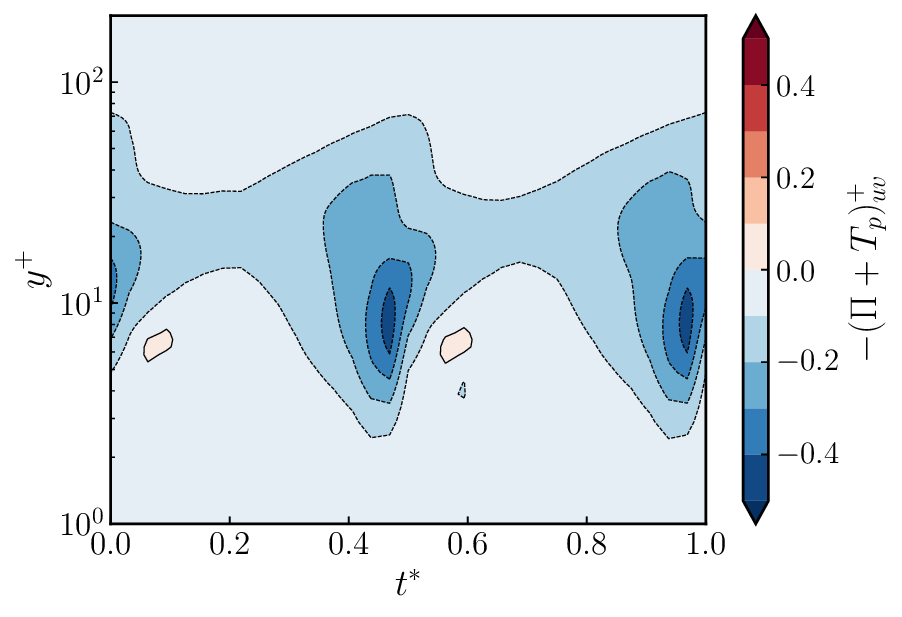}
		\vspace{0.15cm}
		\centerline{(\textit{a})}
	\end{minipage}
	\hfill
	\begin{minipage}{0.48\textwidth}
		\centering
		\includegraphics[width=\textwidth]{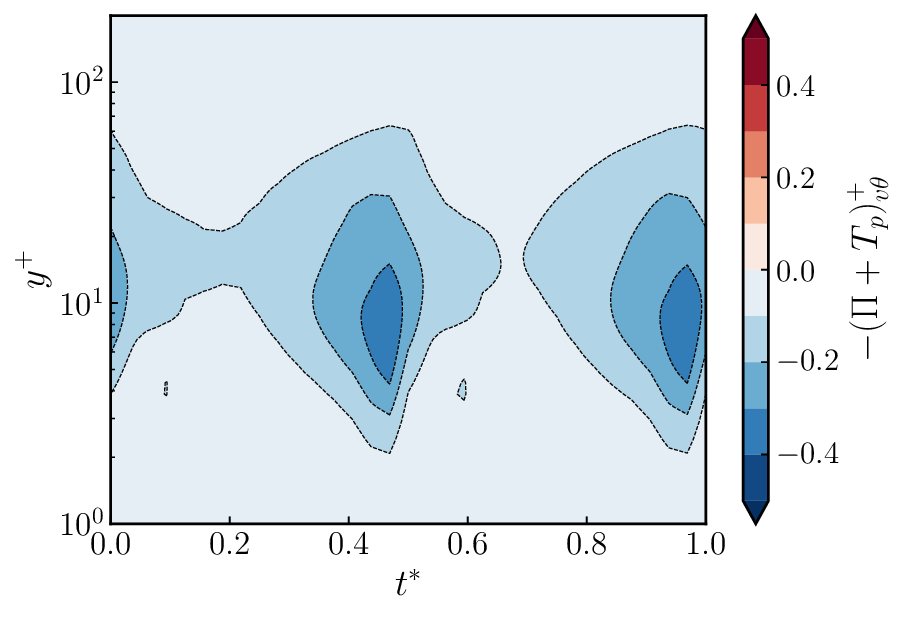}
		\vspace{0.15cm}
		\centerline{(\textit{b})}
	\end{minipage}
	\\[0.25cm]
	\begin{minipage}{0.48\textwidth}
		\centering
		\includegraphics[width=\textwidth]{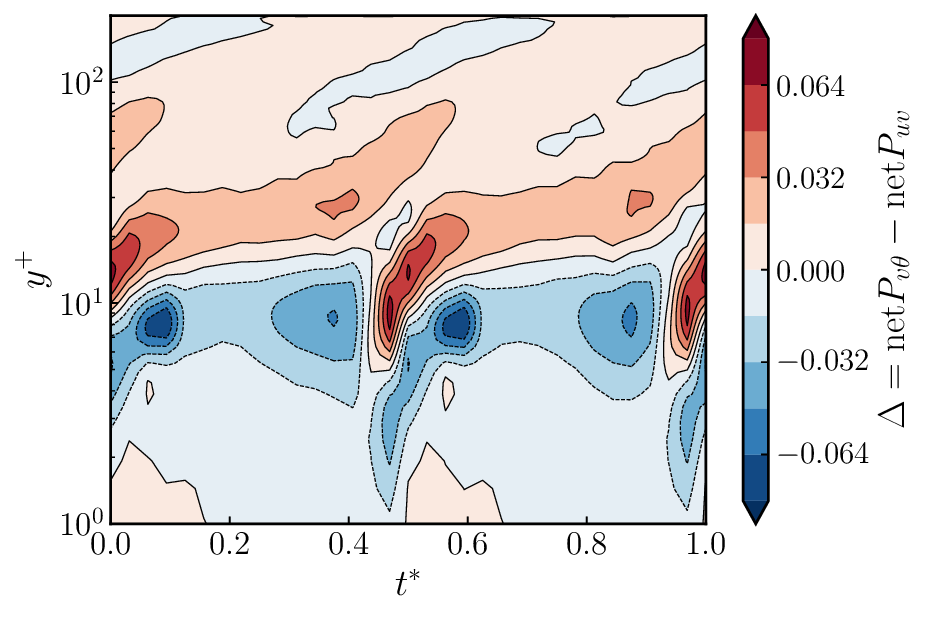}
		\vspace{0.15cm}
		\centerline{(\textit{c})}
	\end{minipage}
	\hfill
	\begin{minipage}{0.48\textwidth}
		\centering
		\includegraphics[width=\textwidth]{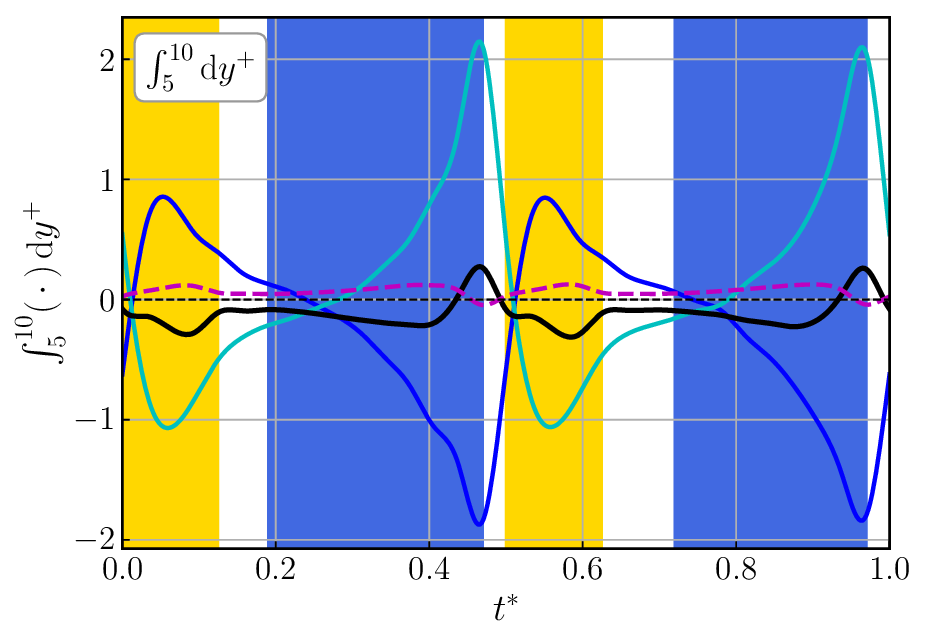}
		\vspace{0.15cm}
		\centerline{(\textit{d})}
	\end{minipage}
	\caption{Phase-resolved net off-diagonal pressure action on the two turbulent fluxes ($T^+=350$, $W^+=30$). (\textit{a},\textit{b}) Maps of the net pressure contribution to the flux-magnitude budgets, $-(\Pi+T_p)^+$, for $-\widetilde{u^{\dprime}v^{\dprime}}$ and $-\widetilde{v^{\dprime}\theta^{\dprime}}$ on a common scale. (\textit{c}) Their difference $\Delta=$\,net$P_{v\theta}-$net$P_{uv}$: red indicates a net pressure action favouring the scalar flux; values above the buffer layer ($y^+\gtrsim15$) lie outside the dynamically active region for DHT. (\textit{d}) Decomposition of $\Delta$ integrated across the lower buffer layer, $\int_5^{10}\mathrm{d}y^+$, into the pressure-strain differential $\Delta_{\mathrm{strain}}$ (blue), the pressure-transport differential $\Delta_{\mathrm{transport}}$ (cyan), their sum $\Delta$ (black), and the production differential (magenta); blue and gold bands mark plateau and reversal phases.}
	\label{fig:offdiag_phase}
\end{figure}